\documentclass[onecolumn]{aa}
\usepackage{psfig}
\def\simlt{\ \raise -2.truept\hbox{\rlap{\hbox{$\sim$}}\raise5.truept   %
\hbox{$<$}\ }}                                                          %
\def\simgt{\ \raise -2.truept\hbox{\rlap{\hbox{$\sim$}}\raise5.truept   %
\hbox{$>$}\ }}                                                          %

\def\be{\begin{equation}}
\def\ee{\end{equation}}
\def\newline{\hfil\break}
\newcommand{\npt}{$n_{e,rel}(\tilde{p}_1)\,$}

\newcommand{\prapp}{$P_{rel}/P_{th}\,$}
\begin{document}

   \title{The Non-Thermal Sunyaev--Zel'dovich Effect in Clusters of Galaxies}

   \author{S. Colafrancesco \inst{1},
           P. Marchegiani\inst{2} and E. Palladino\inst{2} }

   \offprints{S. Colafrancesco, e-mail: cola@coma.mporzio.astro.it}

   \institute{INAF - Osservatorio Astronomico di Roma
              via Frascati 33, I-00040 Monteporzio, Italy.\\
              Email: cola@coma.mporzio.astro.it
    \and
             Dipartimento di Fisica, Universit\`a di Roma ``La Sapienza'', Piazzale
             A. Moro 2, I-000 Roma, Italy
             }

\date{Received 21 May 2002 / Accepted 13 August 2002}

\authorrunning {Sergio Colafrancesco et al.}
\titlerunning {Non-Thermal SZ effect in clusters}

\abstract{ In this paper we provide an general  derivation of the non-thermal
Sunyaev-Zel'dovich (SZ) effect in galaxy clusters which is exact in the Thomson
limit to any approximation order in the optical depth $\tau$. The general
approach we use allows also to obtain an exact derivation of the thermal SZ
effect in a self-consistent framework. Such a general derivation is obtained
using the full relativistic formalism and overcoming the limitations of the
Kompaneets and of the single scattering approximations. We compare our exact
results with those obtained at different approximation orders in $\tau$ and we
give estimates of the precision fit. We verified that the third order
approximation yields a quite good description of the spectral distortion induced
by the Comptonization of CMB photons in the cluster atmosphere. In our general
derivation, we show that the spectral shape of the thermal and non-thermal SZ
effect depends not only from the frequency but also from the cluster parameters,
like the electron pressure and the optical depth and from the energy spectrum of
the electron population. We also show that the spatial distribution of the
thermal and non-thermal SZ effect in clusters depends on a combination of the
cluster parameters and on the spectral features of the effect. To have a
consistent description of the SZ effect in clusters containing non-thermal
phenomena, we also evaluate in a consistent way - for the first time - the total
SZ effect produced by a combination of thermal and non-thermal electron
population residing in the same environment, like is the case in radio-halo
clusters. In this context we show that the location of the zero of the total SZ
effect increases non-linearly with increasing values of the pressure ratio
between the non-thermal and thermal electron populations, and its determination
provides a unique way to determine the pressure of the relativistic particles
residing in the cluster atmosphere. We discuss, in details, both the spectral
and the spatial features of the total (thermal plus non-thermal) SZ effect and
we provide specific predictions for a well studied radio-halo cluster like
A2163. Our general derivation allows also to discuss the overall SZ effect
produced by a combination of different thermal populations residing in the
cluster atmosphere. Such a general derivation of the SZ effect allows to
consider also the CMB Comptonization induced by several electron populations. In
this context, we discuss how the combined observations of the thermal and
non-thermal SZ effect and of the other non-thermal emission features occurring
in clusters (radio-halo, hard X-ray and EUV excesses) provide relevant
constraints on the spectrum of the relativistic electron population and, in
turn, on the presence and on the origin of non-thermal phenomena in galaxy
clusters. We finally discuss how SZ experiments with high sensitivity and wide
spectral coverage, beyond the coming PLANCK satellite, can definitely probe the
presence of a non-thermal SZ effect in galaxy clusters and disentangle this
source of bias from the cosmologically relevant thermal SZ effect.


}

\maketitle


\section{Introduction}

Compton scattering of the Cosmic Microwave Background (CMB)  radiation by hot
Intra Cluster (hereafter IC) electrons -- the Sunyaev Zel'dovich effect
(Zel'dovich and Sunyaev 1969, Sunyaev \& Zel'dovich 1972, 1980) -- is an
important process whose spectral imprint on the CMB can be used as a powerful
astrophysical and cosmological probe (see Birkinshaw 1999 for a review). Such a
scattering produces a systematic shift of the CMB photons from the
Rayleigh-Jeans (RJ) to the Wien side of the spectrum.

An approximate description of the scattering of an isotropic Planckian radiation field by a non-relativistic
Maxwellian electron population can be obtained by means of the solution of the Kompaneets (1957) equation (see,
e.g., Sunyaev \& Zel'dovich 1980). The resulting change in the spectral intensity, $\Delta I_{th}$, due to the
scattering of CMB photons by a {\it thermal} electron distribution can be written as
\be
\Delta I_{th} = 2{(k_BT_0)^3 \over (hc)^2} y_{th} g(x) ~, \ee where $x = h \nu / k_BT_0$ is the a-dimensional
frequency and the spectral shape of the effect is contained in the function
\be
g(x) = {x^4 e^x \over (e^x -1)^2} \bigg[x {e^x +1 \over e^x -1} -4\bigg]
\ee
 which is zero at the frequency $x_0=3.83$ (or $\nu
= 217$ GHz for a value of the CMB temperature $T_0 = 2.726$ K), negative at $x < x_0$ (in the RJ side) and
positive at $x
> x_0$ (in the Wien side). The Comptonization parameter, $y_{th}$, due to the thermal SZ effect  is given by
\be
y_{th}= {\sigma_T \over m_e c^2} \int d\ell ~n_e k_BT_e ~, \label{y_th} \ee
where $n_e$ and $T_e$ are the electron density and temperature of the IC gas,
respectively, $\sigma_T$ is the Thomson cross section, valid in the limit $T_e
\gg T_0$, $k_B$ is the Boltzmann constant and $m_ec^2$ is the rest mass energy
of the electron. The Comptonization parameter is proportional to the integral
along the line of sight $\ell$ of the kinetic pressure, $P_{th} = n_e k_B T_e$,
of the IC gas. Thus, the previous Eq.(\ref{y_th}) can be written as
\be
y_{th} = {\sigma_T \over m_e c^2} \int d\ell ~P_{th} ~, \ee where the relevant
dependence from the total kinetic pressure, $P_{th}$, of the clusters along the
line of sight $\ell$ appears.\\
 The previous description of the thermal SZ effect is obtained under the Kompaneets (1957) approximation and in
the single scattering regime of the true photon redistribution function. As
such, it only provides an approximation of the SZ effect in galaxy clusters for
low temperatures ($T_e \simlt 3$ keV) and low optical depth ($\tau = \sigma_T
\int d\ell n_e \simlt 10^{-3}$).

However, recent X-ray observations have revealed the existence of many
high-temperature clusters (David et al. 1993; Arnaud et al. 1994; Markevitch et
al. 1994; Mushotzky and Scharf 1997) with $k_B T_e$ up to $\sim 17$ keV (see,
e.g., Tucker et al. 1998). Thus, the calculation of the thermal SZ effect from
these hot clusters requires to take into account the appropriate relativistic
corrections (see Birkinshaw 1999 and references therein).\\
 Analytical expressions for
the SZ effect in the relativistic case have been considered by various authors
(Stebbins 1997; Itoh et al. 1998; Challinor \& Lasenby 1998; see also Birkinshaw
1999 and references therein). Some of these calculations (out of the Monte Carlo
simulations) are still approximate since they have been carried out in the
following limits: {\it i)} single scattering of CMB photons against the IC gas
electrons; {\it ii)} diffusion limit in which the use of the Kompaneets equation
is justified. Itoh et al. (1998) obtained higher-order relativistic corrections
to the thermal SZ effect in the form of a Fokker-Planck expansion (generalised
Kompaneets equation) and by direct integration of the Boltzmann collision term.
Such an integration has been carried out in the full relativistic regime. So,
the results of Itoh et al. (1998) can be regarded as exact in the framework of
the single scattering approximation. Analytical fitting formulae of such
derivation can be found in Nozawa et al. (2000) and Itoh et al. (2002).
Such a formalism seems to offer a detailed description of the thermal SZ effect
for $k_B T_e \simlt 15 $ keV, while Monte Carlo simulations describe more
correctly the thermal SZ effect even for $k_B T_e \simgt 20$ keV (see, e.g.,
Challinor and Lasenby 1998). Recently, Dolgov et al. (2001) proposed an approach
based on an analytical reduction of the collision integral which is contained in
the Boltzmann-like collision equation for the SZ effect produced by a thermal
population of electrons. They found a numerical solution for the thermal SZ
effect which is valid for generic values of $\tau$ and $T_e$.
Sazonov and Sunyaev (2000) presented a derivation of the monochromatic
redistribution function in the mildly relativistic limit which considers also
quantum effects and the use of the Klein-Nishina cross-section. However, they
still consider only the single Compton scattering limit and the relativistic
corrections up to some intermediate order due to low-energy photons and
relativistic electrons. Such derivation of the monochromatic redistribution
function in the limit $h \nu \ll k_BT_e$ reproduces the results of Fargion et
al. (1997).
More recently, Itoh et al. (2001) also presented a calculation of the thermal SZ
effect which considers the contribution from multiple scattering in the
relativistic limit. These last authors concluded that the multiple scattering
contribution to the thermal SZ effect is negligible in galaxy clusters.

Another general assumption which is made in the calculation of the SZ effect is
the use of a single population of thermal electrons which constitutes the hot
(with temperature $T_e \sim 10^7-10^8$ K), optically thin (with density $n_e
\sim 10^{-3} - 10^{-2}$ cm$^{-3}$, and size $R\sim$ a few Mpc)  Intra Cluster
Medium (hereafter ICM). This assumption is based on the evidence that the IC gas
is mainly constituted by thermal electrons (and protons) which are responsible
for the X-ray emission observed in many clusters through thermal bremsstrahlung
radiation  (see Sarazin 1988 for a review).

Nonetheless, in addition to the thermal IC gas, many galaxy clusters contain a
population of relativistic electrons which produce a diffuse radio emission
(radio halos and/or relics) via synchrotron radiation in a magnetized ICM (see,
e.g., Feretti 2000 for a recent observational review). The electrons which are
responsible for the radio halo emission must have energies $E_e \simgt$ a few
GeV to radiate at frequencies $\nu \simgt 30$ MHz in order to reproduce the main
properties of the observed radio halos (see, e.g., Blasi \& Colafrancesco 1999;
Colafrancesco \& Mele 2001,  and references therein). A few nearby clusters also
show the presence of an EUV/soft X-ray excess (Lieu et al. 1999, Kaastra et al.
1999, 2002; Bowyer 2000) and of an hard X-ray excess (Fusco-Femiano et al.
1999-2000; Rephaeli et al. 1999; Kaastra et al. 1999; Henriksen 1999) over the
thermal bremsstrahlung radiation. These emission excesses over the thermal X-ray
emission may be produced either by Inverse Compton Scattering (hereafter ICS) of
CMB photons off an additional population of relativistic electrons or by a
combination of thermal (reproducing the EUV excess, Lieu et al. 2000) and
suprathermal (reproducing the hard X-ray excess by non-thermal bremsstrahlung,
Blasi et al. 2000, Dogiel 2000, Sarazin \& Kempner 2000) populations of distinct
origins.
However, since the inefficient non-thermal bremsstrahlung mechanism would
require a large energy input and thus imply an excessive heating of the IC gas
 - which is not observed - a more complex electron population is
required to fit both the radio halo and the EUV/hard X-ray spectra of galaxy
clusters (Petrosian 2001). The high-energy part ($E \simgt 1 $ GeV) of such
spectrum contains non-thermal electrons and produce synchrotron radio emission
which can fit the observed radio-halo features: this is the best studied region
of the non-thermal spectrum in galaxy clusters.

Indeed, in many of the clusters in which the SZ effect has been detected there
is also evidence for radio halo sources (see Colafrancesco 2002). So it is of
interest to assess whether the detected SZ effect is, in fact, mainly produced
by the thermal electron population or there is a relevant contribution from
other non-thermal electron populations. We will show specifically in this paper
that each one of the electron populations which reside in the cluster atmosphere
produces a distinct SZ effect with peculiar spectral and spatial features.

The description of the non-thermal SZ effect produced by a single electron
population with a non-thermal spectrum has been attempted by various authors
(McKinnon et al. 1991; Birkinshaw 1999; Ensslin \& Kaiser 2000; Blasi et al.
2000). Several limits to the non-thermal SZ effect are available in the
literature (see, e.g., Birkinshaw 1999 for a review) from observations of galaxy
clusters which contain powerful radio halo sources (such as A2163) or radio
galaxies (such as A426), but only a few detailed analysis of the results (in
terms of putting limits to the non-thermal SZ effect) have been possible so far.
Only a single devoted search for the SZ effect expected from a relativistic
population of electrons in the lobes of bright radio galaxies has been attempted
to date (McKinnon et al. 1991). No signals were seen and a detailed spectral fit
of the data to separate residual synchrotron and SZ effect signals was not done,
so that the limits on the SZ effect do not strongly constrain the electron
populations in the radio lobes.
Also the problem of detecting the non-thermal SZ effect in radio-halo clusters
is likely to be severe because of the associated synchrotron radio emission. In
fact, at low radio frequencies, such a synchrotron emission could easily
dominate over the small negative signal produced by the SZ effect. At higher
frequencies there is in principle more chance to detect the non-thermal SZ
effect, but even here there are likely to be difficulties in separating the SZ
effect from the flat-spectrum component of the synchrotron emission (Birkinshaw
1999).

From the theoretical point of view, preliminary calculations (Birkinshaw 1999,
Ensslin \& Kaiser 2000, Blasi et al. 2000) of the non-thermal SZ effect have
been carried out in the diffusion approximation ($\tau \ll 1$), in the limit of
single scattering and for a single non-thermal population of electrons.
Specifically, Ensslin \& Kaiser (2000) and Blasi et al. (2000) considered the SZ
effect produced, under the previous approximations, by a supra-thermal tail of
the Maxwellian electron distribution claimed to exist in the Coma cluster and
concluded that the SZ effect, even though of small amplitude, could be
measurable in the sub-mm region by the next coming PLANCK experiment. However,
Petrosian (2001) showed that the suprathermal electron distribution faces with
several crucial problems, the main being the large heating that such electrons
would induce through Coulomb collisions in the ICM. The large energy input of
the suprathermal distribution in the ICM of Coma would heat the IC gas up to
unreasonably high temperatures, $k_BT_e \sim 10^{16}$ K, which are not observed.
In addition, Colafrancesco (2002) noticed that dust obscuration does not allow
any detection of the SZ signal from Coma at frequencies $\simgt 600$ GHz.

Matters are significantly more complicated if the full relativistic formalism is
used. However, this is necessary, since many galaxy clusters show extended radio
halos and the electrons which produce the diffuse synchrotron radio emission are
certainly highly relativistic so that the use of the Kompaneets approximation is
invalid. Moreover, the presence of thermal and non-thermal electrons in the same
location of the ICM  renders the single scattering approximation and the single
population approach unreasonable, so that the treatment of multiple scattering
among different electronic populations coexisting in the same cluster atmosphere
is necessary to describe correctly the overall SZ effect.

Here we derive the spectral and spatial features of the SZ effect using an exact
derivation of the spectral distortion induced by a combination of a thermal and
a non-thermal population of electrons which are present at the same time in the
ICM. Lately, we also consider the case of the combination of several non-thermal
and thermal populations. The plan of the paper is the following: in Sect. 2 we
will provide a general derivation of the SZ effect in an exact formalism within
the framework of our approach, i.e., considering the fully relativistic approach
outlined in Birkinshaw (1999) and the effect induced by multiple scattering. We
work here in the Thomson limit $h \nu \ll m_e c^2$. Within such framework, we
derive the exact SZ effect for a single electron population both in the thermal
and non-thermal cases. In Sect. 3 we derive the SZ effect produced by a single
population of non-thermal electrons providing both exact and approximate (up to
any order in $\tau$) expressions for the non-thermal SZ effect. In Sect. 4 we
derive the total SZ effect produced by a combination of two populations: a
distribution of thermal electrons - like that responsible for the X-ray emission
of galaxy clusters - and a non-thermal electron distribution - like the one
responsible for the radio halo emission which is present in many galaxy
clusters. Our general approach allows to derive the expression for the SZ effect
produced by the combination of any electronic population. So, in Sect. 5 we
consider also the SZ effect generated by the combination of two different
thermal populations. In Sect. 6 we discuss the spatial features associated to
the presence of a non-thermal SZ effect superposed to the thermal SZ effect. In
Sect. 7 we derive limits on the presence and amplitude of the non-thermal SZ
effect discussing specifically the cases of a few clusters in which there is
evidence for the presence of non-thermal, high-energy electrons. We show how the
possible detection of a non-thermal SZ effect can set relevant constraints on
the relative electron population. We summarize our results and discuss our
conclusions in the final Sect. 8. We use $H_0 = 50$ km s$^{-1}$ Mpc$^{-1}$ and
$\Omega_0=1$ throughout the paper unless otherwise specified.

%
%

\section{The SZ effect for galaxy clusters: a generalized approach}

In this section we derive a generalized expression for the SZ effect which is
valid in the Thomson limit for a generic electron population in the relativistic
limit and includes also the effects of multiple scattering. First we consider an
expansion in series for the distorted spectrum $I(x)$ in terms of the optical
depth, $\tau$, of the electron population. Then, we consider an exact derivation
of the spectral distortion using the Fourier Transform method already outlined
in Birkinshaw (1999).

An electron with momentum $p=\beta \gamma$, with $\beta= v/c$ and
$\gamma=E_e/m_ec^2$ increases the frequency $\nu$ of a scattered CMB photon on
average by the factor $t \equiv \nu' / \nu = {4 \over 3} \gamma^2 - {1 \over
3}$, where $\nu'$ and $\nu$ are the photon frequencies after and before the
scattering, respectively. Thus, in the Compton scattering against relativistic
electrons ($\gamma \gg 1$) a CMB photon is effectively removed from the CMB
spectrum and is found at much higher frequencies. We work here in the Thomson
limit, (in the electron's rest frame $\gamma h\,\nu \ll \,m_{\rm e}\,c^2$),
which is valid for the interesting range of frequencies at which SZ observations
are feasible.\\
 The redistribution function of the CMB photons scattered once by the IC
electrons writes in the relativistic limit as,
 \be \label{p1s}
 P_1(s)= \int_0^{\infty} dp f_e(p) P_s(s;p)~,
 \ee
where $f_e(p)$ is the electron momentum distribution and $P_s(s;p)$ is the
redistribution function for a mono-energetic electron distribution, with $s
\equiv ln(t)$. An analytical expression for the redistribution function of CMB
photons which suffer a single scattering, $P_s(s;p)$, has been given by Ensslin
\& Kaiser (2000). The expression for such a redistribution function has been
also given analytically by Fargion et al. (1997) and by Sazonov \& Sunyaev
(2000). Once the function $P_1(s)$ is known, it is possible to evaluate the
probability that a frequency change $s$ is produced by a number $n$ of repeated,
multiple scattering. This is given by the repeated convolution
 \be
 \label{pn.generica}
P_n(s) = \int_{-\infty}^{+\infty} ds_1 \ldots ds_{n-1} P_1(s_1) \ldots P_1(s_{n-1}) P_1(s-s_1- \ldots -s_{n-1})
 \equiv \underbrace{P_1(s) \otimes \ldots \otimes P_1(s)}_{\mbox{n times}}
\ee (see Birkinshaw 1999), where the symbol $\otimes$ indicates each convolution
product. As a result, the location of the maximum of the function $P_n(s)$ moves
towards higher values of $s$ for higher values of $n$ and the distribution
$P_n(s)$ widens a-symmetrically towards high values of $s$ giving thus higher
probabilities to have large frequency shifts. The resulting total redistribution
function $P(s)$ can be written as the sum of all the functions $P_n(s)$, each
one weighted by the probability that a CMB photon can suffer $n$ scatterings,
which is assumed to be Poissonian with expected value $\tau$:
\begin{eqnarray} \label{pstot}
 P(s)&=&\sum_{n=0}^{+\infty} \frac{e^{-\tau} \tau^n}{n!} P_n(s)
\nonumber \\
 &=&e^{-\tau} \left[P_0(s)+ \tau P_1(s) + \frac{1}{2} \tau^2 P_2(s)+
\ldots\right] \nonumber \\
 &=&e^{-\tau} \left[\delta(s) + \tau P_1(s) + \frac{1}{2} \tau^2 P_1(s)
\otimes P_1(s) + \ldots \right] ~.
\end{eqnarray}
The spectrum of the Comptonized radiation is then given by
\be
I(x) = \int_{- \infty}^{+\infty} ds~I_0(x e^{-s}) P(s) ~,
 \label{idx}
 \ee
 where
\begin{equation}
\label{spettro_inc} I_0(x)=2\frac{(k_B T_0)^3}{(hc)^2} \frac{x^3}{e^x-1}
\end{equation}
is the incident CMB spectrum in terms of the a-dimensional frequency $x$.

In the following we will derive the expression for the distorted spectrum using
first an expansion in series of $\tau$ and then the exact formulas obtained with
the Fourier Transform (FT) method.

\subsection{High order $\tau$-expansion}

In the calculation of the SZ effect in galaxy clusters it is usual to use the
expression of $P(s)$ which is derived in the single scattering approximation and
in the diffusion limit, $\tau \ll 1$. In these limits, the distorted spectrum
writes, in our formalism, as:
\be
I(x) = J_0(x) + \tau \bigg[  J_1(x) - J_0(x) \bigg] ~,
\ee
where
\begin{eqnarray} \label{j1.j0}
J_0(x)&=&\int_{-\infty}^{+\infty} I_0(xe^{-s}) P_0(s) ds =
\int_{-\infty}^{+\infty} I_0(xe^{-s}) \delta(s) ds=I_0(x) ~,\\
 J_1(x)&=&\int_{-\infty}^{+\infty} I_0(xe^{-s}) P_1(s) ds~.
\end{eqnarray}
To evaluate the SZ distorted spectrum $I(x)$ up to higher order in $\tau$, we
make use of the general expression of the series expansion of the function
$P(s)$
\begin{equation}
\label{sviluppo generale} P(s)=\sum_{n=0}^{+\infty} a_n(s) \tau^n ~,
\end{equation}
which can be written, using Eq.(\ref{pstot}), as
\begin{equation}
P(s)=\sum_{k=0}^{+\infty} \frac{(-\tau)^k}{k!} \sum_{k'=0}^{+\infty} \frac{\tau^{k'}}{k'!} P_{k'}(s) ~.
\label{sviluppo_pds}
\end{equation}
The general $n$-th order term is obtained by selecting the terms in the double
summation which contain the optical depth $\tau$ up to the $n$-th power. These
terms are obtained for $k'=n-k$, and provide the following expression for the
series expansion coefficients:
\begin{equation}
\label{coeff.svil} a_n(s)=\sum_{k=0}^n \frac{(-1)^k}{k!(n-k)!} P_{n-k}(s) = \frac{1}{n!} \sum_{k=0}^n
\left(\begin{array} {c} n \\ k
\end{array} \right) (-1)^k P_{n-k}(s).
\end{equation}
Inserting the coefficients $a_n(s)$ given in Eq.(\ref{coeff.svil}) in
Eq.(\ref{sviluppo generale}) and this last one in Eq.(8), the resulting spectral
distortion due to the SZ effect can be written as:
\begin{equation}
\label{espansione_I}
I(x)=\sum_{n=0}^{+\infty} b_n(x) \tau^n ~,
\end{equation}
where the coefficients $b_n(x)$ are given by
\begin{equation} \label{coeff.bn}
b_n(x)=\frac{1}{n!} \sum_{k=0}^n  \left(\begin{array} {c} n \\ k
\end{array} \right) (-1)^k J_{n-k}(x) ~,
\end{equation}
and
\begin{equation} \label{jn}
 J_{n-k}(x)=\int_{-\infty}^{+\infty} I_0(xe^{-s}) P_{n-k}(s) ds.
\end{equation}
In this general approach, we can write the corrections to the distorted spectrum
$I(x)$ up to any order $n$ in the quantity $\tau$.

\subsection{Exact derivation of the SZ effect}

The redistribution function $P(s)$ can also be obtained in an exact form
considering all the terms of the series expansion in Eq.(\ref{sviluppo_pds}). In
fact, since the Fourier transform (hereafter FT) of a convolution product of two
functions is equal to the product of the Fourier transforms of the two
functions, the FT of $P(s)$ writes as
\begin{equation}
\label{ptrasf} \tilde{P}(k)=e^{-\tau} \left[1+\tau \tilde{P}_1(k)+\frac{1}{2} \tau^2
\tilde{P}_1^2(k)+\ldots\right]=e^{-\tau}e^{\tau \tilde{P}_1(k)}=e^{-\tau[1-\tilde{P}_1(k)]} ~,
\end{equation}
 where
\begin{equation}
\tilde{P}_1(k)=\int_{-\infty}^{+\infty} P_1(s) e^{-iks} ds~
\end{equation}
(Taylor \& Wright 1989). The exact form of the Comptonized spectrum $I(x)$ is
then given by Eq.(\ref{idx})
in terms of the exact redistribution function
 \be
P(s) = {1 \over 2 \pi} \int_{-\infty}^{+\infty} \tilde{P}(k) e^{iks} dk
 \ee
which is obtained as the anti Fourier transform of $\tilde{P}(k)$ given in
Eq.(\ref{ptrasf}). To compare the exact calculations of the SZ spectral
distortion with those obtained in the non-relativistic limit [see eqs.(1-2) as
for the thermal case], it is useful to write the distorted spectrum in the form
\begin{equation}
  \Delta I(x)=2\frac{(k_B T_0)^3}{(hc)^2}y \tilde{g}(x) ~,
\end{equation}
where $\Delta I(x) \equiv I(x) - I_0(x)$.
 The spectral shape of the SZ effect is contained in the function
\begin{equation}
 \tilde{g}(x)=\bigg( \frac{\Delta I}{I_0}\bigg) \frac{1}{y} \frac{x^3}{e^x-1}=\frac{\Delta i(x)}{y}
\end{equation}
where $\Delta i\equiv \Delta I \frac{(hc)^2}{2(k_B T_0)^3}$. The Comptonization
parameter $y$ is defined, in our general approach, in terms of the pressure $P$
of the considered electron population:
\begin{equation}
\label{ygen}
  y=\frac{\sigma_T}{m_e c^2}\int P d\ell ~.
\end{equation}

\subsection{The case of a single thermal electron population}

For a thermal electron population in the non-relativistic limit with momentum distribution
 $f_{e,th} \propto p^2 exp(-\eta \sqrt{1+p^2})$ with $\eta= m_e c^2/k_B T_e$,
one writes the pressure as
\begin{equation}\label{press termica}
  P_{th}=n_e k_B T_e
\end{equation}
and it is easy from Eq.(\ref{ygen}) to re-obtain the Compton parameter in Eq.(3)
as
\begin{equation}
  y_{th}=\frac{\sigma_T}{m_e c^2}\int n_e k_B T_e d\ell =\tau \frac{k_B T_e}{m_e
  c^2} ~
\end{equation}
(we consider here, for simplicity, an isothermal cluster).
The relativistically correct expression of the function $\tilde{g}(x)$ for a
thermal population of electrons writes, at first order in $\tau$, as:
\begin{equation} \label{gtilde1}
 \tilde{g}(x)=\frac{\Delta i}{y_{th}}=\frac{\tau [j_1-j_0]}{\tau \frac{k_B T_e}{m_e
 c^2}}= \frac{m_e c^2}{k_B T_e}[j_1-j_0] ~,
\end{equation}
where $j_i \equiv J_i \frac{(hc)^2}{2(k_B T_0)^3}$ and the functions $J_i$ are
given in Eq.(18).

\begin{figure}[h]
\begin{center}
\psfig{file=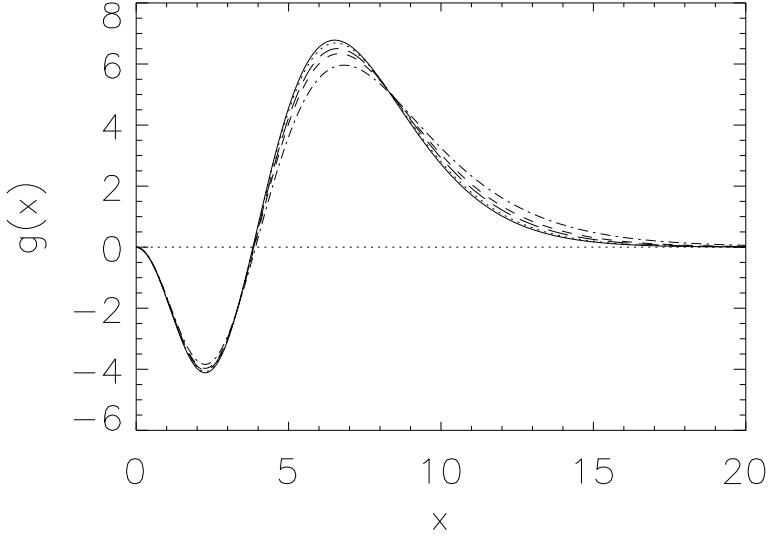,width=12.cm,angle=0.}
  \caption{\footnotesize{The function $g(x)$ (solid line) is compared with the function $ \tilde{g}(x)$
  for thermal electron populations with $k_B T_e=10$ (dot-dashed), $5$ (dashes),
  $3$ (long dashes) and  $1$ (dotted) keV, respectively. }}
  \label{fig.gtilde}
\end{center}
\end{figure}

\noindent In the same line, it is  possible to write the expression of
$\tilde{g}(x)$ up to higher orders in $\tau$. For example, limiting the series
expansion in Eq.(14) at third order in $\tau$, we found:
\begin{equation}
  \tilde{g}(x)=\frac{m_e c^2}{ k_B T_e }
  \left[(j_1-j_0)+\frac{1}{2}\tau(j_2-2j_1+j_0)+\frac{1}{6}\tau^2
  (j_3-3j_2+3j_1-j_0)\right].
\end{equation}
Notice that while the expression derived for $\tilde{g}(x)$ at first order
approximation in $\tau$ is independent of $\tau$, the expression for
$\tilde{g}(x)$ at higher order approximation in $\tau$ depends directly on
$\tau$. This is even more the case for the exact expression of $\tilde{g}(x)$
given in the following. In fact, using the exact form of the function  $P(s)$
given in Eq.(22), it is possible to write the exact form of the function
$\tilde{g}(x)$ as:
\begin{equation}
\tilde{g}(x)=\frac{m_e c^2}{k_B T_e } \left\{ \frac{1}{\tau} \left[\int_{-\infty}^{+\infty} i_0(xe^{-s}) P(s) ds-
i_0(x)\right] \right\}. \label{gtilde_esatta}
\end{equation}
The expression of $\tilde{g}(x)$ approximated at first order in $\tau$ as given
by Eq.(\ref{gtilde1}) is the one to compare directly with the expression of
$g(x)$ obtained from the Kompaneets (1957) equation, since both are evaluated
under the assumption of single scattering suffered by a CMB photon against the
IC electrons. Fig.\ref{fig.gtilde} shows how the function  $ \tilde{g}(x)$ tends
to $g(x)$ for lower and lower IC gas temperatures $T_e$. This confirms that the
distorted spectrum obtained from the Kompaneets equation is the non-relativistic
limit of the exact spectrum. Notice that the function $\tilde{g}(x)$ given in
 Eq.(\ref{gtilde_esatta}) is
the spectral shape of the SZ effect obtained in the exact calculation while the function $g(x)$ refers to the
1$^{st}$ order approximated case of a single, thermal, non-relativistic population of electrons.

It is worth to notice that while in the non-relativistic case it is possible to
separate the spectral dependence of the effect [which is contained in the
function $g(x)$] from the dependence on the cluster parameters [which are
contained in Compton parameter $y_{th}$, see eqs. (1-3)], this is no longer
valid in the relativistic case in which the function $J_1$ depends itself also
on the cluster parameters. Specifically, for a thermal electron distribution,
$J_1$ depends non-linearly from the electron temperature $T_e$ through the
function $P_1(s)$. This means that, even at first order in $\tau$, the spectral
shape $\tilde{g}(x)$ of the SZ effect depends on the cluster parameters, and
mainly from the electron pressure $P_{th}$.

To evaluate the errors done by using the non-relativistic expression $g(x)$
instead of the relativistic, correct function $ \tilde{g}(x)$ given in
Eq.(\ref{gtilde_esatta}) we calculate the fractional error $\varepsilon=\left|
[g(x)- \tilde{g}(x)] / \tilde{g}(x) \right|$ for thermal populations with $k_B
T_e=$ 10, 5, 3, 2 and 1 keV and for three representative frequencies (see Table
1). The fractional errors in Table 1 tend to decrease systematically at each
frequency with decreasing temperature. Note, however, that the error found in
the high frequency region, $x=15$, is much higher than the errors found at lower
frequencies and produces uncertainty levels of $\simgt 50 \%$ for $T_e \simgt 5$
keV. This indicates that the high frequency region of the SZ effect is more
affected by the relativistic corrections and by multiple scattering effects.
\begin{table}[htb]
\begin{center}
\begin{tabular}{|l|*{5}{c|}}
\hline
 & $k_BT_e=10\,keV$ & $k_BT_e=5\,keV$ & $k_BT_e=3\,keV$ & $k_BT_e=2\,keV$ & $k_BT_e=1\,keV$\\
 \hline
 $x=2.3$ & 7.20 & $3.68$ & $2.26$ & $1.57$ & $0.93$\\
 $x=6.5$ & 14.81 & $7.28$ & $4.33$ & $2.87$ & $1.40$\\
 $x=15$ & 64.84 & $45.99$ & $32.67$ & $23.87$ & $13.15$\\
 \hline
 \end{tabular}
 \end{center}
 \caption{\footnotesize{The fractional error $\varepsilon=\left| [g(x)- \tilde{g}(x)] / \tilde{g}(x)
\right|$ computed for thermal electron populations
 with $k_B T_e=$10, 5, 3, 2, 1 keV is shown for different values of the frequency $x$.
 The quantity $\varepsilon$ is given in units of $10^{-2}$.}}
 \label{tab.diffg}
 \end{table}
In Table 2 we show the fractional error $\varepsilon$ done when considering the
exact calculation of the thermal SZ effect and those at first, second and third
order approximations in $\tau$ for two values of the optical depth, $\tau =
10^{-2}$ and $\tau = 10^{-3}$, as reported in the Table caption. Even for the
highest cluster temperatures here considered, $k_BT_e \sim 20$ keV, the
difference between the exact and approximated calculations is $\simlt 0.25 \%$
at the minimum of the SZ effect ($x \sim 2.3$) and is $\simlt 2.23 \%$ in the
high-frequency tail ($x \sim 15$), the two frequency ranges where the largest
deviations are expected and could be measurable, in principle. For
high-temperature clusters, the third-order approximated calculations of the SZ
effect ensures a precision $\simlt 2 \%$ at any interesting frequency.
\begin{table}[htb]
\begin{center}
\begin{tabular}{|l|*{3}{c|}}
\hline
                        & $First\:order$ & $Second\:order$ & $Third\:order$\\
\hline
  $k_BT_e=20$ keV       &                &                 &               \\
  $\tau=10^{-2}$        &                &                 &               \\
 \hline
 $x=2.3$ & $0.31$ & $0.22$ & $0.22$\\
 $x=6.5$ & $0.12$ & $0.27$ & $0.27$\\
 $x=15$ & $2.23$ & $1.71$ & $1.71$\\
 \hline
  $k_BT_e=20$ keV       &                &                 &               \\
  $\tau=10^{-3}$        &                &                 &               \\
 \hline
 $x=2.3$ & $0.24$ & $0.23$ & $0.23$\\
 $x=6.5$ & $0.21$ & $0.23$ & $0.23$\\
 $x=15$ & $0.23$ & $0.18$ & $0.18$\\
 \hline
  $k_BT_e=10$ keV       &                &                 &              \\
  $\tau=10^{-2}$        &                &                 &               \\
 \hline
 $x=2.3$ & $0.02$ & $0.02$ & $0.02$\\
 $x=6.5$ & $0.7$ & $0.02$ & $0.02$\\
 $x=15$ & $0.74$ & $0.21$ & $0.21$\\
 \hline
  $k_BT_e=10$ keV       &                &                 &               \\
  $\tau=10^{-3}$        &                &                 &               \\
 \hline
 $x=2.3$ & $0.01$ & $0.01$ & $0.01$\\
 $x=6.5$ & $0.01$ & $0.02$ & $0.01$\\
 $x=15$ & $0.31$ & $0.25$ & $0.25$\\
 \hline
 \end{tabular}
 \end{center}
 \caption{\footnotesize{The fractional error $\varepsilon$ done considering the
 first, second and third order approximation in $\tau$ compared with the exact
 calculation of the thermal SZ effect is reported for three interesting values
 of the frequency $x$.
 The assumed values of $k_B T_e$ and $\tau$ are shown in the
first column of the Table. The values of $\varepsilon$ are given in units of
$10^{-2}$.}}
 \label{Tab2}
 \end{table}

\begin{figure}[htp]
\begin{center}
\psfig{file=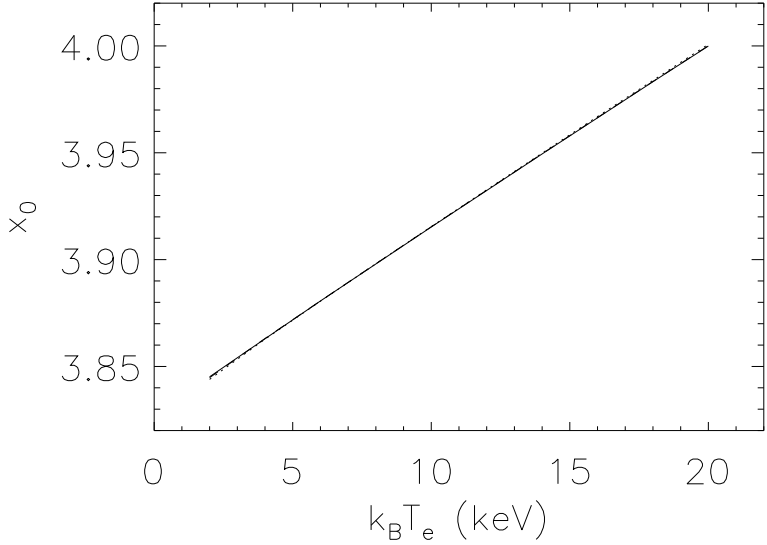,width=12.cm,angle=0.}
  \caption{\footnotesize{The location of $x_0$ for a thermal electron population  is shown as a function of
  $k_BT_e$.  The exact calculation for $\tau=10^{-2}$ (solid line) and for $\tau=10^{-3}$ (dashed line) are shown together
  with the approximated calculation to first order in $\tau$ (dotted line).}}
  \label{fig.zeri.term}
\end{center}
\end{figure}

Using the general, exact relativistic approach discussed in this section, we
evaluate the frequency location of the zero of the thermal SZ effect, $x_0$, for
different cluster temperatures in the range $2-20$ keV. The frequency location
of $x_0$ depends on the IC gas temperature (or more generally on the IC gas
pressure) and in Fig. \ref{fig.zeri.term} we compare the temperature dependence
of $x_0$ evaluated at first order in $\tau$ (which actually does not depend on
$\tau$) with that evaluated in the exact approach for values $\tau =10^{-3}$ and
$\tau =10^{-2}$.
%
%
%
The location of the null of the SZ effect increases  by $\sim 4.4 \%$ for
$k_BT_e $ up to $\sim 20$ keV. Assuming a quadratic fit $x_0 \approx a+b
\theta_e + c \theta_e^2 $, where $\theta_e=k_BT_e/m_ec^2$, we calculate the
coefficients $a, b$ and $c$ which fit the temperature dependence of $x_0$. These
are given in Table \ref{tab.coeffparab}.
 \begin{table}[htb]
\begin{center}
\begin{tabular}{|l|*{3}{c|}}
\hline
 & $a$ & $b$ & $c$ \\
 \hline
 $Exact \; \tau=0.001$ & $3.8271$ & $4.6059$ & $-4.8221$ \\
 $Exact \; \tau=0.01$ & $3.8262$ & $4.6704$ & $-5.4895$ \\
 $First \; order$ & $3.8294$ & $4.4304$ & $-1.7805$ \\
 \hline
 \end{tabular}
 \end{center}
 \caption{\footnotesize{The coefficients of the quadratic fit of the behaviour
 $x_{0,th}$ as a function of $\theta_e=k_BT_e/m_ec^2$ (see text for details).}}
 \label{tab.coeffparab}
 \end{table}
In particular the  coefficients of the last row in Table \ref{tab.coeffparab}
can be compared with those given by Itoh et al. (1998). At first order in $k_B
T_e$, and for IC gas temperatures up to 50 keV, these last authors found $ x_0
\approx  3.830+4.471\theta_e-3.268\theta_e^2$. The coefficients given in the
first two rows of Table \ref{tab.coeffparab} can be also compared with those
found by Dolgov et al. (2000), who gave the expression of $x_0$ evaluated in an
exact way for $\tau$, and for a temperature range $0\leq k_B T_e \leq 50$ keV
and $0 \leq \tau \leq 0.05$: $x_0=\alpha(T_e)+\tau \beta(T_e)$ with
$\alpha(T_e)= 3.830(1+1.162\theta_e -0.8144\theta_e^2)$ and $ \beta(T_e)=
3.021\theta_e -8.672\theta_e^2$.\\
 For $\tau=0.01$, the expression for $x_0$ derived by Dolgov et al. (2000)
 writes as
 \be
 x_0=3.830+4.481\theta_e-3.206\theta_e^2,
 \ee
 while for $\tau=0.001$ one gets
 \be
x_0=3.830+4.453\theta_e-3.128\theta_e^2 ~.
 \ee
The remaining differences with our coefficients given in Table
\ref{tab.coeffparab} are due to the different $T_e$ and $\tau$ ranges used for
the quadratic fit.

\section{The SZ effect generated from a non-thermal electron population}

Using the same general analytical approach derived in Sect. 2 above we derive
here the exact spectral features of the SZ effect produced by a single
non-thermal population of electrons. Such an exact derivation has been not done
so far. For such a population we consider here two different phenomenological
spectra: {\it i)} a single power-law energy spectrum, $n_{rel} = n_0
E^{-\alpha}$, like that which is able to fit the radio-halo spectra of many
clusters, and {\it ii)} a double power-law spectrum which is able to fit both
the radio halo spectrum and the EUV and hard X-ray excess spectra observed in
some nearby clusters (see Petrosian 2001). Our formalism is, nonetheless, so
general that it can be applied to any electron distribution so far considered to
fit both the observed radio-halo spectra and those of the EUV/Hard X-ray
excesses (see, e.g., Sarazin 1999, Blasi \& Colafrancesco 1999, Colafrancesco \&
Mele 2001, Petrosian 2001).\\
 The single power-law electron population is described by the momentum spectrum
\begin{equation} \label{leggep1}
f_{e,rel}(p;p_1,p_2,\alpha)=A(p_1,p_2,\alpha) p^{-\alpha} ~; \qquad p_1 \leq p
\leq p_2
\end{equation}
where the normalization term $A(p_1,p_2,\alpha)$ is given by:
\begin{equation} \label{normal1}
A(p_1,p_2,\alpha) = \frac{(\alpha-1)} {p_1^{1-\alpha}-p_2^{1-\alpha}}
\end{equation}
with $\alpha \approx 2.5$. This is the simplest electron distribution which is
consistent with the spectra of the radio halos observed in many galaxy clusters
(see, e.g., Feretti 2001 for a recent review).
In the calculation of the non-thermal SZ effect we consider the minimum
momentum, $p_1$, of the electron distribution as a free parameter since it is
not constrained by the available observations, while the specific value of $p_2
\gg 1$ is irrelevant for power-law indices $\alpha > 2$ which are required by
the radio halo spectra observed in galaxy clusters.\\
 The distribution in Eq.(\ref{leggep1}) can be however affected by a crucial
problem: if the electron spectrum is extended at energies below 20 MeV (i.e., at
$p\simlt 40$) with the same spectral slope $\alpha \approx 2.5$ of the high
energy tail, the heating rate of the IC gas produced by Coulomb collisions of
the relativistic electrons becomes larger than the bremsstrahlung cooling rate
of the IC gas (see, e.g., Petrosian 2001). This would produce unreasonably and
unacceptably large heating of the IC gas.\\
 For this reason, we consider also a double power-law electron spectrum which has
a flatter slope below a critical value $p_{cr}$:
\begin{equation}
 \label{leggep2}
f_{e,rel}(p;p_1,p_2,p_{cr},\alpha_1,\alpha_2)= K(p_1,p_2,p_{cr},\alpha_1,\alpha_2) \left\{
\begin{array}{ll}
p^{-\alpha_1} & p_1<p<p_{cr} \\ p_{cr}^{-\alpha_1+\alpha_2}\, p^{-\alpha_2} & p_{cr}<p<p_2
\end{array}
\right.
\end{equation}
with a normalization factor given by:
\begin{equation} \label{normal2}
K(p_1,p_2,p_{cr},\alpha_1,\alpha_2)= \left[ \frac{p_1^{1-\alpha_1}-p_{cr}^{1-\alpha_1}}{\alpha_1-1} +
p_{cr}^{-\alpha_1+\alpha_2} \, \frac{p_{cr}^{1-\alpha_2}-p_2^{1-\alpha_2}}{\alpha_2-1} \right]^{-1}.
\end{equation}
If $p_{cr} \sim 400$, the electron distribution of Eq.(\ref{leggep2}) with
$\alpha_1 \sim 0.5$ at $p \simlt p_{cr}$ can be extended down to very low
energies without violating any constraint set by the IC gas heating. Hence, we
assume here the following parameter values: $\alpha_1=0.5$, $\alpha_2=2.5$,
$p_{cr}=400$ e $p_2 \rightarrow \infty$. We again consider $p_1$ as a free
parameter.

A crucial quantity in the calculation of the non-thermal SZ effect is given by
the number density of relativistic electrons, $n_{e,rel}$. The quantity
$n_{e,rel}$ in galaxy clusters can be estimated from the radio halo spectrum
intensity, but this estimate depends on the assumed value of the IC magnetic
field and on the model for the evolution of the radio-halo spectrum (see, e.g.
Sarazin 1999 for time-dependent models and Blasi \& Colafrancesco 1999,
Colafrancesco \& Mele 2001 for stationary models). For the sake of illustration,
the radio halo flux $J_{\nu}$ for a power-law spectrum is, in fact, given by
$J_{\nu} \propto n_{e,rel} B^{\alpha_r-1} \nu ^{-\alpha_r}$, where $\alpha_r$ is
the radio halo spectral slope. Because of the large intrinsic uncertainties
existing in the density of the relativistic electrons which produce the cluster
non-thermal phenomena, a value $n_{e,rel}=10^{-6}$ cm$^{-3}$ for $p_1=100$ has
been assumed here.
Our results on the amplitude of the non-thermal SZ effect can be easily rescaled
to different values of $n_{e,rel}$: in fact, decreasing (increasing) $n_{e,rel}$
will produce smaller (larger) amplitudes of the SZ effect, as can be seen from
eqs.(22,24,28,29).\\
 The density $n_{e,rel}$ increases for decreasing values of $p_1$. In fact, multiplying the electron distribution in Eq.(\ref{leggep1}) by
the quantity $n_{e,rel}(p_1)$ one obtains
\begin{equation}
N_e(p;p_1)\equiv n_{e,rel}(p_1) A(p_1) p^{-\alpha}
\end{equation}
where the function $A(p_1)$ is given by Eq.(\ref{normal1}). Thus, the electron
density scales as
\begin{equation} \label{dens.p1}
n_{e,rel}(p_1)=n_{e,rel}( \tilde{p_1}) \frac{A(\tilde{p_1})}{ A( p_1)}~,
\end{equation}
where we normalized the electron density at a fixed value of $\tilde{p_1}=100$.

\begin{figure}[tbp]
\begin{center}
  \psfig{file=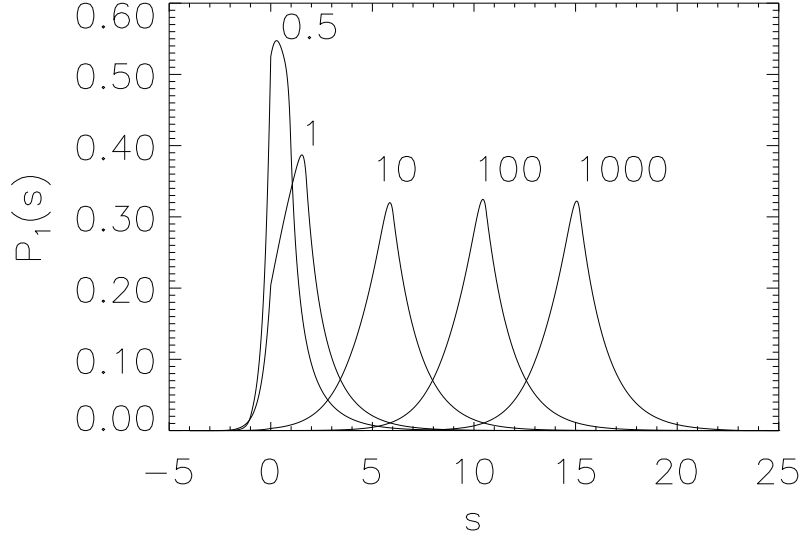,width=12.truecm,angle=0.}
  \caption{\footnotesize{The distribution $P_1(s)$ for
  a non-thermal population as in Eq.(\ref{leggep1}) with $\alpha=2.5$, $p_2\rightarrow\infty$
  and $p_1=$ 0.5, 1,  10, 100, 1000, as indicated.
  }}
  \label{fig.p1sleggep1}
\end{center}
\end{figure}

The value of $p_1$ sets the value of the electron density as well as the value
of the other relevant quantities which depend on it, namely the optical depth
$\tau_{rel}$ and the pressure $P_{rel}$ of the non-thermal population.
 In particular, the pressure $P_{rel}$ for the case of an electron
distribution in Eq.(\ref{leggep1}), is given by 
\begin{eqnarray}\label{press rel}
 P_{rel}&=&n_e \int_0^\infty dp f_e(p) \frac{1}{3} p v(p) m_e c
 \nonumber \\
  & =& \frac{n_e m_e c^2 (\alpha
  -1)}{6[p^{1-\alpha}]_{p_2}^{p_1}}
  \left[B_{\frac{1}{1+p^2}}\left(\frac{\alpha-2}{2},
   \frac{3-\alpha}{2}\right)\right]_{p_2}^{p_1}
\end{eqnarray}
(see, e.g., Ensslin \& Kaiser 2000), where $B_x$ is the incomplete Beta function
\begin{equation}\label{beta inc}
  B_x(a,b)=\int_0^x t^{a-1} (1-t)^{b-1} dt
\end{equation}
(see, e.g., Abramowitz \& Stegun 1965). The optical depth of the non-thermal
electron population is given by
\begin{equation} \label{tau_p1}
\tau_{rel}(p_1) = 2\cdot 10^{-6} ~\frac{A(\tilde{p_1})}{ A( p_1)}
\left[\frac{n_{e,rel}( \tilde{p_1}=100) }{10^{-6}\,cm^{-3}}\right] \left[
\frac{\ell}{1\,Mpc}\right] ~.
\end{equation}
For an electron population as in Eq.(\ref{leggep2}) we obtain analogous results.
 The electron density with the same normalization as before is given by
\begin{equation} \label{dens.p1.lp2}
n_{e,rel}(p_1)=n_{e,rel}( \tilde{p_1}) \frac{K(\tilde{p_1})}{ K( p_1)}.
\end{equation}
The optical depth is again given by  Eq.(\ref{tau_p1}), where we substitute the
ratio $K(\tilde{p_1})/K(p_1)$ to $A(\tilde{p_1})/A(p_1)$. The electron pressure
of this last population has instead a more complicated expression given by:
\begin{eqnarray}
\label{press.p1.lp2}
P_{rel}(p_1)&=&\frac{K(p_1) n_e(p_1) m_e c^2}{6} \left \{ \left [B_{
\frac{1}{1+p^2}} \left( \frac{\alpha_1-2}{2}, \frac{3-\alpha_1}{2} \right)
\right]_{p_{cr}}^{p_1} + \right. \nonumber \\ & &+  \left. \left [B_{
\frac{1}{1+p^2}} \left( \frac{\alpha_2-2}{2}, \frac{3-\alpha_2}{2} \right)
\right]^{p_{cr}}_{p_2} p_{cr}^{-\alpha_1+\alpha_2} \right\}.
\end{eqnarray}
To evaluate the SZ effect induced by a single non-thermal electron population we
use the relativistically correct formalism  described in Sect. 2 which also
takes into account the effects of multiple scattering. Again, we are working
here in the Thomson limit. The general expression for the distorted spectrum
$I(x)$ is given again by Eq.(8). The only change in our general formulation
consists in inserting the non-thermal distribution $f_{e,non-th}$ in the
function $P_1(s)$. The functions $P_n(s)$, which determine the probabilities to
have a frequency shift $s$ due to $n$ scattering are still given by the
convolution of $n$ distributions $P_1(s)$ (see Eq.6). Fig.\ref{fig.p1sleggep1}
shows the function $P_1(s)$ for different values of $p_1$ in the range $0.5 -
1000$.
For low values of $p_1$, the function $P_1(s)$ is centered around $s=0$ and does
not produce, hence, large frequency changes. For higher values of $p_1$ (i.e.,
for increasing values of the minimum energy of the electrons) the function
$P_1(s)$ becomes wider and it is centered at higher and higher values of $s$.
For $p_1=1000$ the redistribution function $P_1(s)$ is centered around $s=15$,
which corresponds to a frequency change $\nu' / \nu = e^s \sim 3\cdot 10^6$.
Thus, the SZ effect caused by electronic populations with very high values of
$p_1$ moves the photon frequencies from the CMB region up to very different
regions causing a depletion of scattered CMB photons at sub-mm wavelengths where
the SZ effect is usually studied. We report in Table \ref{tab.datileggep1} the
values of the pressure and of the density of a single power-law population as a
function of $p_1$.
\begin{table}[htb]
\begin{center}
\begin{tabular}{|c|*{2}{c|}}
\hline
 $p_1$ & $P_{rel}$ (keV cm$^{-3}$) & $n_{e,rel}$ (cm$^{-3}$)\\
 \hline
 $0.5$ & $0.59$ & $2.83\cdot10^{-3}$ \\
 $1$ & $0.47$ & $1\cdot10^{-3}$ \\
 $10$ & $0.16$ & $3.16\cdot10^{-5}$ \\
 100 & $5.11\cdot10^{-2}$& $1\cdot10^{-6}$\\
 1000 & $1.61\cdot10^{-2}$& $3.16\cdot10^{-8}$\\
 \hline
 \end{tabular}
 \end{center}
 \caption{\footnotesize{Values of the pressure and of the density of a single power-law population
 as in Eq.(\ref{leggep1}) with $\alpha=2.5$, $p_2\rightarrow\infty$ and $n_{e,rel}( \tilde{p}_1=100)=1\cdot
 10^{-6}$ cm$^{-3}$ for different values of $p_1$.
 }}
 \label{tab.datileggep1}
 \end{table}

\begin{figure}[htbp]
\begin{center}
\hbox{
  \psfig{file=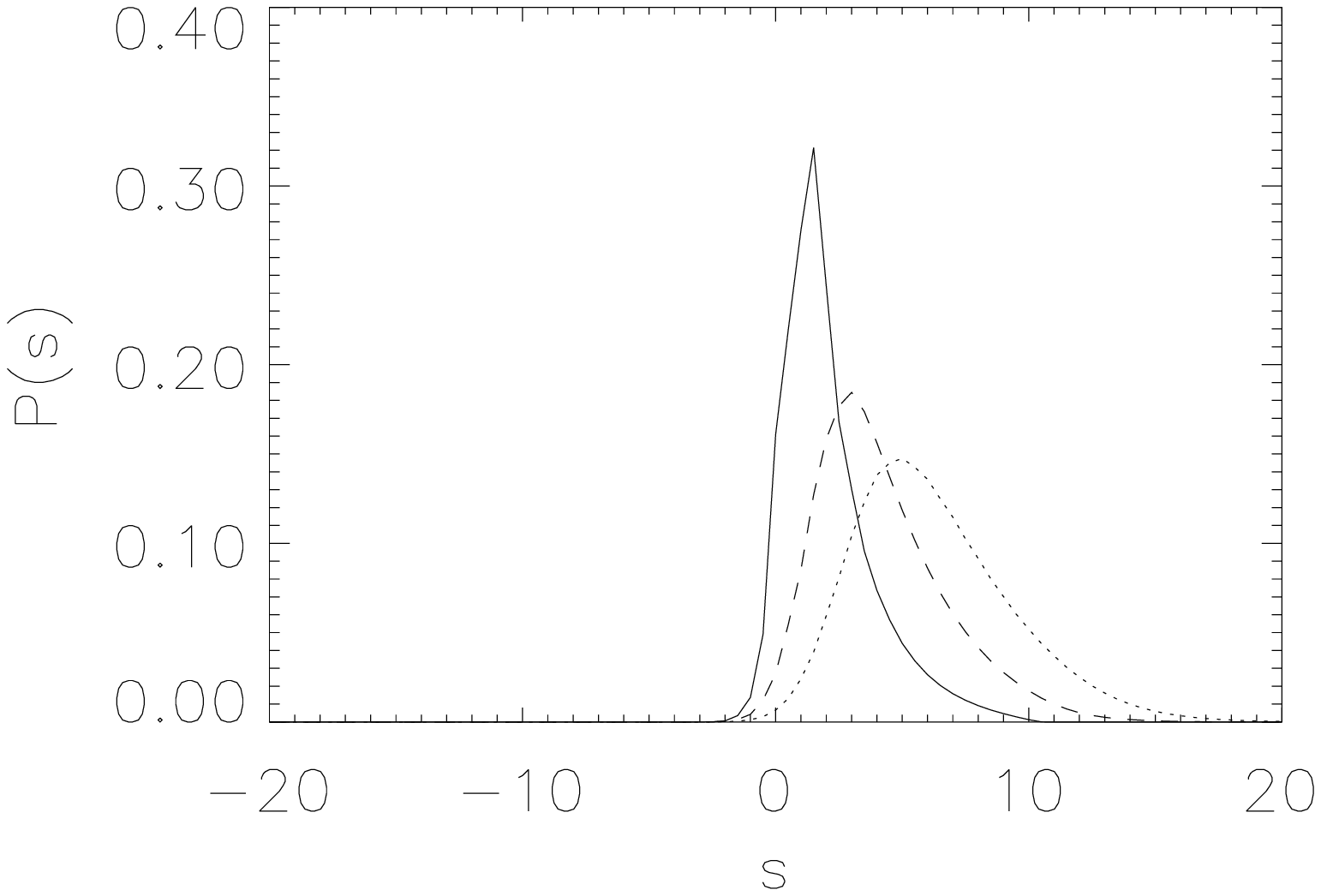,width=9.truecm,height=8.truecm,angle=0.}
  \psfig{file=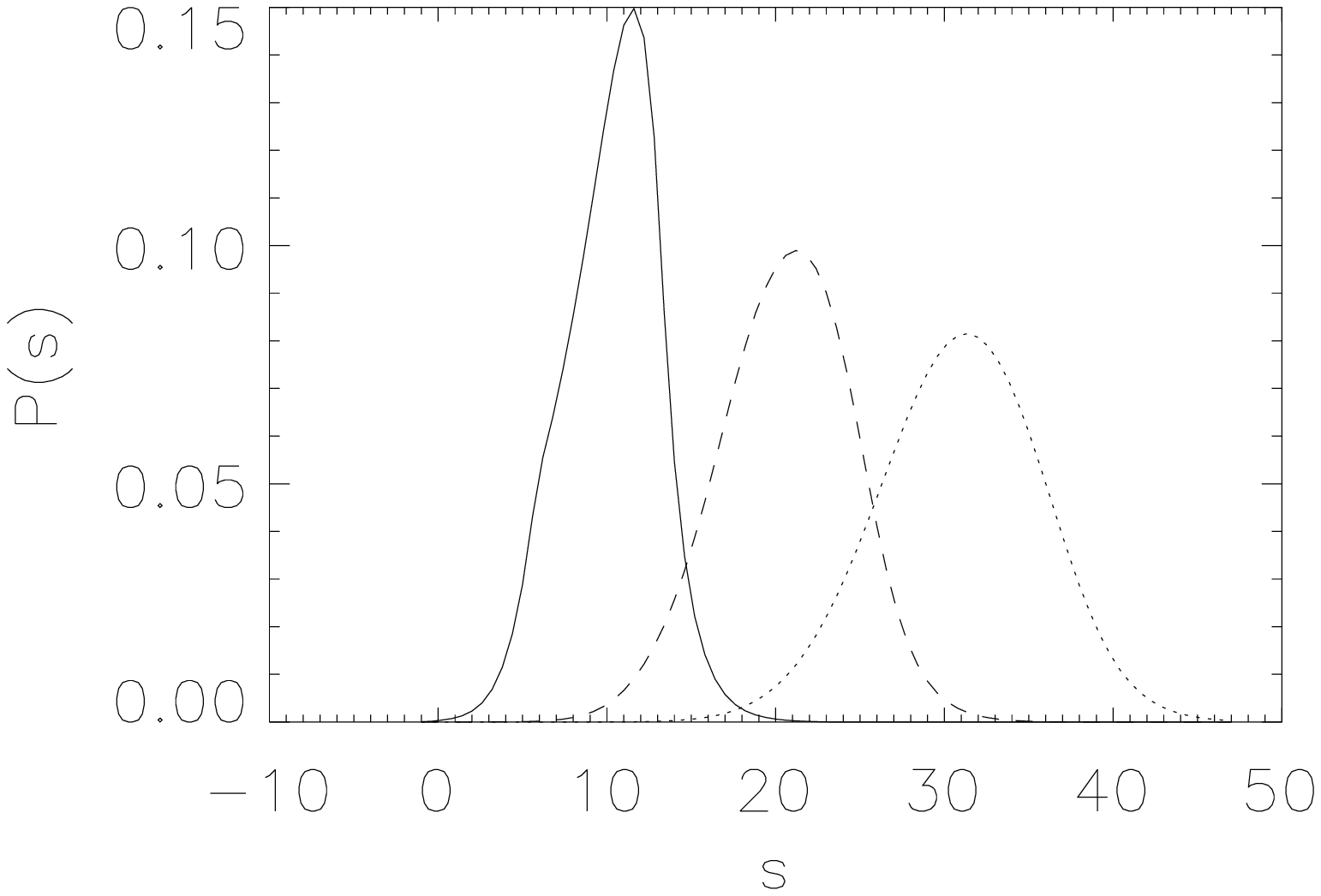,width=9.truecm,height=8.truecm,angle=0.}
 }
  \caption{\footnotesize{The distributions $P_1(s)$ (solid line), $P_2(s)$ (dashed) and
  $P_3(s)$ (dotted) evaluated for non-thermal electron populations with a single
  power-law spectrum given in Eq.(\ref{leggep1}) with $\alpha=2$, $p_1=1$
  and $p_2=100$ (left panel) and with a double power-law spectrum given in
  Eq.(\ref{leggep2}) (right panel).
  Note that these distributions are wider and
  more skewed towards larger values of $s$ than the corresponding distributions
  evaluated in the case of a thermal electron population.
  }}
  \label{fig.p3lp}
\end{center}
\end{figure}

To emphasize the changes in the photon redistribution function produced by a
non-thermal electron distribution, we show in Fig.\ref{fig.p3lp} the functions
$P_1(s), P_2(s)$ and $P_3(s)$ for a single and double power-law electron
distributions. It is evident that these functions are wider and more
a-symmetrically skewed towards large and positive values of $s$ with respect to
the same distributions evaluated in the case of a single thermal electron
population.

\begin{figure}[htbp]
\begin{center}
  \psfig{file=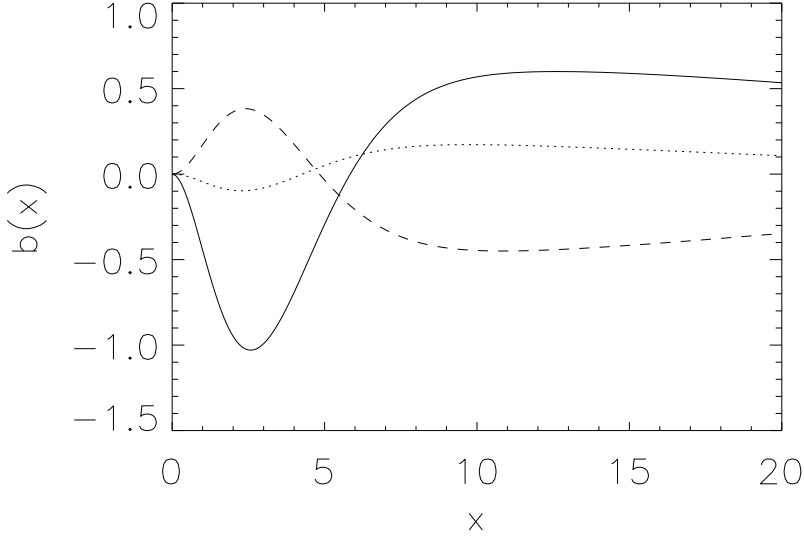,width=12.truecm,angle=0.}
  \caption{\footnotesize{The functions $b_1(x)$ (solid), $b_2(x)$ (dashed) and
  $b_3(x)$ (dotted) evaluated for a non-thermal electron population with a single power-law
  spectrum as in Eq.(32).
  }}
  \label{fig.contrlp}
\end{center}
\end{figure}

Once the exact redistribution function $P(s)$ is known, it is possible to
evaluate the exact spectral distortion $\Delta I(x)$ through Eq.(\ref{idx}).\\
 It is also possible to evaluate the distorted spectrum using Eq.(16) which
gives the series expansion up to the $n$-th order in $\tau$ and to compare the
exact and approximated shapes for $I(x)$. We show in Fig.\ref{fig.contrlp} the
first three terms $b_1,b_2, b_3$ of the series expansion in Eq.(16) as a
function of the a-dimensional frequency $x$.
The approximated expression for $I(x)$ up to third order in $\tau$ is an
excellent fit of its exact behaviour at frequencies around the minimum of the
effect ($x \approx 2.3$) and it shows a difference $\sim 3 \%$ at $x \approx
6.5$ and a difference $\approx 0.1\% $ at high frequencies, $x \approx 15$ (see
Table\ref{tab.lp001}). The precision of the approximation depends also on the
value of $\tau$ and we verified that the precision of the second and third order
approximations does not increase for values $\tau \simlt 10^{-3}$ at any
frequency between $x \approx 2$ and $x \approx 15$.
 \begin{table}[htb]
\begin{center}
\begin{tabular}{|l|*{3}{c|}}
\hline
             & $First\:order$ & $Second\:order$ & $Third\:order$\\
\hline
 $\tau=0.01$ &  &   &  \\
 \hline
 $x=2.3$ & $0.82$ & $0.44$ & $0.44$\\
 $x=6.5$ & $1.42$ & $2.90$ & $2.90$\\
 $x=15$ & $0.83$ & $0.12$ & $0.12$\\
 \hline
 $\tau=0.001$ &   &        &       \\
 \hline
 $x=2.3$ & $0.49$ & $0.45$ & $0.45$\\
 $x=6.5$ & $2.73$ & $2.88$ & $2.88$\\
 $x=15$ & $0.19$ & $0.12$ & $0.12$\\
 \hline
 \end{tabular}
 \end{center}
 \caption{\footnotesize{Fractional difference (in units of $10^{-2}$)
 between the exact spectral distortion
 and that evaluated at first, second and third order in $\tau$.
 We report here the cases for values of the optical depth $\tau=0.01$ and $\tau=0.001$.}}
 \label{tab.lp001}
 \end{table}

\subsection{The function $ \tilde{g}(x)$ for a single non-thermal population}

Also for a non-thermal electron population it is possible to write the spectral
distortion in the general form of Eq.(22) as:
\begin{equation}
\Delta I_{non-th}(x)=2\frac{(k_B T_0)^3}{(hc)^2}y_{non-th} ~\tilde{g}(x) ~,
\end{equation}
where the Comptonization parameter is given by the general expression
\begin{equation}
y_{non-th}=\frac{\sigma_T}{m_e c^2}\int P_{rel} d\ell ~,
\end{equation}
in terms of the electronic pressure $P_{rel}$ which, for a single and double
power-law populations, is given by Eq.(\ref{press rel}) and by Eq.(42),
respectively.

\begin{figure}[htbp]
\begin{center}
    \psfig{file=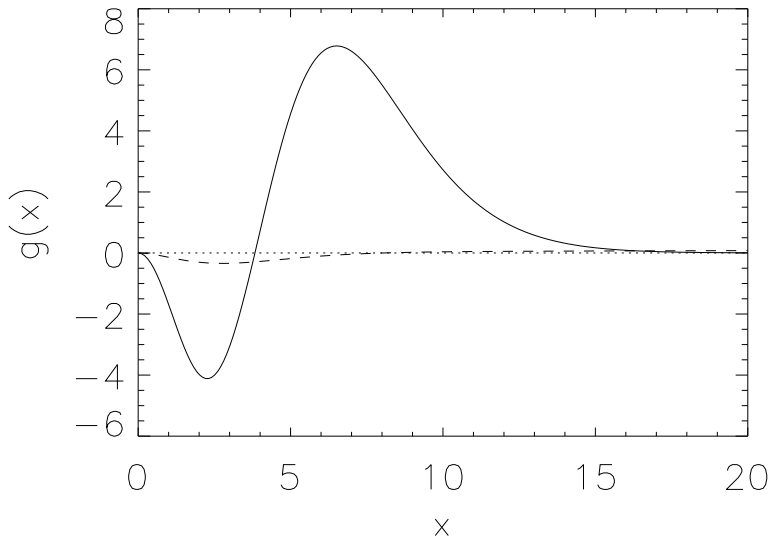,width=12.truecm,angle=0.}
  \caption{\footnotesize{The spectral function $g(x)$ for a thermal population
  in the Kompaneets limit (solid line) is compared with the function $\tilde{g}(x)$
  approximated at first order in $\tau$ for a non-thermal population with a single
  power-law spectrum
  with parameters $p_1=4$, $p_2=10^{10}$ and $\alpha=2.5$ (dashed line), evaluated from
  Eq.(\ref{gnonterm1ord}). The dotted line represents the zero reference value. }}
  \label{fig.gtnonterm}
\end{center}
\end{figure}

\begin{figure}[htbp]
\begin{center}
    \psfig{file=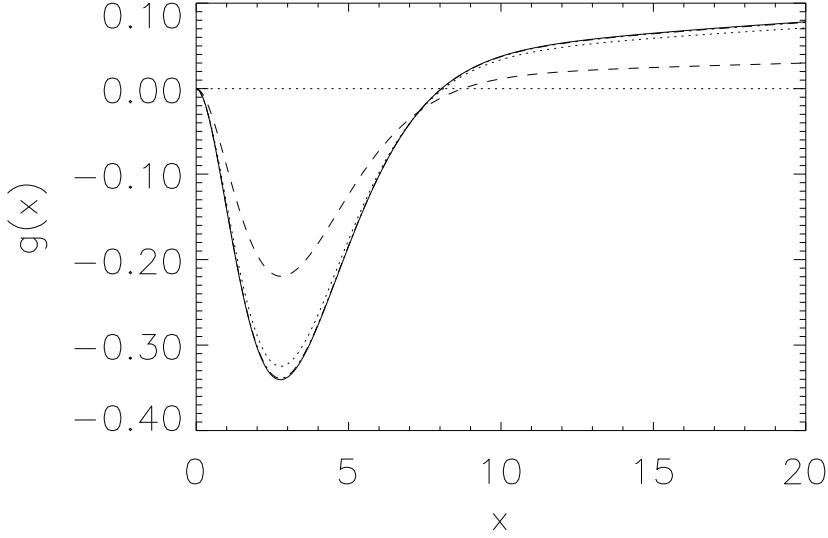,width=12.truecm,angle=0.}
  \caption{\footnotesize{The function $\tilde{g}(x)$ for a non-thermal population
  with single power-law spectrum with parameters $p_1=4$, $p_2=10^{10}$
  and $\alpha=2.5$. We show the first order approximation
  in $\tau$ (solid line) and the exact case for $\tau=1$ (dashed line), $\tau=0.1$
  (dotted line) and $\tau=0.01$ (dot-dashed line).
  Note that the first order approximation is a reasonable approximation of the exact
  case only for small values of $\tau$.
  }}
  \label{fig.gtnonterm2}
\end{center}
\end{figure}

\begin{figure}[tbp]
\begin{center}
\hbox{
  \psfig{file=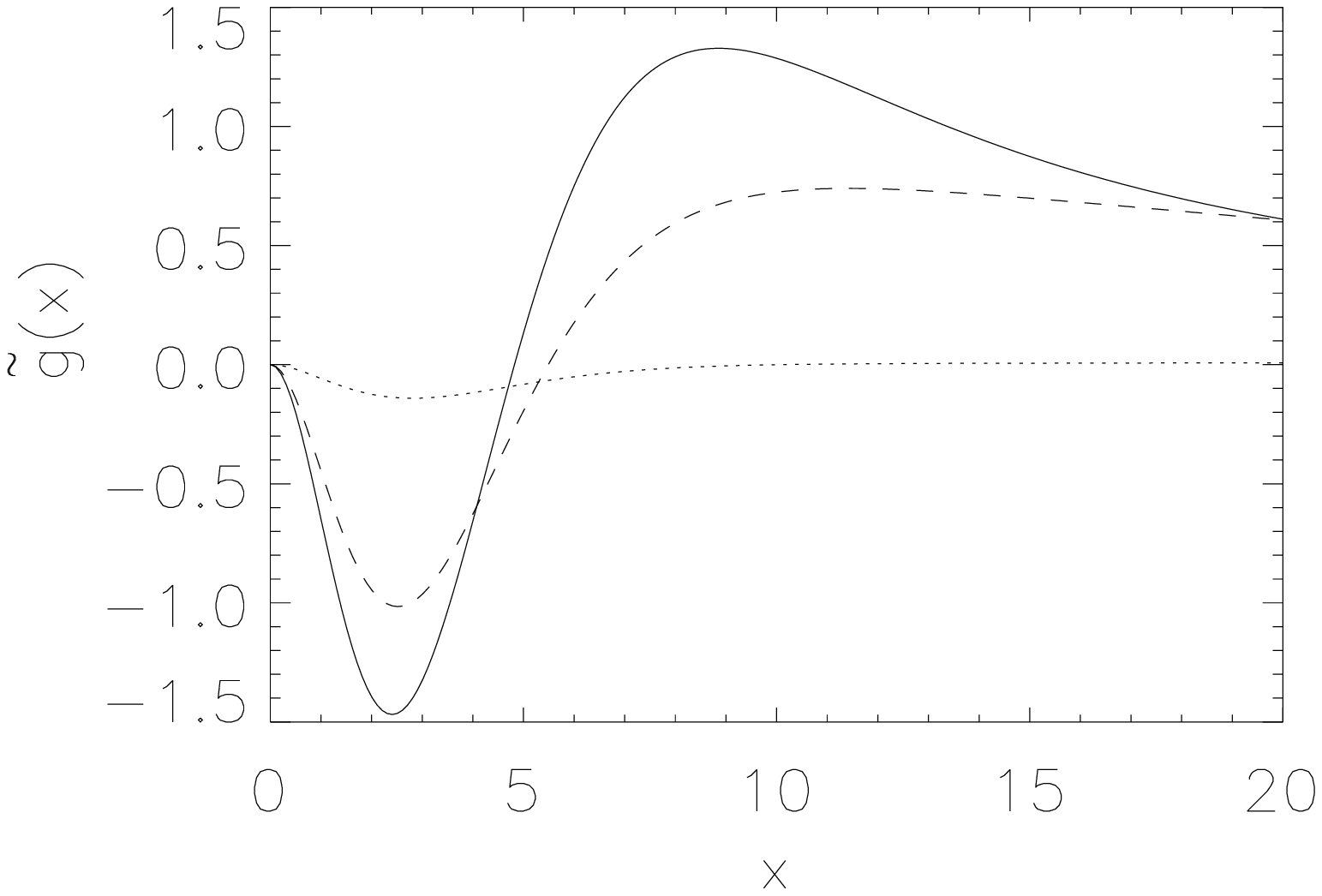,width=9.truecm,height=8.truecm,angle=0.}
  \psfig{file=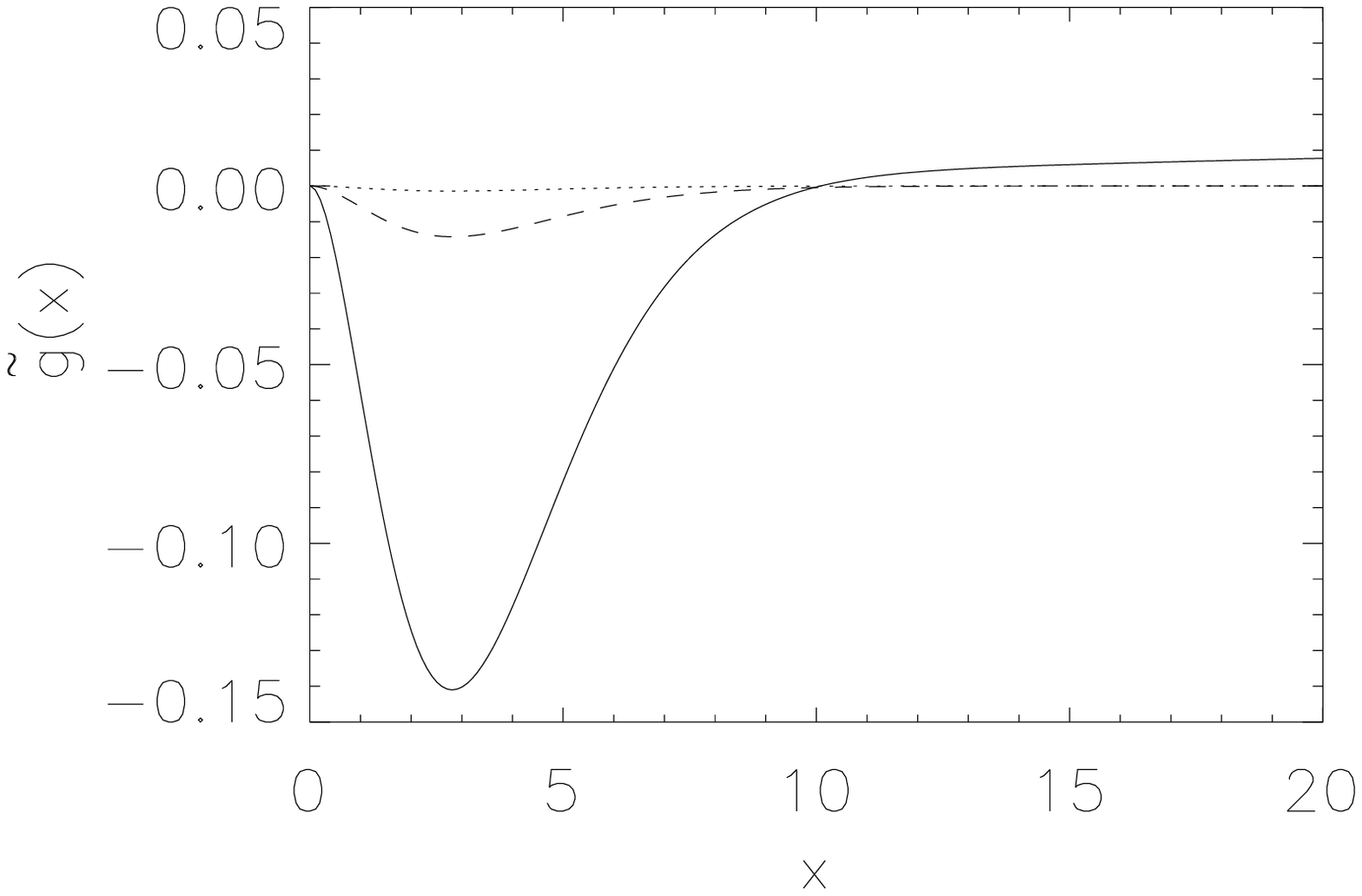,width=9.truecm,height=8.truecm,angle=0.}
}
  \caption{\footnotesize{In the left panel we show the function $\tilde{g}(x)$
  (see Eq.\ref{gnonterm1ord}) for a
  non-thermal population as in Eq.(\ref{leggep1}) with $p_1=$ 0.5 (solid line),
  1 (dashed) and  10 (dotted). The right panel shows the same function for
  $p_1=$ 10 (solid line), 100 (dashed) and 1000 (dotted).
  }}\label{fig.gtilde1leggep1}
\end{center}
\end{figure}

\begin{figure}[tbp]
\begin{center}
\hbox{
  \psfig{file=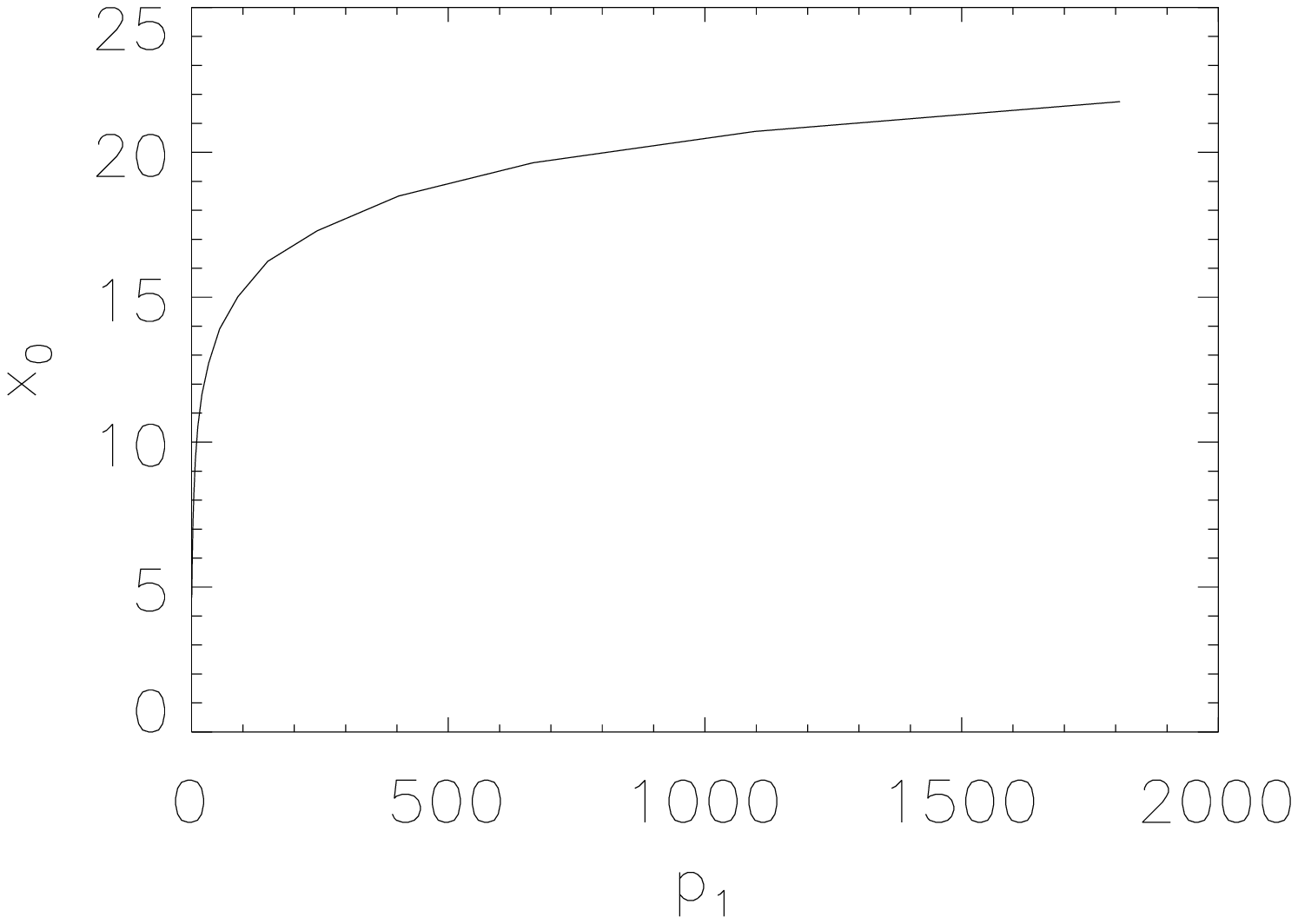,width=9.truecm,height=8.truecm,angle=0.}
  \psfig{file=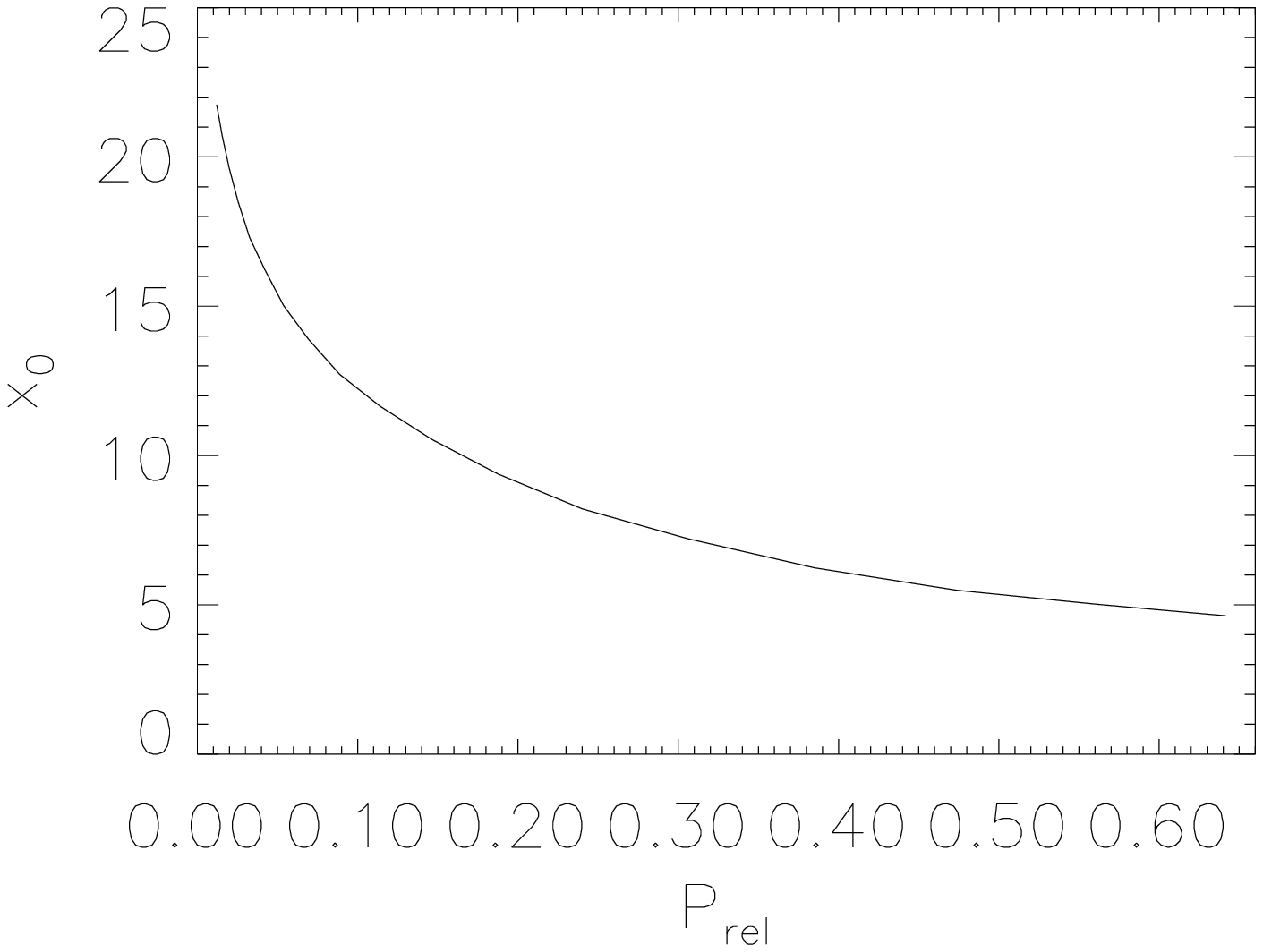,width=9.truecm,height=8.truecm,angle=0.}
}
  \caption{\footnotesize{The behaviour of the zero of the SZ effect for a non-thermal population
  given in Eq.(\ref{leggep1}) as a function of $p_1$ (left panel)
  and of the pressure $P_{rel}$ (right panel) expressed in keV cm$^{-3}$. Note
  that the pressure $P_{rel}$ decreases for increasing values of $p_1$, as
  discussed in Sect.3
    }}
  \label{fig.zeri.leggep1}
\end{center}
\end{figure}

\begin{figure}[tbp]
\begin{center}
  \psfig{file=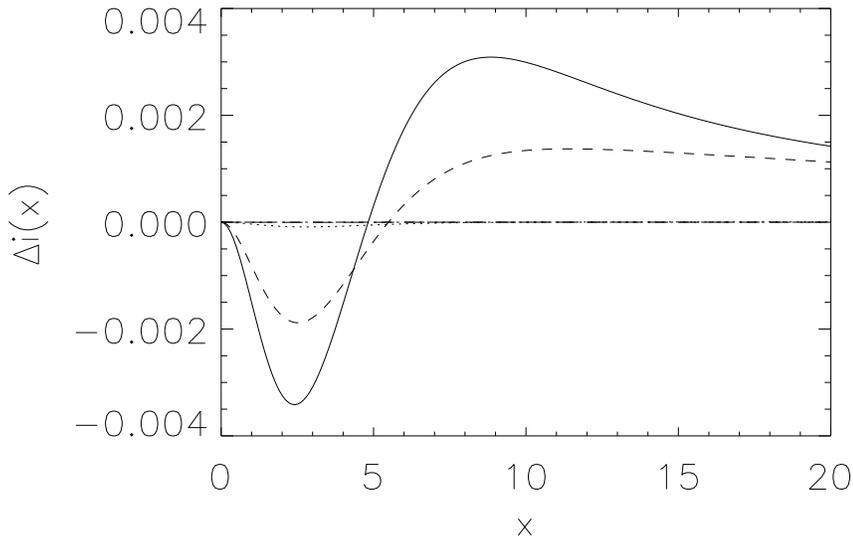,width=12.truecm,angle=0.}
  \caption{\footnotesize{The spectral distortion $\Delta i(x)$ in units of
  $2(k_B T_0)^3/(hc)^2$ for a non-thermal population as in Eq.(\ref{leggep1})
  evaluated for $p_1=$ 0.5 (solid line),
  1 (dashes), 10 (dotted), 100 (long dashes) and 1000
  (dot-dashes).}}
  \label{fig.distleggep1}
\end{center}
\end{figure}

\noindent
At first order in $\tau$ we can write the spectral function
$\tilde{g}(x)$ of the non-thermal SZ effect as
\begin{equation}
\label{gnonterm1ord}
 \tilde{g}(x)=\frac{\Delta i}{y}=\frac{\tau [j_1-j_0]}{\frac{\sigma_T}{m_e c^2}\int P
 d\ell} \equiv \frac{m_e c^2}{\langle k_B T_e \rangle} [j_1-j_0]
\end{equation}
where we defined the quantity
\begin{eqnarray} \label{temp.media}
\langle k_B T_e \rangle& \equiv &\frac{\sigma_T}{\tau}\int P d\ell=\frac{\int P
d\ell}{\int n_e d\ell}= \nonumber
\\
&=&\int_0^\infty dp f_e(p) \frac{1}{3} p v(p) m_e c ~,
\end{eqnarray}
which is the analogous of the average temperature for a thermal population.
 In fact, comparing Eq.(\ref{temp.media}) with Eq.(\ref{press termica}),
$\langle k_B T_e \rangle = k_B T_e$ obtains. For a non-thermal population with a
single power-law distribution one gets
\begin{equation}
\langle k_B T_e \rangle= \frac{ m_e c^2 (\alpha
  -1)}{6[p^{1-\alpha}]_{p_2}^{p_1}}
  \left[B_{\frac{1}{1+p^2}}\left(\frac{\alpha-2}{2},
   \frac{3-\alpha}{2}\right)\right]_{p_2}^{p_1} ~,
\end{equation}
while in the case of a double power-law distribution we find instead
\begin{eqnarray}
 \langle k_B T_e \rangle
 &=&\frac{K(p_1)  m_e c^2}{6} \left \{ \left [B_{ \frac{1}{1+p^2}} \left(
\frac{\alpha_1-2}{2}, \frac{3-\alpha_1}{2} \right) \right]_{p_{cr}}^{p_1} +
\right. \nonumber \\ & &+  \left. \left [B_{ \frac{1}{1+p^2}} \left(
\frac{\alpha_2-2}{2}, \frac{3-\alpha_2}{2} \right) \right]^{p_{cr}}_{p_2}
p_{cr}^{-\alpha_1+\alpha_2} \right\}.
\end{eqnarray}
It is possible to write the Comptonization parameter $y_{non-th}$ as a function
of the quantity $\langle k_B T_e \rangle$ and of the optical depth $\tau$:
\begin{equation} \label{y.tmedia}
y_{non-th}=\frac{\sigma_T}{m_e c^2}\int P_{rel} d\ell= \sigma_T \frac{\langle
k_B T_e \rangle}{m_e c^2} \int n_{e,rel} d\ell= \frac{\langle k_B T_e
\rangle}{m_e c^2} \tau
\end{equation}
(note that here $\tau = \tau_{rel}$ holds).\\
 We can also write the $n$-th order approximations of the distorted
spectrum using Eq.(16). For example, at third order in $\tau$, the following
explicit expression founds:
\begin{equation}
  \tilde{g}(x)=\frac{m_e c^2}{\langle k_B T_e \rangle}
  \left[(j_1-j_0)+\frac{1}{2}\tau(j_2-2j_1+j_0)+\frac{1}{6}\tau^2
  (j_3-3j_2+3j_1-j_0)\right] ~.
\end{equation}
Finally, the exact form of the function $\tilde{g}(x)$ for a non-thermal
electron population writes as:
\begin{equation} \label{gnontermesatta}
\tilde{g}(x)=\frac{m_e c^2}{\langle k_B T_e \rangle} \left\{ \frac{1}{\tau} \left[\int_{-\infty}^{+\infty}
i_0(xe^{-s}) P(s) ds- i_0(x)\right] \right\}.
\end{equation}
We show in Fig.\ref{fig.gtnonterm} a comparison between the function $g(x)$
obtained under the Kompaneets approximation [see Eq.(2)] and the function
$\tilde{g}(x)$, approximated to first order in $\tau$, and evaluated for a
single power-law population. A major difference between the two functions is the
different position of the zero of the SZ effect which is moved to higher
frequencies in the case of the non-thermal population with respect to the case
of the thermal one.
 In Fig.\ref{fig.gtnonterm2} we compare the function
$\tilde{g} (x)$ for a non-thermal population evaluated at first order in $\tau$
(which is actually independent from the value of $\tau$) with the exact function
for values $\tau=1$, $\tau=0.1$ and $\tau=0.01$. We notice that the first order
approximation is a good approximation of the exact case for values $\tau <
10^{-3}$.
Fig.\ref{fig.gtilde1leggep1} shows the function $\tilde{g}(x)$ (evaluated at
first order in $\tau$) for a single power-law population and considering
different values of $p_1$. For increasing values of $p_1$ the maximum of the
function $\tilde{g}(x)$ moves towards higher and higher frequencies. As a
consequence, also the zero of the non-thermal SZ effect moves to higher values
of $x$ as shown also in Fig.\ref{fig.zeri.leggep1}. The location of $x_0$
changes rapidly with increasing values of $p_1$ and reaches a value $x_0 \approx
22$ for $p_1 = 2 \cdot 10^3$. At such values of $p_1$ the pressure $P_{rel}$
takes values $\approx 0.02$ keV cm$^{-3}$, as shown in Fig.9.
An important point to notice is that the electron density of the non-thermal
population decreases for increasing values of $p_1$ as was already shown in
Table \ref{tab.datileggep1}. For this reason the pressure $P_{rel}$ and the
resulting spectral distortion decreases in amplitude for increasing values of
$p_1$. This result is shown more specifically in Fig.\ref{fig.distleggep1} where
we plot the quantity $\Delta i(x) = \Delta I(x)[(hc)^2/2(k_BT_0)^3]$ for values
of $p_1$ in the range $0.5-10^3$.

For an electron population with a double power-law spectrum as in
Eq.(\ref{leggep2}) the values of the density and pressure are also reported in
Table \ref{tab.datileggep2} as a function of $p_1$ (note that the case
$p_1=1000$ is the same of the single power-law spectrum). The values of
$P_{rel}$ and $n_{e,rel}$ for low values of $p_1$ are now smaller than those of
the single power-law case (see Table4), thus avoiding the problem of an
excessive heating of the cluster IC gas for low values of $p_1$.
\begin{table}[htb]
\begin{center}
\begin{tabular}{|c|*{2}{c|}}
\hline
 $p_1$ & $P_{rel}$ (keV cm$^{-3}$) & $n_{e,rel}$ (cm$^{-3}$)\\
 \hline
 $0.5$ & $1.26\cdot10^{-3}$ & $1.56\cdot10^{-6}$ \\
 $1$ & $1.26\cdot10^{-3}$ & $1.54\cdot10^{-6}$ \\
 $10$ & $1.23\cdot10^{-3}$ & $1.41\cdot10^{-6}$ \\
 100 & $9.88\cdot10^{-4}$& $1\cdot10^{-6}$\\
 \hline
 \end{tabular}
 \end{center}
 \caption{\footnotesize{Values of the pressure and of the density of a double power-law population
 as in Eq.(\ref{leggep2}) with $\alpha_1=0.5$, $\alpha_2=2.5$, $p_{cr}=400$,
 $p_2\rightarrow\infty$ and $n_{e,rel}( \tilde{p}_1=100)=1\cdot
 10^{-6}$ cm$^{-3}$ for different values of $p_1$.
 }}
 \label{tab.datileggep2}
 \end{table}

\begin{figure}[tbp]
\begin{center}
  \psfig{file=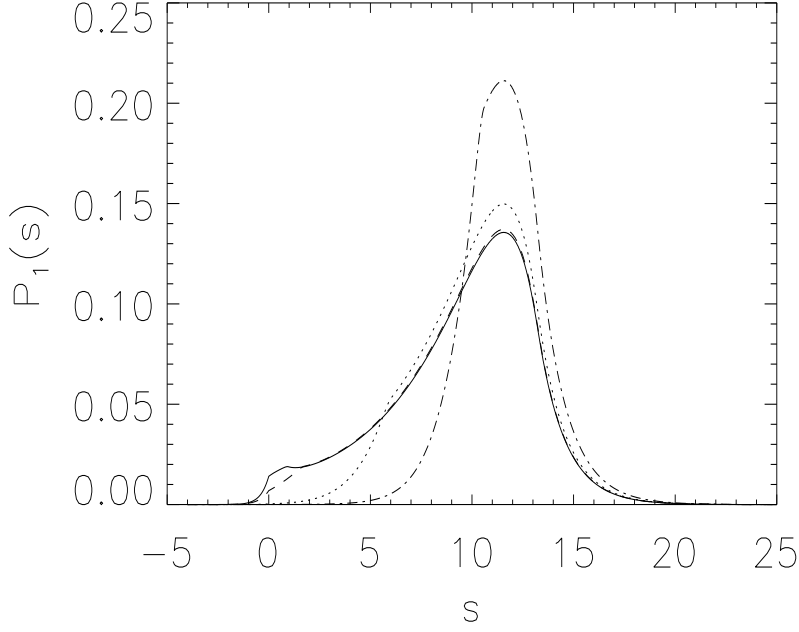,width=12.truecm,height=10.truecm,angle=0.}
  \caption{\footnotesize{The function $P_1(s)$ [see Eq.(\ref{p1s})] is shown for
  non-thermal populations given in  Eq.(\ref{leggep2}) with $p_1=$ 0.5 (solid),
  1 (dashes), 10 (dotted), 100 (dot-dashes).
}}
  \label{fig.p1s.lp2}
\end{center}
\end{figure}

\begin{figure}[tbp]
\begin{center}
  \psfig{file=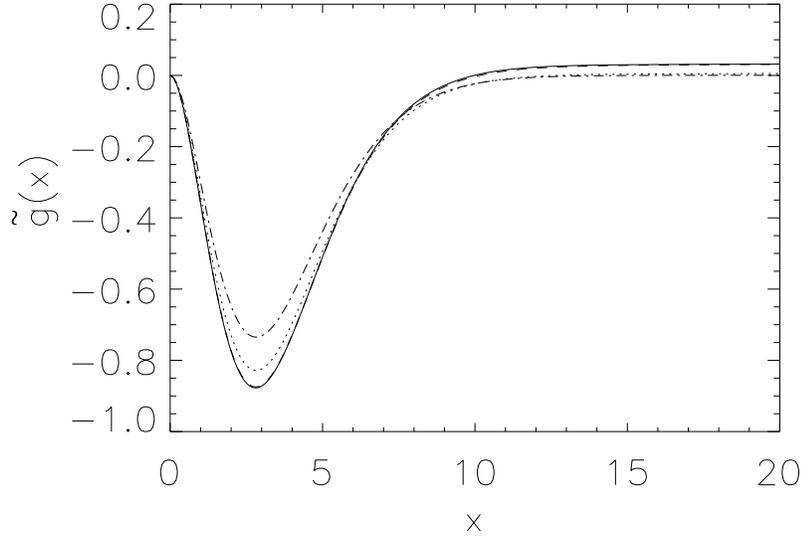,width=12.truecm,angle=0.}
  \caption{\footnotesize{The function $\tilde{g}(x)$ [see
  Eq.(\ref{gnonterm1ord})] for non-thermal populations as in Eq.(\ref{leggep2}) with
  $p_1=$ 0.5 (solid line), 1 (dashes), 10 (dotted) and 100 (dot-dashes).}}
  \label{fig.gt.lp2}
\end{center}
\end{figure}

\begin{figure}[tbp]
\begin{center}
\hbox{
  \psfig{file=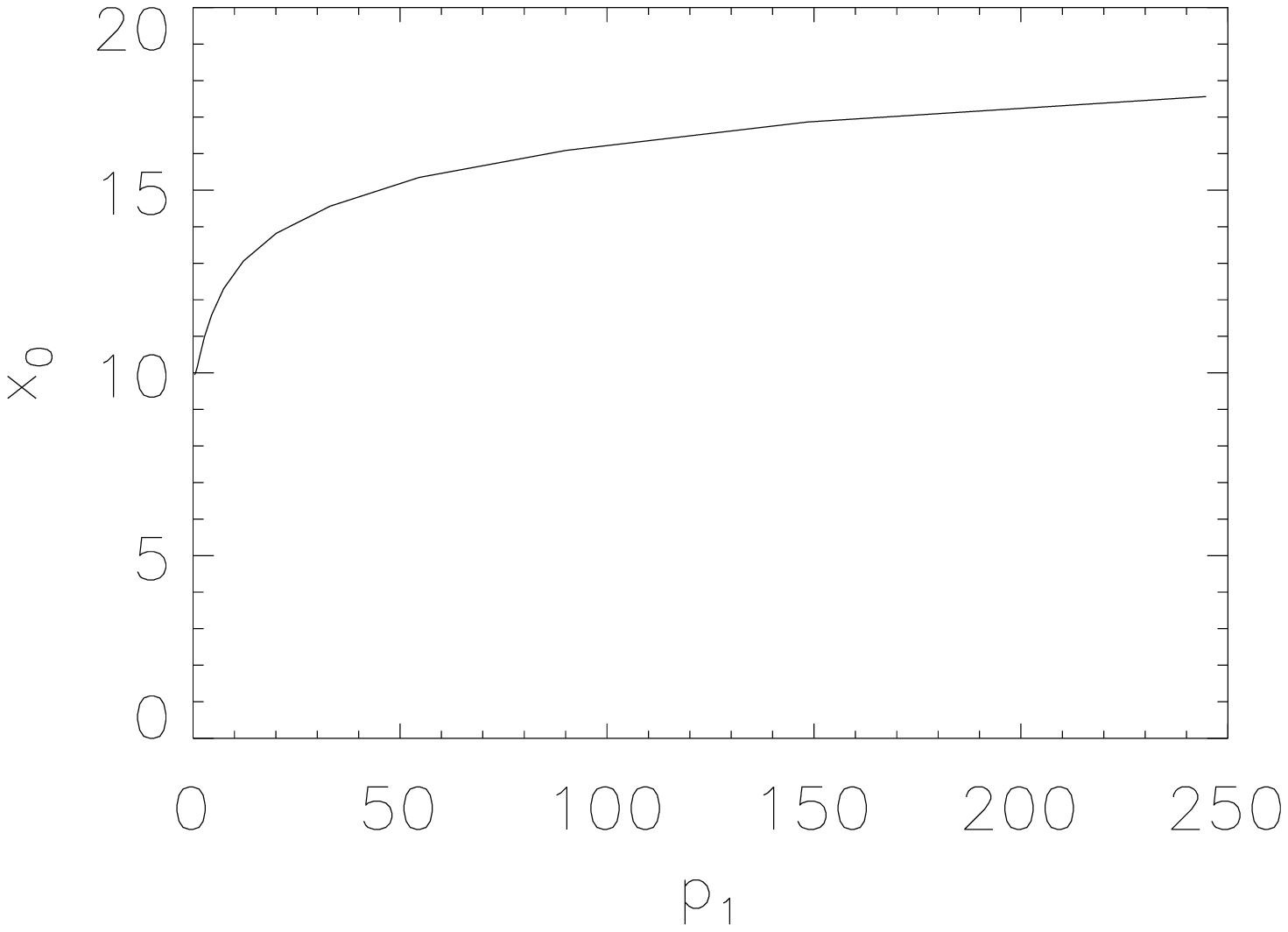,width=9.truecm,height=8.truecm,angle=0.}
  \psfig{file=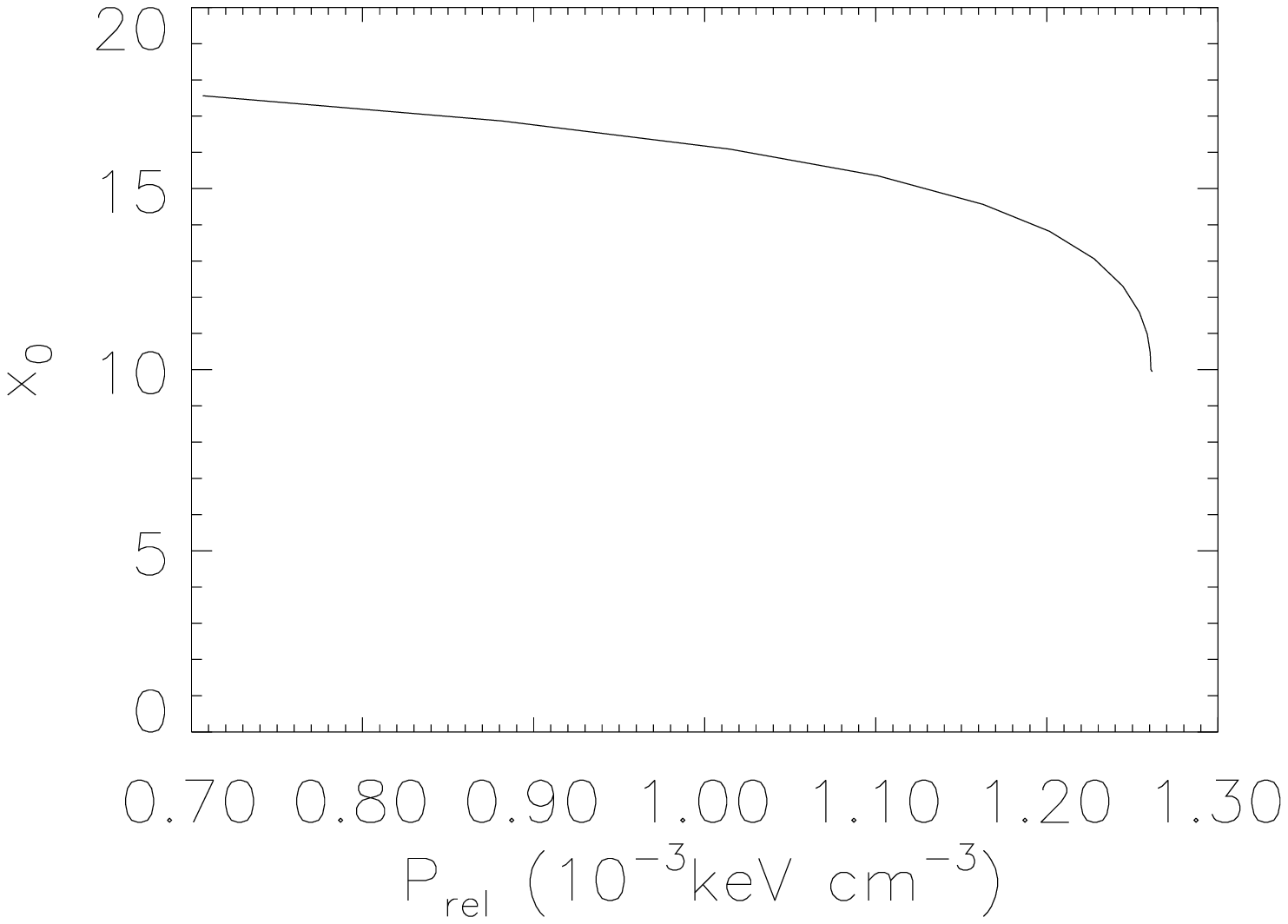,width=9.truecm,height=8.truecm,angle=0.}
}
  \caption{\footnotesize{The behaviour of the zero of the SZ effect for a
  non-thermal population with spectrum
  given by Eq.(\ref{leggep2}) as a function of $p_1$ (left panel)
  and of the pressure $P_{rel}$ (right panel) in units of keV cm$^{-3}$.
As in Fig.9, the pressure $P_{rel}$ decreases for increasing values of $p_1$ but
with a different rate depending on the different spectrum of the electron
distribution. }}
  \label{fig.zeri.leggep2}
\end{center}
\end{figure}

\begin{figure}[tbp]
\begin{center}
  \psfig{file=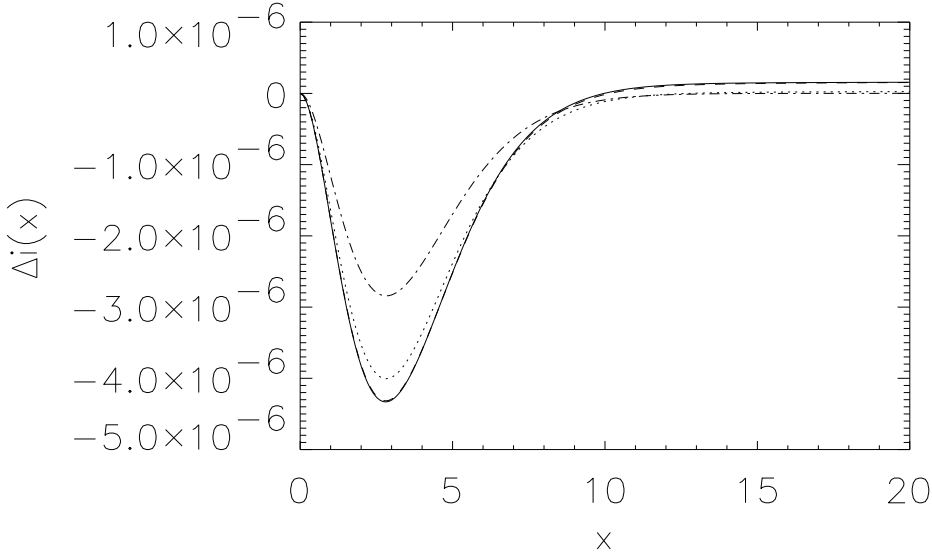,width=12.truecm,angle=0.}
  \caption{\footnotesize{The spectral distortion in units of $2(k_B T_0)^3/(hc)^2$
  for a non-thermal population given by Eq.(\ref{leggep2}) computed for $p_1=$ 0.5
  (solid),  1 (dashes), 10 (dotted) and 100 (dot-dashes).}}
  \label{fig.dist.lp2}
\end{center}
\end{figure}

In Fig.\ref{fig.p1s.lp2} we show the function $P_1(s)$ evaluated for an electron
population with double power-law spectrum (see Eq.34) and its dependence on the
lower momentum cutoff $p_1$. The behaviour of $P_1(s)$ in this case is quite
different with respect to the single power-law case (see
Fig.\ref{fig.p1sleggep1}) and depends mainly on the flatness of the electron
spectrum below the momentum cutoff $p_{cr}$. We notice that {\it i)} even for
low values of $p_1 \sim 0.5$ small frequency changes are unlikely, as shown by
the strongly a-symmetric shape of $P_1(s)$; {\it ii)} there are small
differences in the function $P_1(s)$ for different values of $p_1 \simlt p_{cr}$
because the electron density in this range is lower than in the single power-law
case and its variation is mild.
We also show in Fig.\ref{fig.gt.lp2} the function $ \tilde{g}(x)$ evaluated at
first order in $\tau$ for a double power-law electron population as in
Eq.(\ref{leggep2}). In this case, there is no well defined maximum of the SZ
effect since low frequency changes are unlikely. Moreover, the function $
\tilde{g}(x)$ is less sensitive to different values of the lower momentum cutoff
$p_1$ for $p_1 < p_{cr}$ because of the flatness of the electron spectrum in
this region. This fact is also confirmed by the variation of the zero location
of the non-thermal SZ effect (see Fig.\ref{fig.zeri.leggep2}). The location of
$x_0$ changes now less rapidly  than in the case of a single power-law
distribution (see Fig.11) due to the different shape of $\tilde{g}(x)$, and
hence of $P(s)$, caused by  the lower density of the low-energy electrons with
respect to the case of the single power-law spectrum. The zero of the
non-thermal SZ effect produced by a single non-thermal population is also
shifted to frequencies $x_0 \simgt 10$ (much higher than the value
$x_{0,th}=3.83$) due to the large frequency shifts experienced by the CMB
photons scattering the high energy non-thermal electrons.\\
 We finally show in
Fig.\ref{fig.dist.lp2} the spectral distortion $\Delta i(x)$ for different
values of $p_1$. Note, finally,  that the amplitude of $\Delta I(x)$ produced by
an electron population with a double power-law spectrum as in Eq.(\ref{leggep2})
is much lower than that produced by a population with a single power-law
spectrum as in Eq. (\ref{leggep1}).

The case of a system dominated by non-thermal electrons is not applicable
straightforwardly to galaxy clusters but rather to radio galaxies whose jets
inject in the ICM large quantities of relativistic non-thermal electrons. We
want to stress that our general approach is able to describe the Compton
scattering distortions even in the cases of the radio lobes and of the jets of
radio galaxies or AGNs with large optical depths and ultra-relativistic electron
energies. We will address more specifically this issue in a forthcoming paper.

\section{The total SZ effect produced by a combination of thermal and non-thermal populations}

For galaxy clusters which  contain two (or more) different electronic
populations, like radio-halo clusters which contain a thermal population
emitting X-rays and a non-thermal population producing the radio halo emission,
one has to evaluate the spectral distortion produced simultaneously by both
populations on the CMB radiation. Such a derivation has been not done so far. We
assume here that the two populations are independent and that no change on the
thermal population is induced by the non-thermal electrons. This condition is
reasonable for non-thermal electrons with energies $\simgt 150$ MeV which  do
not appreciably loose energy through Coulomb collisions in the ICM (see, e.g.,
Blasi \& Colafrancesco 1999; Colafrancesco \& Mele 2001) and hence do not
produce a substantial heating of the IC gas. Electrons with $E_e \simgt 150$ MeV
loose energy mainly through ICS and synchrotron losses. The radio halo emission
in galaxy clusters is, in fact, produced by electrons with energy $E_e \approx
16.4 ~{\rm GeV}~ (B/\mu G)^{-1/2} (\nu_r/GHz)^{1/2}$ (see, e.g., Colafrancesco
\& Mele 2001) which yield $E_e \sim 1.6 - 52$ GeV for a typical  IC magnetic
field at the level of $B = 1 \mu$G at the typical frequencies, $\nu_r \sim
10^{-2} - 10$ GHz, at which radio halo spectra are observed. For such energies
the non-thermal electrons do not strongly affect the thermal IC gas. However, we
discuss in the following the impact of different  non-thermal population spectra
on both the total SZ effect and on the cluster IC gas.

Under the hypothesis of independence between the thermal and non-thermal
electron populations, the probability that a CMB photon is scattered by an
electron of population A (say thermal) is not affected by the fact that the same
photon has been scattered by an electron of population B (say non-thermal).
Thus, at first order in $\tau$ we expect that the total SZ effect is given by
the sum of the first order SZ effects due to each single population.
 At higher order approximation in $\tau$, however, one has to consider the effect
of repeated scattering (see Sect.2).  It is intuitive to think that in this last
case the total SZ effect is not simply given by the sum of the separate SZ
effects at higher orders in $\tau$  and the terms describing the
cross--scattering between CMB photons and the electrons of populations A and B
have to be taken into account. In the following, we will derive in details the
general expression for the total SZ distortion caused by a combination of two
electronic populations.

Let us consider two electron populations with momentum distributions $f_A(p)$
and $f_B(p)$ with optical depths $\tau_A$ and $\tau_B$, respectively. We also
assume that the two electron distributions are both normalized to 1. The total
distribution of the electron momenta is
\begin{equation}
\label{ftot} f_e(p)=c_Af_A(p)+c_Bf_B(p) ~,
\end{equation}
with appropriate normalization coefficients $c_A$ and $c_B$. Since the total
distribution $f_e(p)$ is also normalized to 1, the coefficients $c_A$ and $c_B$
are related by the condition $c_A+c_B=1$. The coefficients $c_A$ and $c_B$ yield
the weight of each population with respect to the other and thus their ratio is
given by
\begin{equation}
\frac{c_A}{c_B}=\frac{\tau_A}{\tau_B}.
\end{equation}
Such a  condition, together with the normalization condition, yields:
\begin{equation} \label{coeffab}
c_A=\frac{\tau_A}{\tau_A+\tau_B}   \qquad     c_B=\frac{\tau_B}{\tau_A+\tau_B}.
\end{equation}
The total distribution $f_e(p)$ has an optical depth given by
\begin{equation} \label {tautot}
\tau=\tau_A + \tau_B.
\end{equation}
In fact, the probability that a CMB photon can suffer $n$ scatterings by the
electrons of each distribution is given by the sum of all the possible
combinations of the probability to suffer $n_A$ scatterings from the electrons
of the distribution $f_A$  and the probability to suffer $n_B$ scattering from
the electrons of the distribution $f_B$, with $n_A+n_B=n_e$. Since the two
distributions are independent, this probability is given by the product of the
two separate probabilities:
\begin{eqnarray}
p_n&=&\sum_{n_A+n_B=n}p_{n_A}\cdot p_{n_B}=\sum_{n_A+n_B=n}\frac{e^{-\tau_A} \tau_A^{n_A}}{n_A!}\cdot
\frac{e^{-\tau_B} \tau_B^{n_B}}{n_B!}= \nonumber \\
 &=&\sum_{n_A+n_B=n}
 \frac{e^{-(\tau_A+\tau_B)}\tau_A^{n_A}\tau_B^{n_B}}{n_A!n_B!}=\nonumber
 \\
 &=&\sum_{n_A=0}^n
 \frac{e^{-(\tau_A+\tau_B)}\tau_A^{n_A}\tau_B^{n-n_A}}{n_A!(n-n_A)!}=
 \nonumber \\
 &=&e^{-(\tau_A+\tau_B)}\sum_{n_A=0}^n
 \frac{1}{n!}\frac{n!}{n_A!(n-n_A)!}\tau_A^{n_A}\tau_B^{n-n_A}=
 \nonumber \\
 &=&\frac{e^{-(\tau_A+\tau_B)}(\tau_A+\tau_B)^n}{n!}\nonumber \\
 &\equiv&\frac{e^{-\tau}\tau^n}{n!}
\end{eqnarray}
(we assume here a Poisson probability distribution). This means that the electrons of the distribution $f_B$
simply add to the electrons of the distribution $f_A$. In fact, from the definition of optical depth, one gets:
\begin {eqnarray}
\tau&=&\sigma_T \int n_e d\ell \nonumber \\ &\equiv&\tau_A+\tau_B=\sigma_T \int (n_A+n_B) d\ell.
\end{eqnarray}
The exact expression for the spectral distortion produced by two electron populations can be evaluated from eqs.
(\ref{ptrasf})-(21), using the appropriate expression for $P_1(s)$.
 The general  expression for the distribution $P_1(s)$ is given by:
\begin{equation} \label{p1per2}
P_1(s)=\int_0^\infty dp f_e(p) P_s(s;p).
\end{equation}
Inserting Eq.(\ref{ftot}) in Eq.(\ref{p1per2}) one obtains the function $P_1(s)$
for two populations as a function of the distributions $P_1(s)$ of each single
population:
\begin{eqnarray} \label{p1stot}
P_1(s)&=&c_A \int_0^\infty dp f_A(p) P_s(s;p)+ c_B \int_0^\infty dp f_B(p) P_s(s;p) \nonumber \\ & \equiv & c_A
P_{1A}(s) + c_B P_{1B}(s).
\end{eqnarray}
The Fourier transform of Eq.(\ref{p1stot}) is
\begin{equation}
\tilde{P}_1(k)=c_A\tilde{P}_{1A}(k)+c_B\tilde{P}_{1B}(k)
\end{equation}
and from this expression one obtains the Fourier transform of the total
redistribution function:
\begin{eqnarray}
\tilde{P}(k)&=&e^{-\tau[1-\tilde{P}_1(k)]}= \nonumber \\ & =& e^{-(\tau_A+\tau_B)\left[1-
\frac{\tau_A}{\tau_A+\tau_B}\tilde{P}_{1A}(k)- \frac{\tau_B}{\tau_A+\tau_B}\tilde{P}_{1B}(k) \right]}=
\nonumber
\\
&=&e^{-\tau_A[1-\tilde{P}_{1A}(k)]-\tau_B[1-\tilde{P}_{1B}(k)]}= \nonumber \\ &=& \tilde{P}_A(k) \tilde{P}_B(k).
\end{eqnarray}
The exact total redistribution function $P_{tot}(s)$ is given by the convolution
of the redistribution functions of the separate electron populations:
\begin{equation}
P_{tot}(s)=P_A(s)\otimes P_B(s).
\end{equation}
Thus, the exact spectral distortion produced by two electron populations on the
CMB radiation is given by:
\begin{equation}
I_{tot}(x)=\int_{-\infty}^{+\infty} I_0(xe^{-s}) P_{tot}(s) ds,
\end{equation}
in terms of the exact total redistribution function $P_{tot}(s)$ given in
Eq.(62). We also give in the Appendix the analytic expressions for the
approximations at first and second order in $\tau$ of the total spectral
distortion $I_{tot}(x)$.

\subsection{The function $\tilde{g}(x)$ for a combination of two electron populations}
\label{par.gt2pop}

The total SZ effect produced by the combination of two electron populations can
be written again in terms of a Comptonization parameter $y_{tot}$ and of a
spectral function $\tilde{g}(x)$, in the form of Eq.(22). For two electron
populations, we can write the average temperature in  Eq.(\ref{temp.media}) as:
\begin{eqnarray}
\langle k_B T_e \rangle&=&\int_0^\infty dp f_e(p) \frac{1}{3} p
v(p) m_e c= \nonumber \\ &=&\int_0^\infty dp
\left[c_Af_A(p)+c_Bf_B(p) \right] \frac{1}{3} p v(p) m_e c=
\nonumber \\ &=&c_A \langle k_B T_e \rangle_A +c_B \langle k_B T_e
\rangle_B.
\end{eqnarray}
The Comptonization parameter $y_{tot}$ can be evaluated using
Eq.(\ref{y.tmedia}) as:
\begin{eqnarray}
y_{tot}&=&\frac{1}{m_ec^2}[c_A \langle k_B T_e \rangle_A + c_B \langle k_B T_e \rangle_B] (\tau_A+\tau_B)=
\nonumber \\ &=&\frac{1}{m_ec^2}[\tau_A \langle k_B T_e \rangle_A + \tau_B \langle k_B T_e \rangle_B]= \nonumber
\\ &=& y_A + y_B.
\end{eqnarray}
At first order in $\tau$, the function $ \tilde{g}(x)$ for two populations can be expressed as
\begin{eqnarray}
\tilde{g}(x)&=& \frac{\Delta i(x)}{y_{tot}}= \frac{\tau_A [j_{1A}(x)-j_0(x)] + \tau_B [j_{1B}(x)-j_0(x)]}{y_A+y_B}
\nonumber
\\
&=& \frac{y_A \tilde{g}_A(x)+ y_B \tilde{g}_B(x)}{y_A+y_B} ~,
\end{eqnarray}
in terms of the functions  $\tilde{g}(x)$ for each single electron population.
We notice that in this case the function $ \tilde{g}(x)$ depends from the
Comptonization parameters, and hence from the optical depths, of the two
electron populations even at the first order approximation in $\tau$. Moreover,
at higher orders in $\tau$, the function $ \tilde{g}(x)$ cannot be expressed
simply in terms of the spectral functions of the single populations because in
the expression for $\Delta i(x)$ [see Eq. (\ref{ord2tot})], cross-correlation
terms appear.

\subsection{The SZ effect produced by a combination of a thermal plus a
non-thermal populations}
\label{calcolo-term-nonterm}

Here we apply the previous derivation of the total SZ effect to a galaxy cluster
which contains both a termal and a non-thermal electron populations. In the
following computation of the total SZ effect for a Coma-like cluster we choose a
thermal population with the following parameters: $k_B T_e=8.5$ keV,
$n_{e,th}\simeq 3\cdot 10^{-3}$ cm$^{-3}$ and a cluster radius of $1\,
h_{50}^{-1}$ Mpc. With these parameters the optical depth of the thermal
population is
\begin{equation}
 \tau_{th}\simeq 6\cdot 10^{-3}
\left[\frac{n_{e,th}}{3\cdot10^{-3}\,cm^{-3}}\right] \left[
\frac{\ell}{1\,h_{50}^{-1}~Mpc}\right],
\end{equation}
and its pressure is:
\begin{equation}
P_{th}\simeq 2.55\cdot10^{-2} \,keV \,cm^{-3} \left[\frac{n_{e,th}}{3\cdot10^{-3}\,cm^{-3}}\right] \left[
\frac{k_B T_e}{8.5\, keV}\right].
\end{equation}
As for the spectrum of the non-thermal population we consider both the
phenomenological cases of a single and double power-law populations [see
Eq.(\ref{leggep1}) and Eq.(\ref{leggep2}) respectively]. The parameters we
consider are specifically: $\alpha=2.5$ and $p_2\rightarrow\infty$ for the
single power-law population and $\alpha_1=0.5$, $\alpha_2=2.5$, with
$p_{cr}=400$ and $p_2\rightarrow\infty$ for the double power-law population. We
consider $p_1$ as a free parameter in both cases. The relativistic electron
density has been normalized at $n_{e,rel}( \tilde{p}_1)=10^{-6}$ cm$^{-3}$ for $
\tilde{p}_1=100$ in both cases. The final SZ effect can be re-scaled simply to
different values of $n_{e,rel}$ as discussed in Sect. 3.

\begin{figure}[tbp]
\begin{center}
  \psfig{file=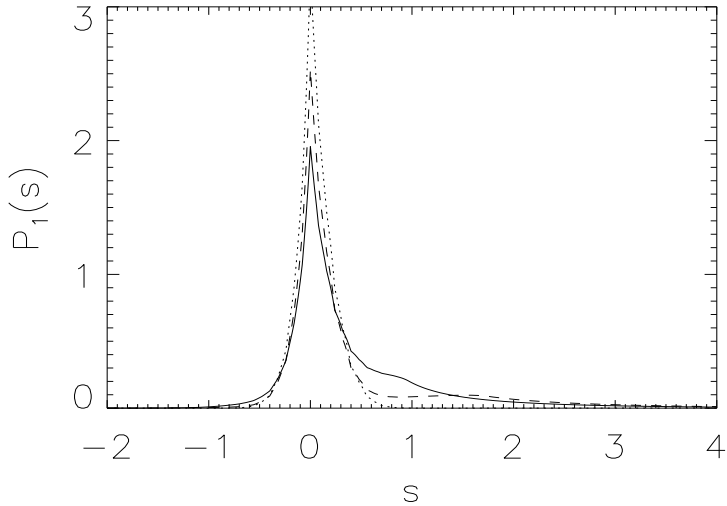,width=12.truecm,height=8.truecm}
  \caption{\footnotesize{The function $P_1(s)$ [see Eq. (\ref{p1stot})] for a combination
  of thermal population and non-thermal population as in Eq.(\ref{leggep1}) with
  $p_1=$ 0.5 (solid), 1 (dashes) and 10 (dotted).}}
  \label{fig.p1s.a.tlp1}
\end{center}
\end{figure}

The amplitude of the non-thermal the contribution to the total SZ effect
increases with decreasing value of $p_1$. For the case of a single non-thermal
power-law population it becomes appreciable for $p_1 \simlt 10$ while for $p_1
\simgt 100$ the total SZ effect is indistinguishable from the pure thermal
effect. In fact, for $p_1 \simgt 10$ the function $P_1(s)$ is nearly coincident
with that of a single thermal population as shown in Fig.\ref{fig.p1s.a.tlp1},
while for $p_1 <10$ the function $P_1(s)$ is more a-symmetric towards large and
positive values of $s$. The peaks of the non-thermal distributions at large $s$
are present in the total distribution $P_1(s)$ but their amplitude decreases
with $\tau_{rel}$ and becomes lower and lower with increasing values of $p_1$.

The amplitude and the shape of the total SZ effect in a cluster which contains
non-thermal phenomena depends on the spectral and spatial distribution of the
non-thermal population, given that the cluster IC gas properties are well known
from X-ray observations. We can take advantage of this fact to use the
non-thermal SZ effect as a tool to constrain the physical properties of the
non-thermal population. We show in Fig.16 the total spectral distortion due to
the SZ effect in a cluster with non-thermal population as a function of $p_1$.
Low values of $p_1 \simlt 1$ for the non-thermal population induce large
distortions with respect to the pure thermal SZ effect. However, for low values
of $p_1$ the pressure of the non-thermal population with single power-law
distribution exceeds substantially the thermal pressure $P_{th}$ (see Table
\ref{tab.dati.tlp1}); as a consequence, the low energy electrons produce an
exceedingly large pressure unbalance and also a large heating of the IC gas,
which are not acceptable.
\begin{table}[htb]
\begin{center}
\begin{tabular}{|c|*{2}{c|}}
\hline
 $p_1$ & $P_{rel}/P_{th}$ & $n_{e,th}/n_{e,rel}$\\
 \hline
 $0.5$ & $23.3$ & $1.06$ \\
 $1$ & $18.6$ & $3.00$ \\
 $10$ & $6.33$ & $94.87$ \\
 100 & $2.00$& $3000$\\
 1000 & $0.63$& $94868$\\
 \hline
 \end{tabular}
 \end{center}
 \caption{\footnotesize{Values of pressure and density ratios
 for a combination of thermal population and a non-thermal
 population with single power-law spectrum
 for different values of $p_1$ (see text for details).
}}
 \label{tab.dati.tlp1}
 \end{table}
\begin{figure}[tbp]
\begin{center}
\hbox{
  \psfig{file=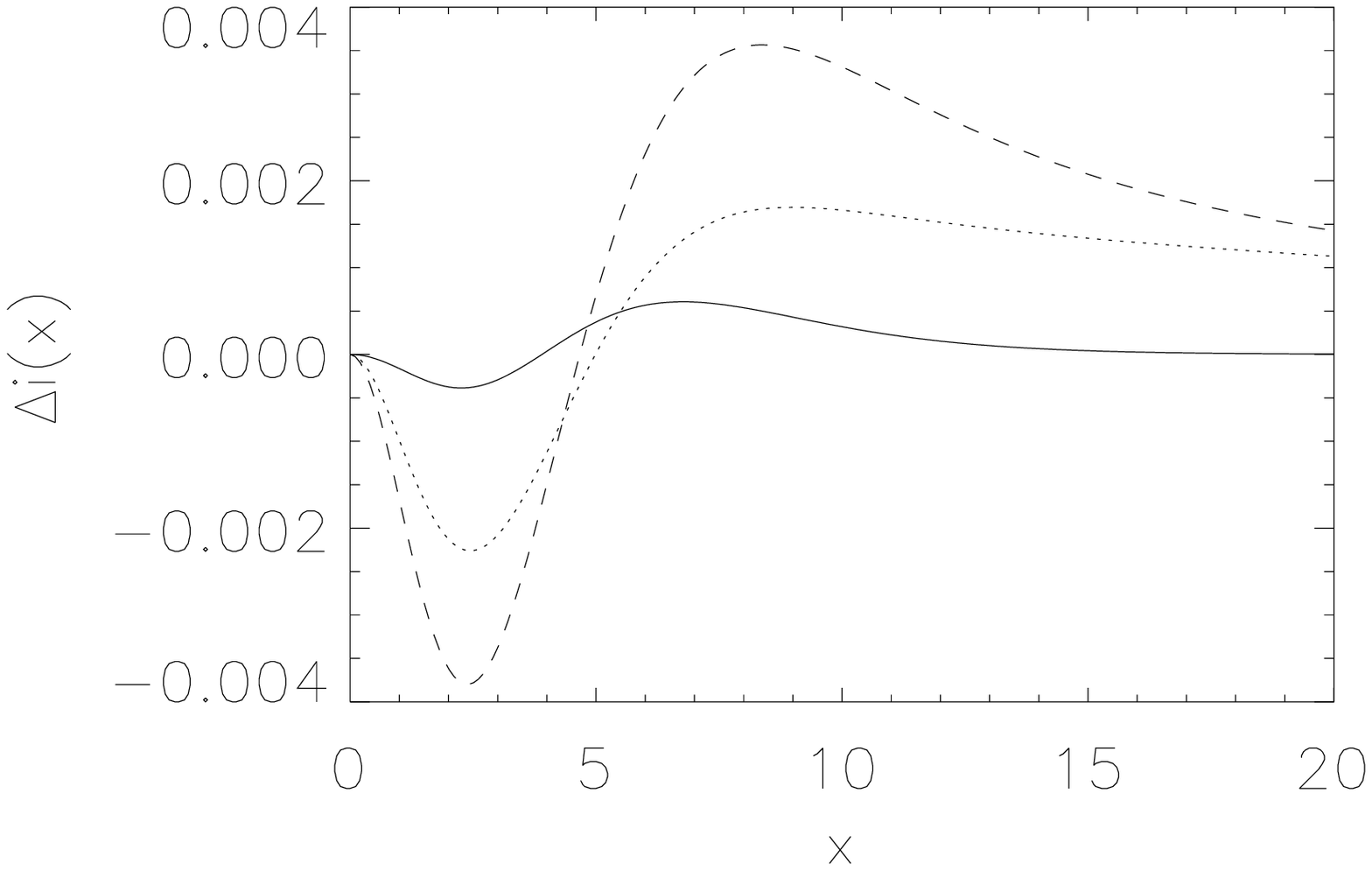,width=9.cm,height=8cm}
  \psfig{file=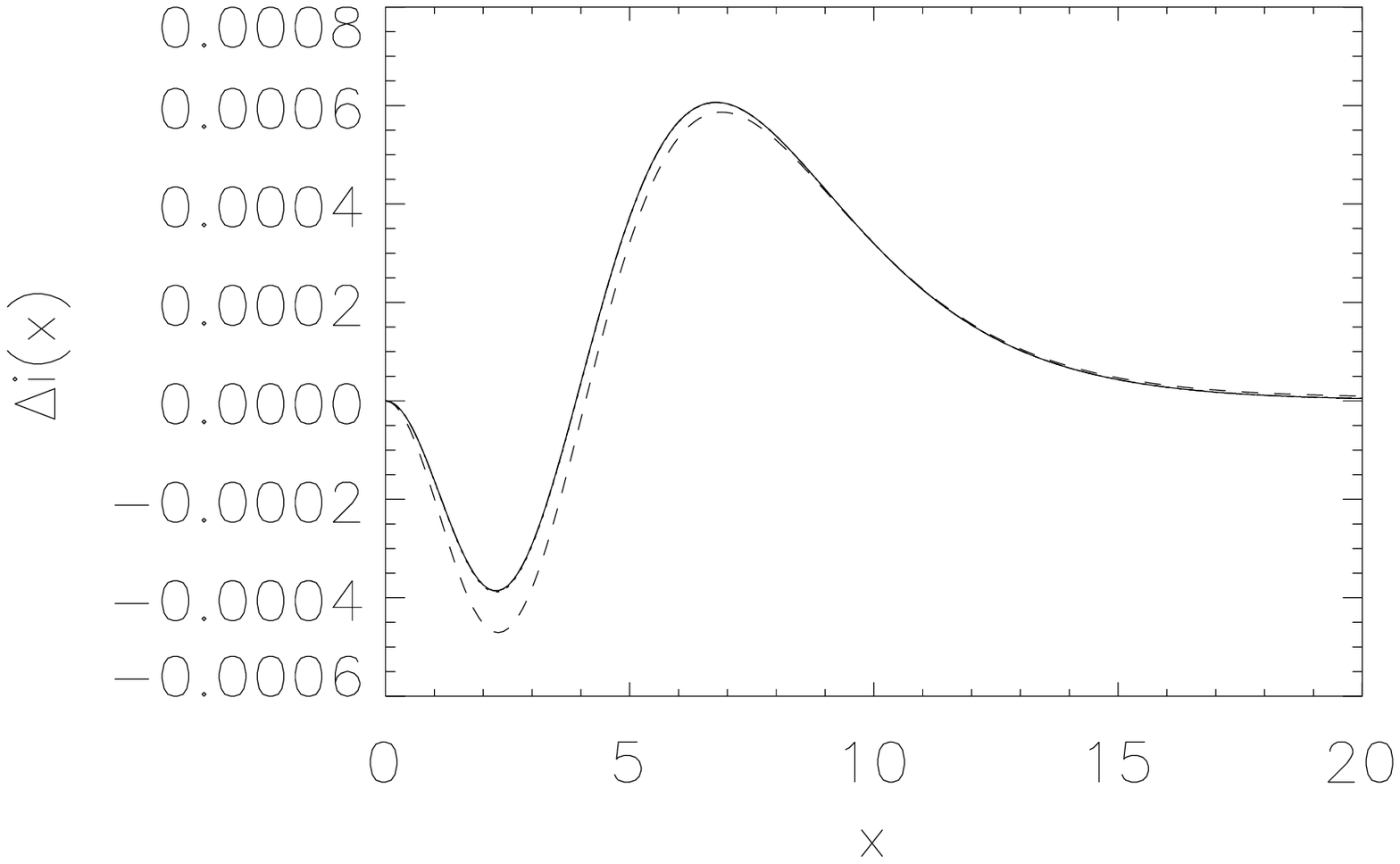,width=9.cm,height=8cm}
}
  \caption{\footnotesize{The spectral distortion,
  in units of $2(k_B T_0)^3/(hc)^2$, of a single thermal population of a
  Coma-like cluster (solid) is compared with the spectral distortion evaluated at first order in
  $\tau$ (see Eq. (\ref{ord1tot})), produced by a combination of thermal and
  non-thermal population given in Eq.(\ref{leggep1}) with $p_1=$ 0.5 (dashes) and 1
  (dotted) (left panel) and for
  $p_1=$ 10 (dashes), 100 (dotted) and 1000 (dot-dashes) (right panel).
  }}
  \label{fig.distors.a.tlp1}
\end{center}
\end{figure}

\begin{figure}[tbp]
\begin{center}
\hbox{
  \psfig{file=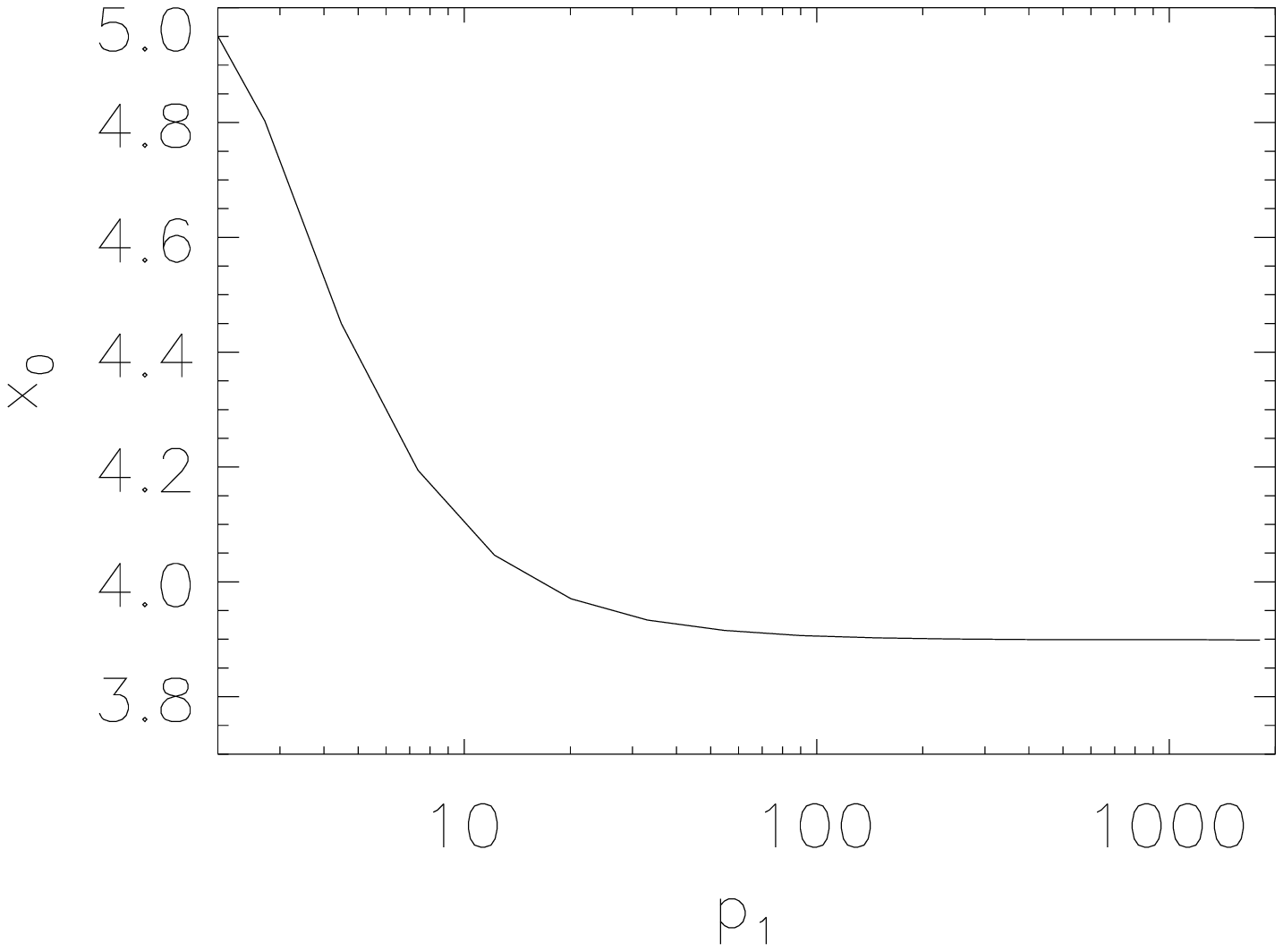,width=9.cm,height=8cm}
  \psfig{file=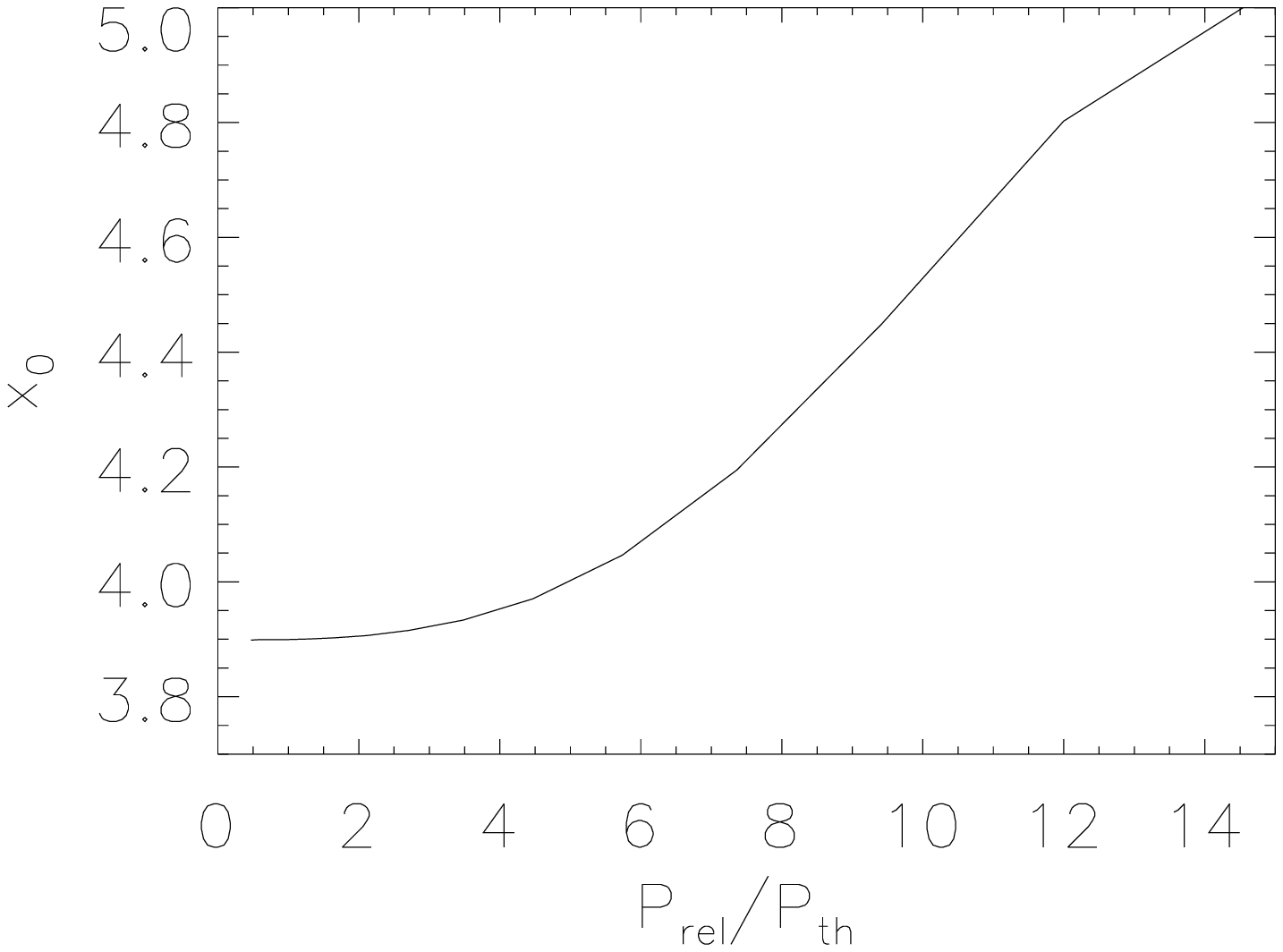,width=9.cm,height=8cm}
}
  \caption{\footnotesize{The behaviour of the zero of the total SZ effect evaluated
  for a combination of thermal and non-thermal
  population given by Eq.(\ref{leggep1}) is shown as a function of minimum momentum,
  $p_1$, of the non-thermal population
  (left) and of the pressure ratio $P_{rel}/P_{th}$ (right).}}
  \label{fig.zeri.tlp1}
\end{center}
\end{figure}

The frequency location of the zero of the total SZ effect, $x_0$,  is also a
powerful diagnostic of the presence and of the nature of the non-thermal
population. The frequency location of $x_0$ increases for decreasing values of
$p_1$ as the impact of the non-thermal population becomes larger (see
Fig.\ref{fig.zeri.tlp1}). Consistently, $x_0$ increases with increasing values
of the pressure ratio $P_{rel}/P_{th}$, as also shown in
Fig.\ref{fig.zeri.tlp1}.

\begin{figure}[tbp]
\begin{center}
  \psfig{file=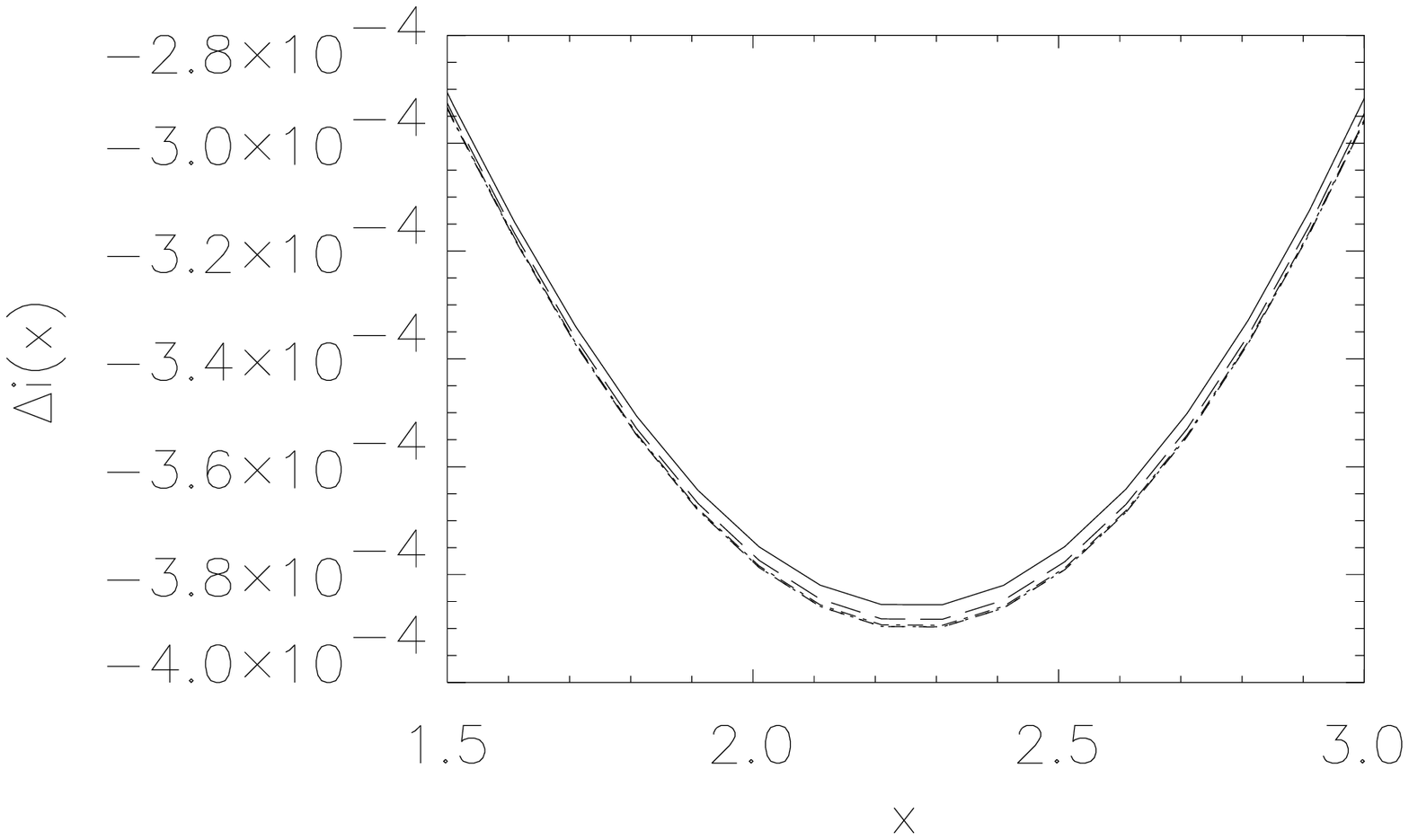,width=10cm,height=8cm}
  \psfig{file=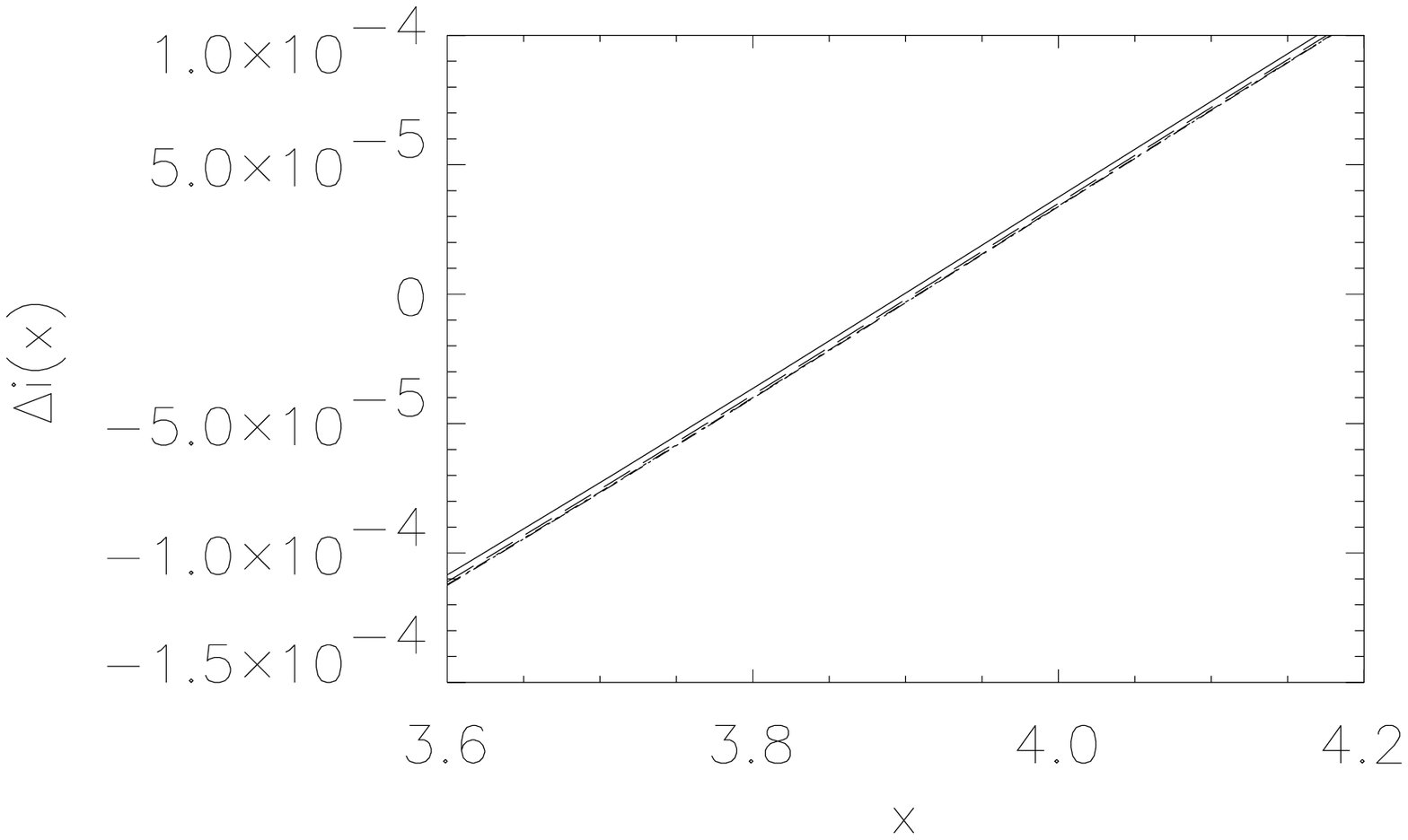,width=10cm,height=8cm}
  \psfig{file=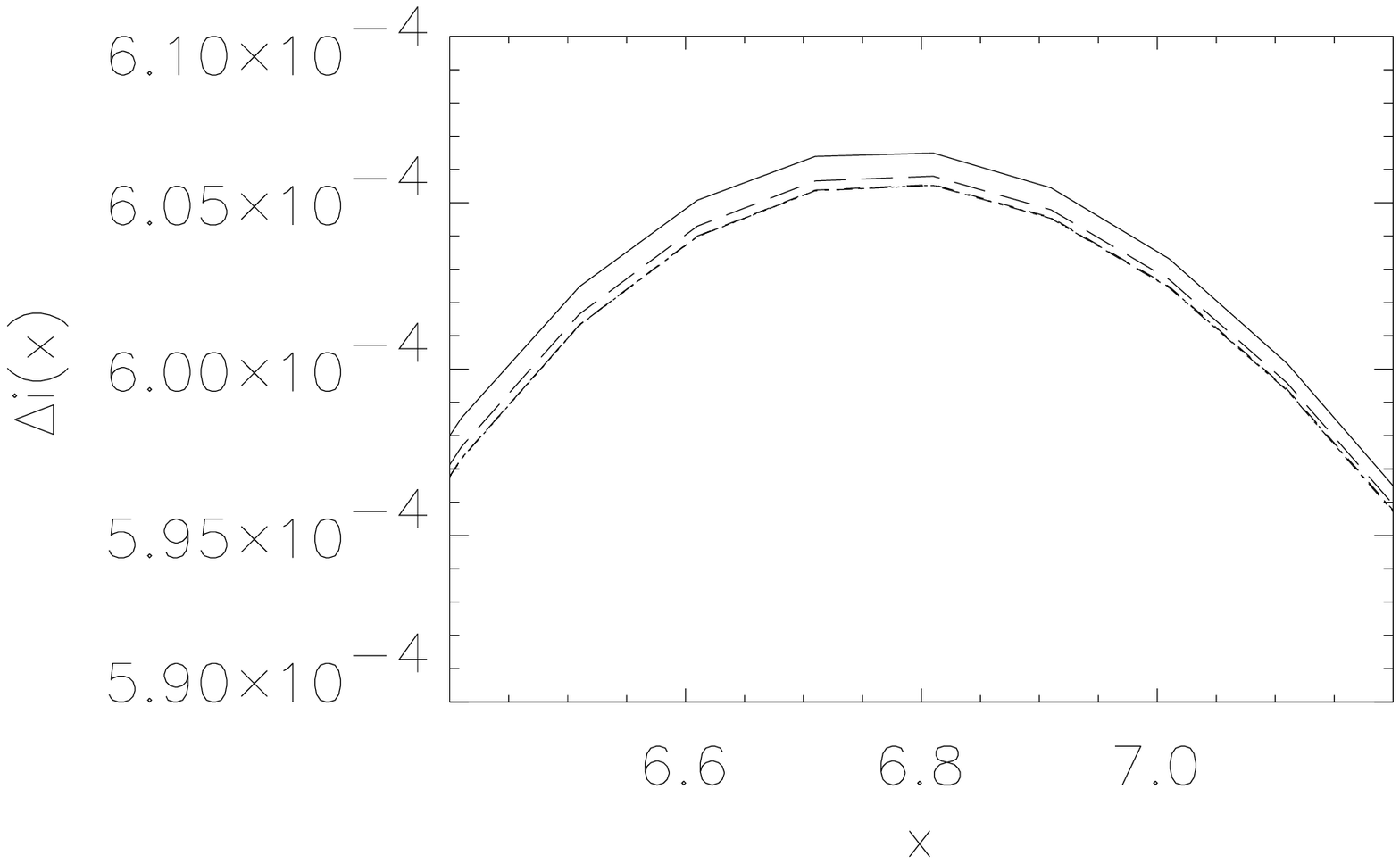,width=10cm,height=8cm}
  \caption{\footnotesize{The spectral distortion,
  in units of $2(k_B T_0)^3/(hc)^2$, of a single thermal population
  (solid) is compared with the spectral distortion evaluated at first
  order in $\tau$
   produced by a combination of thermal and
  non-thermal population given in Eq.(\ref{leggep2}) with
  $n_{e,rel}( \tilde{p}_1=100)=1\cdot10^{-6}$
  and $p_1=$ 0.5 (dashes), 1 (dotted), 10 (dot-dashes) and 100
  (long dashes). We show the enlargement in the regions of the
  minimum (top), of the zero (mid) and of the maximum
  (bottom) of the total SZ effect.
}}
  \label{fig.distors.tlp2a}
\end{center}
\end{figure}

\begin{figure}[tbp]
\begin{center}
  \psfig{file=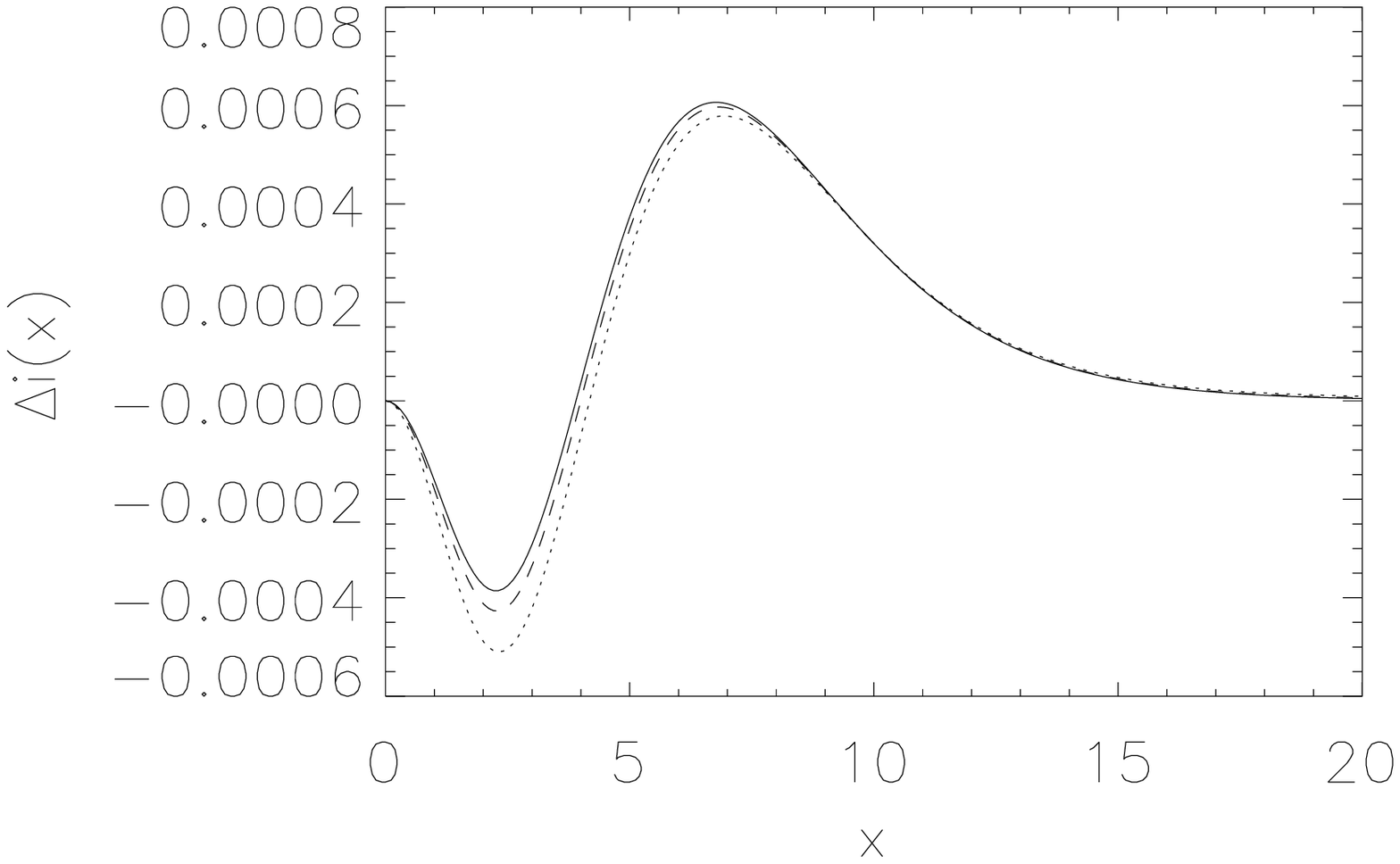,width=12cm,height=10cm}
  \caption{\footnotesize{The spectral distortion,
  in units of $2(k_B T_0)^3/(hc)^2$, of a single thermal population
  (solid) is compared with the spectral distortion evaluated at first order in $\tau$
   produced by a combination of thermal and
  non-thermal population given in Eq.(\ref{leggep2}) with $p_1=$ 0.5 and
  $n_e( \tilde{p}_1=100)=$
  10$^{-5}$ (dashes) and $3\cdot10^{-5}$ cm$^{-3}$ (dotted).}}
  \label{fig.distors.bc.tlp2}
\end{center}
\end{figure}

\begin{figure}[tbp]
\begin{center}
\hbox{
  \psfig{file=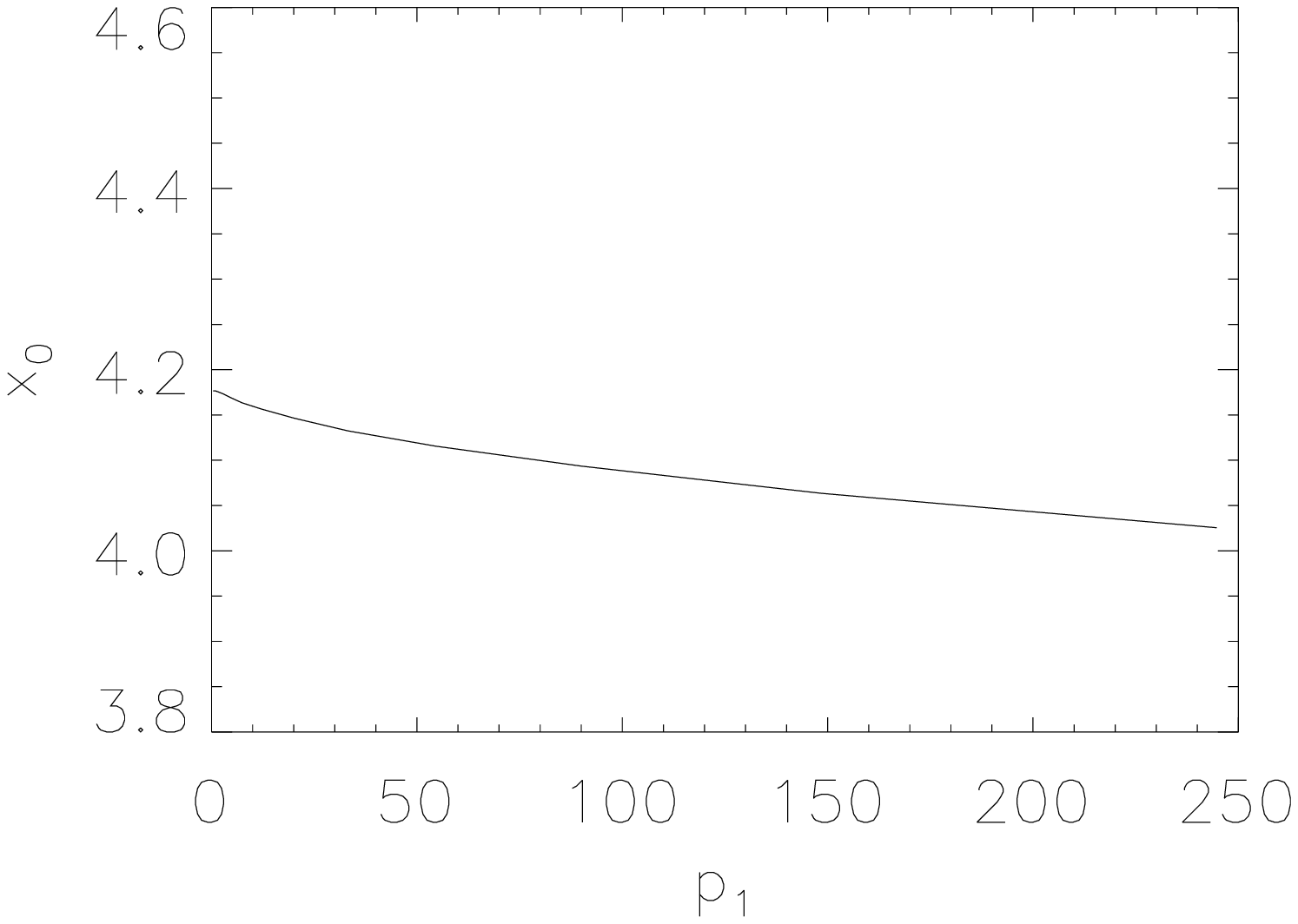,width=9.cm,height=8cm}
  \psfig{file=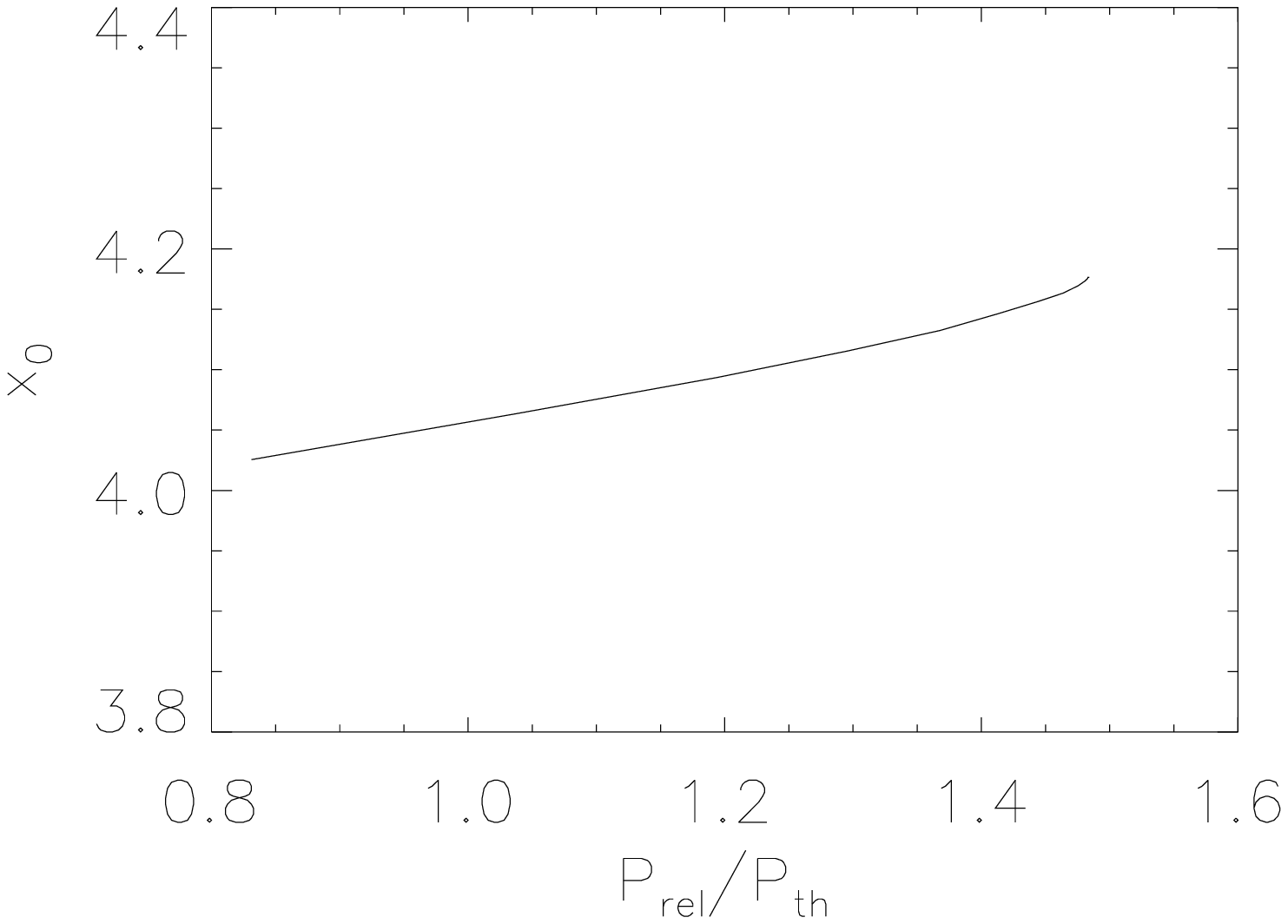,width=9.cm,height=8cm}
}
  \caption{\footnotesize{The behaviour of the zero of the total SZ effect
 produced by a combination of a thermal and non-thermal population given in Eq.(\ref{leggep2}) with
  $n_{e,rel}(\tilde{p}_1=100)=3\cdot10^{-5}$ cm$^{-3}$ as a function of $p_1$ (left) and
 of the pressure ratio $P_{rel}/P_{th}$ (right).}}
\label{fig.zeri.tlp2}
\end{center}
\end{figure}

\begin{figure}[tbp]
\begin{center}
  \psfig{file=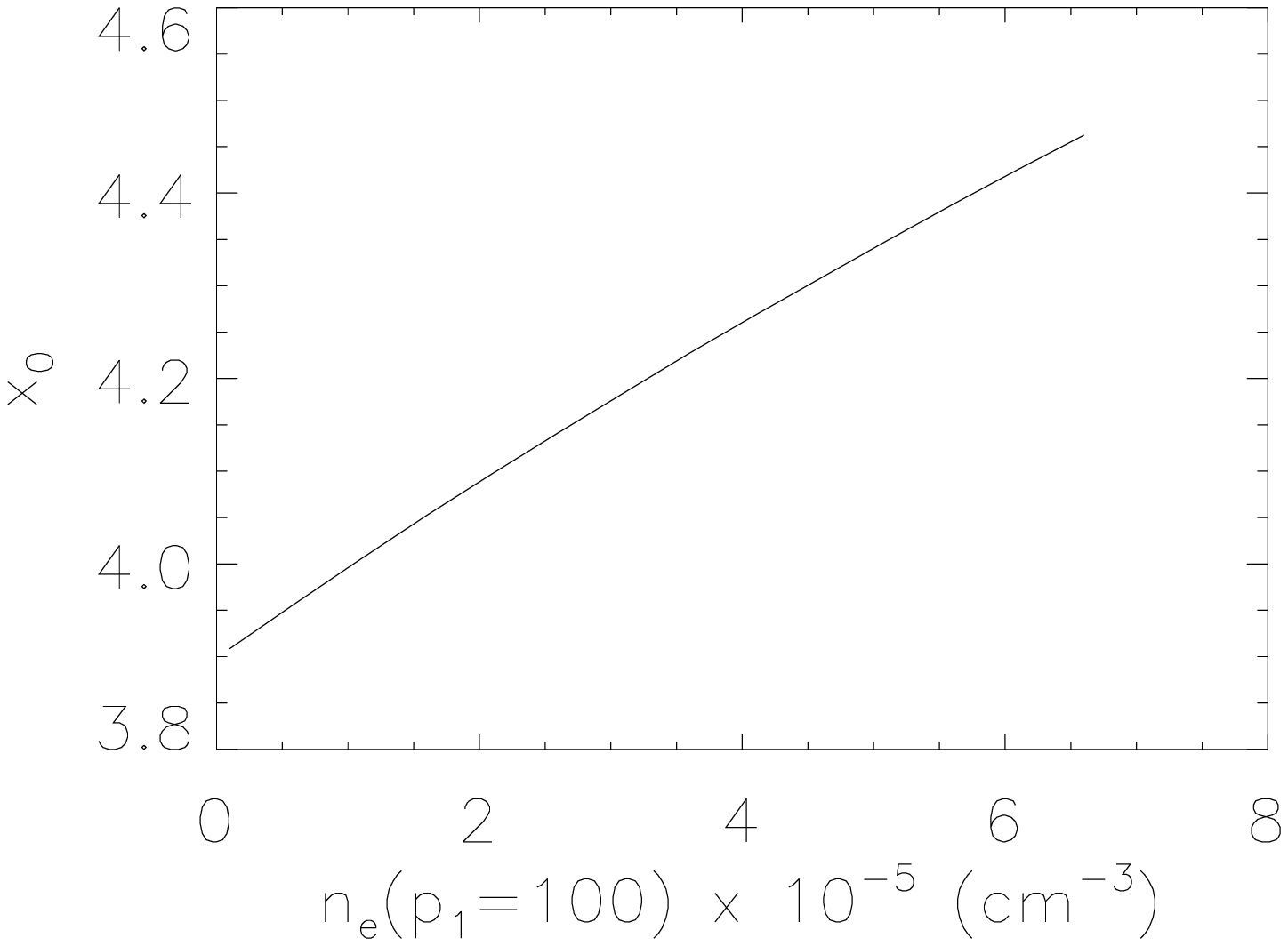,width=12cm,height=8cm}
  \caption{\footnotesize{The behaviour of the zero of the total SZ effect
 produced by a combination of a thermal and non-thermal population given in Eq.(\ref{leggep2})
 with $p_1=0.5$ is shown as a function of $n_{e,rel}( \tilde{p}_1=100)$.
  }}\label{fig.zeri2.tlp2}
\end{center}
\end{figure}

In the case of a double power-law non-thermal population the ratio
$P_{rel}/P_{th}$ always remains of  order $10^{-2}$ even for vary low values
$p_1 \sim 0.5$ (see Table \ref{tab.dati.tlp2}), and in this case there is a
negligible dynamical and thermal influence of the non-thermal population on the
thermal one.
\begin{table}[htb]
\begin{center}
\begin{tabular}{|c|*{2}{c|}}
\hline
 $p_1$ & $P_{rel}/P_{th}$ & $n_{e,th}/n_{e,rel}$\\
 \hline
 $0.5$ & $4.94\cdot10^{-2}$ & $1926$ \\
 $1$ & $4.94\cdot10^{-2}$ & $1948$ \\
 $10$ & $4.84\cdot10^{-2}$ & $2127$ \\
 100 & $3.87\cdot10^{-2}$& $3000$\\
 \hline
 \end{tabular}
 \end{center}
 \caption{\footnotesize{ Values of pressure and density ratio
 for a thermal population and a non-thermal  population with double power-law
 spectrum for different values of $p_1$ (see text for details).
}}
 \label{tab.dati.tlp2}
 \end{table}
In this case, also the total distribution $P_1(s)$ and the total spectral
distortion are little affected by the value of $p_1$. We show in
Fig.\ref{fig.distors.tlp2a} the spectral distortion of a single thermal
population compared with the spectral distortion produced by a combination of
thermal and non-thermal population with double power-law spectrum.
The contribution of the non-thermal population to the total SZ effect depends on
the normalization of the density $n_{e,rel}$ (which is highly uncertain). For
increasing values of the density of the non-thermal population the non-thermal
SZ effect becomes more relevant and produces substantial changes to the SZ
distortion produced by a single thermal population at frequencies $x \simlt 10$
(see Fig.\ref{fig.distors.bc.tlp2}). Much milder changes are present in the high
frequency range $x \simgt 15$ due to the low amplitude of the non-thermal signal
in this region (see, e.g., Fig.14).
\begin{table}[htb]
\begin{center}
\begin{tabular}{|c|*{3}{c|}}
\hline
 $n_{e,rel}( \tilde{p}_1=100)$ (cm$^{-3}$)& $P_{rel}/P_{th}$ & $n_{e,th}/n_{e,rel}$ & $x_0$\\
 \hline
  $1\cdot10^{-6}$ & $4.94\cdot10^{-2}$ & $1926$ &3.9085\\
 $1\cdot10^{-5}$ & $0.49$ & $192$ &3.9965\\
  $3\cdot10^{-5}$ & $1.48$ & $64$ &4.1765\\
 \hline
 \end{tabular}
 \end{center}
 \caption{\footnotesize{ Values of pressure and density ratio
 and the corresponding position of the zero
 for a thermal population and a non-thermal double power-law
 population for $p_1=0.5$
 as a function of $n_{e,rel}( \tilde{p}_1=100)$ (see text for details).
}}
 \label{tab.dati2.tlp2}
 \end{table}
We also show in Fig.\ref{fig.zeri.tlp2} the behaviour of $x_0$ as a function of
$p_1$ and of the pressure ratio for a density normalization
$n_{e,rel}(\tilde{p}_1=100)=3\cdot10^{-5}$ cm$^{-3}$.
High values of the ratio $P_{rel}/P_{th}$ produce a substantial displacement of
$x_0$ up to values $x_0 \simgt 4$ which correspond to frequencies $\simgt 227$
GHz.
In Fig.\ref{fig.zeri2.tlp2} we also show the behaviour of  $x_0$ a a
function of $n_{e,rel}(\tilde{p}_1=100)$ for $p_1=0.5$. We also report in Table
\ref{tab.dati2.tlp2} the values of the density and the corresponding values of
$P_{rel}/P_{th}$ and of  $n_{e,th}/n_{e,rel}$ as well as the frequency position
of the zero of the total SZ effect.
It is interesting to note in this context that the precise detection of the
frequency location of $x_0$ provides a direct measure of the ratio \prapp and,
in turn, of the energy density and of the properties of the spectrum of the
non-thermal population once the pressure of the thermal population is known from
X-ray observations.
In fact, all these physical information contributes to set the specific
amplitude and the spectral shape of the total SZ effect (see Figs.17 and 20 and
Figs. 16, 18 and 20 for the different SZ effect features produced by two
different non-thermal electron populations).
This is a relevant and unique feature of the non-thermal SZ effect in clusters
since it yields a measure of the total pressure in relativistic non-thermal
particles in the cluster atmosphere, an information which is not easily
accessible from the study of other non-thermal phenomena like radio halos and/or
EUV or hard X-ray excesses. Thus, the detailed observations of the frequency
shift of $x_0$ in clusters provides unambiguously relevant constraints to the
relativistic particle content in the ICM and to their energy distribution. Such
a measurement is also crucial to determine the true amount of kinematic SZ
effect due to the cluster peculiar velocity since it is usually estimated from
the residual SZ signal at the location of the zero of the thermal (relativistic)
SZ effect. Due to the steepness of the SZ spectral shape in the region of the
null of the overall SZ effect,  the possible additional non-thermal SZ signal
must be determined precisely in order to derive reliable limits to the kinematic
SZ effect. We will present a more detailed analysis of the kinematic SZ effect
in a forthcoming paper.

\section{The SZ effect produced by a combination of different thermal populations.}

\begin{figure}[tbp]
\begin{center}
  \psfig{file=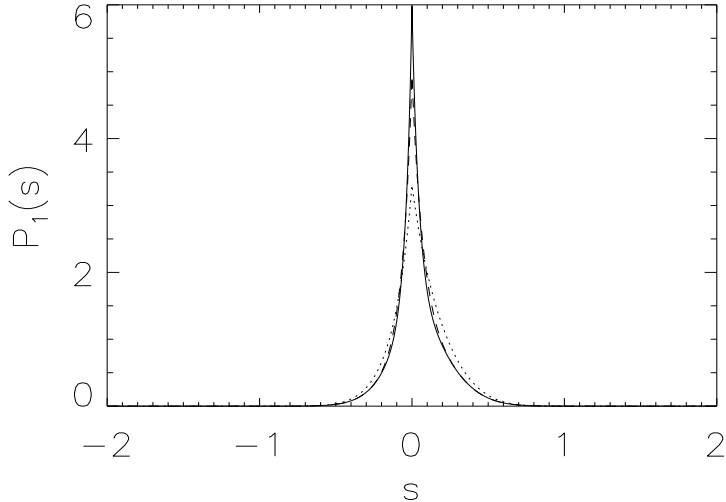,width=12cm,height=8cm}
  \caption{\footnotesize{The function $P_1(s)$ computed for a combination of two
 thermal populations with $k_B T_{e,1} =8.5$ and $ k_B T_{e,2} =0.5$ keV (solid line)
 and with $k_B T_{e,1} =8.5$ and $k_B T_{e,2} = 1$ keV (dashed line) is compared with
 that of the single thermal population with  $k_B T_e=$ 8.5 keV (dotted line).
 }}
  \label{fig.p1s.tt1}
\end{center}
\end{figure}

\begin{figure}[tbp]
\begin{center}
  \psfig{file=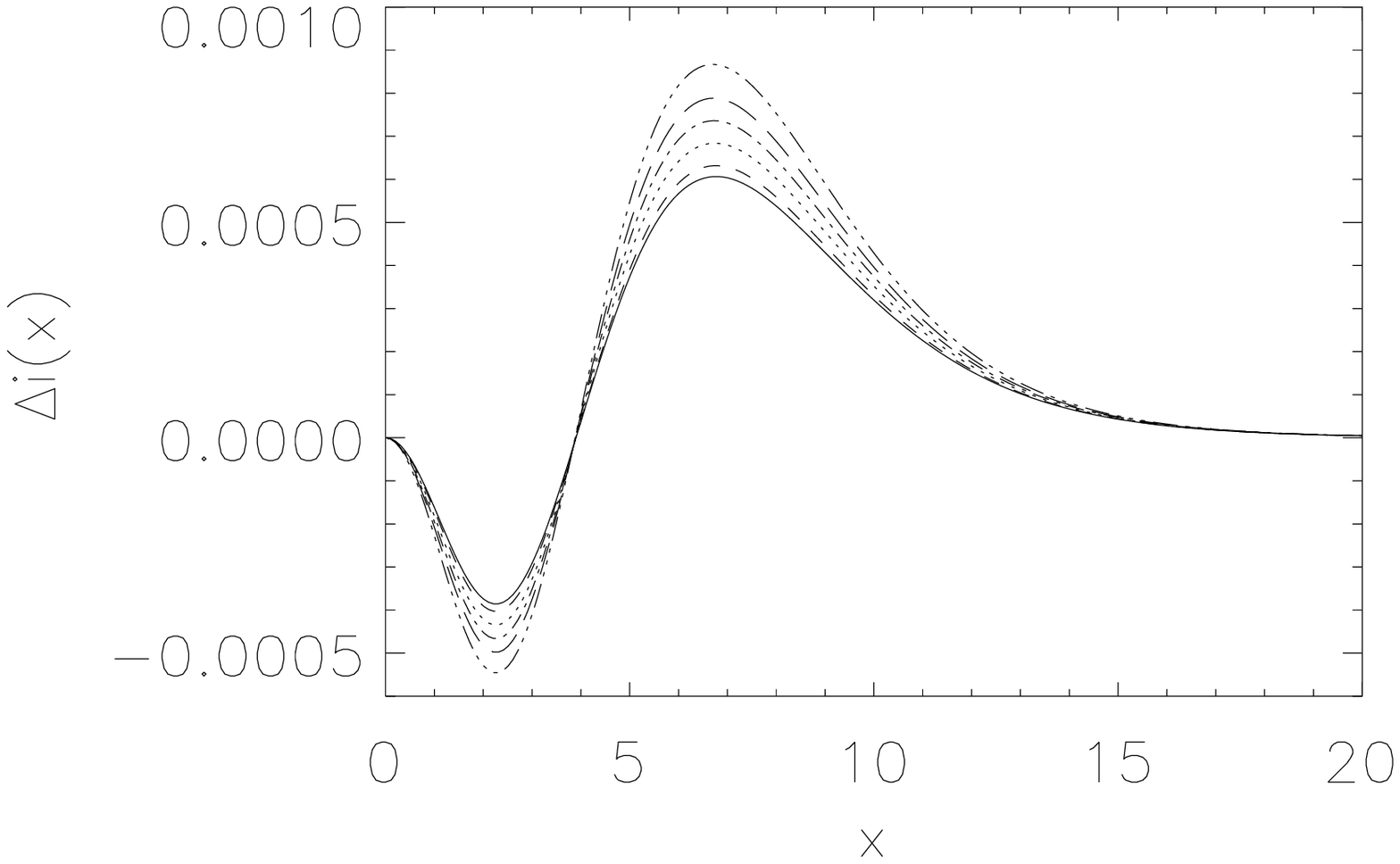,width=12cm,height=10cm}
  \caption{\footnotesize{The spectral distortion in units of
  $2(k_B T_0)^3/(hc)^2$ produced by a single thermal population
  with $k_BT_{e,1} = 8.5$ keV and $n_{e,1}=3 \cdot 10^{-3} cm^{-3}$ (solid line) is
  compared with that produced by a
  combination of two thermal populations with  $k_B T_{e,2}=1$ keV and
  $n_{e,2}=10^{-3}$ (dashed line), $3\cdot10^{-3}$ (dotted),
  $5\cdot10^{-3}$ (dot-dashed), $7\cdot10^{-3}$ (long dashes) and $10^{-2}$
  (dot-dot-dashes) cm$^{-3}$.
  }}
  \label{fig.distors.ae.tt1}
\end{center}
\end{figure}

The general derivation of the total SZ effect produced by different electron
populations described in Sect. 4 above can be generally applied also to the case
of a combination of different thermal populations. So, in this Section we
consider the SZ effect produced by two thermal populations with different
temperatures and densities residing in the same cluster. The presence of an
additional thermal IC population of quite low temperature, $T \simlt 1$ keV, has
been considered as a possibility to explain the origin of the EUV/soft X-ray
excess observed in some nearby clusters (Coma and Virgo for instance, see e.g.
Lieu et al. 1999) and it is also expected in the outer regions of the clusters
due to infall of warm gas from large scale filaments. Here we want to explore
the modifications that such additional warm component can introduce into the
overall SZ effect in galaxy clusters. More interestingly, we want to show that
is possible to obtain constraints on any additional thermal component in galaxy
clusters using SZ observations. We consider for this additional thermal
population temperatures in the range $k_B T_{e,2}=0.5-1$ keV and densities
$n_{e,2}$ in the range $10^{-3} - 10^{-2}$ cm$^{-3}$.  We also assume here, for
simplicity, that the two thermal populations have the same spatial distribution.
Moreover, we report in Tables \ref{tab.dati.tt1} and \ref{tab.dati.tt05} the
pressure ratio, the density ratio and the frequency location of the zero of the
total SZ effect for two values of the temperature of the warm population: $k_B
T_{e,2}=0.5$ and $1$ keV, respectively.
\begin{table}[htb]
\begin{center}
\begin{tabular}{|c|*{3}{c|}}
\hline
 $n_{e,2}$ (cm$^{-3}$)& $P_2/P_1$ & $n_{e,1}/n_{e,2}$ & $x_0$\\
 \hline
  $1\cdot10^{-3}$ & $3.92\cdot10^{-2}$ & $3$ &3.8955\\
 $3\cdot10^{-3}$ & $0.12$ & $1$ &3.8945\\
  $5\cdot10^{-3}$ & $0.20$ & $0.6$ &3.8895\\
 $7\cdot10^{-3}$ & $0.27$ & $0.43$ &3.8855\\
  $1\cdot10^{-2}$ & $0.39$ & $0.3$ &3.8805\\
 \hline
 \end{tabular}
 \end{center}
 \caption{\footnotesize{ Values of pressure and density ratios
 and the corresponding position of the zero
 for a combination of two thermal populations with $k_B
 T_{e,1}=8.5$ and $k_B T_{e,2}=1$ keV
 for different values of $n_{e,2}$. A value $n_{e,1} = 3 \cdot 10^{-3} cm^{-3}$
 is adopted here.
}}
 \label{tab.dati.tt1}
 \end{table}
 \begin{table}[htb]
\begin{center}
\begin{tabular}{|c|*{3}{c|}}
\hline
 $n_{e,2}$ (cm$^{-3}$)& $P_2/P_1$ & $n_{e,1}/n_{e,2}$ & $x_0$\\
 \hline
  $1\cdot10^{-3}$ & $1.96\cdot10^{-2}$ & $3$ &3.9005\\
 $3\cdot10^{-3}$ & $5.88\cdot10^{-2}$ & $1$ &3.8965\\
  $5\cdot10^{-3}$ & $9.80\cdot10^{-2}$ & $0.6$ &3.8935\\
 $7\cdot10^{-3}$ & $0.14$ & $0.43$ &3.8895\\
  $1\cdot10^{-2}$ & $0.20$ & $0.3$ &3.8855\\
 \hline
 \end{tabular}
 \end{center}
 \caption{\footnotesize{ Values of pressure and density ratio
 and the corresponding position of the zero
 for a combination of two thermal populations with $k_B
 T_{e,1}=8.5$ and $k_B T_{e,2}=0.5$ keV
 for different values of $n_{e,2}$. A value $n_{e,1} = 3 \cdot 10^{-3} cm^{-3}$
 is adopted here.
}}
 \label{tab.dati.tt05}
 \end{table}
The presence of additional cooler electrons produces a tightening of the photon
re-distribution function $P_1(s)$ (see Fig.\ref{fig.p1s.tt1}) around $s=0$. At
the same time, the total optical depth increases and, as a consequence, the
total spectral distortion shows a deeper minimum and a higher maximum with
respect to the single thermal population case (see
Fig.\ref{fig.distors.ae.tt1}). It is evident that high values of the density of
the cooler population can introduce substantial changes to the total SZ effect.

\begin{figure}[tbp]
\begin{center}
\hbox{
  \psfig{file=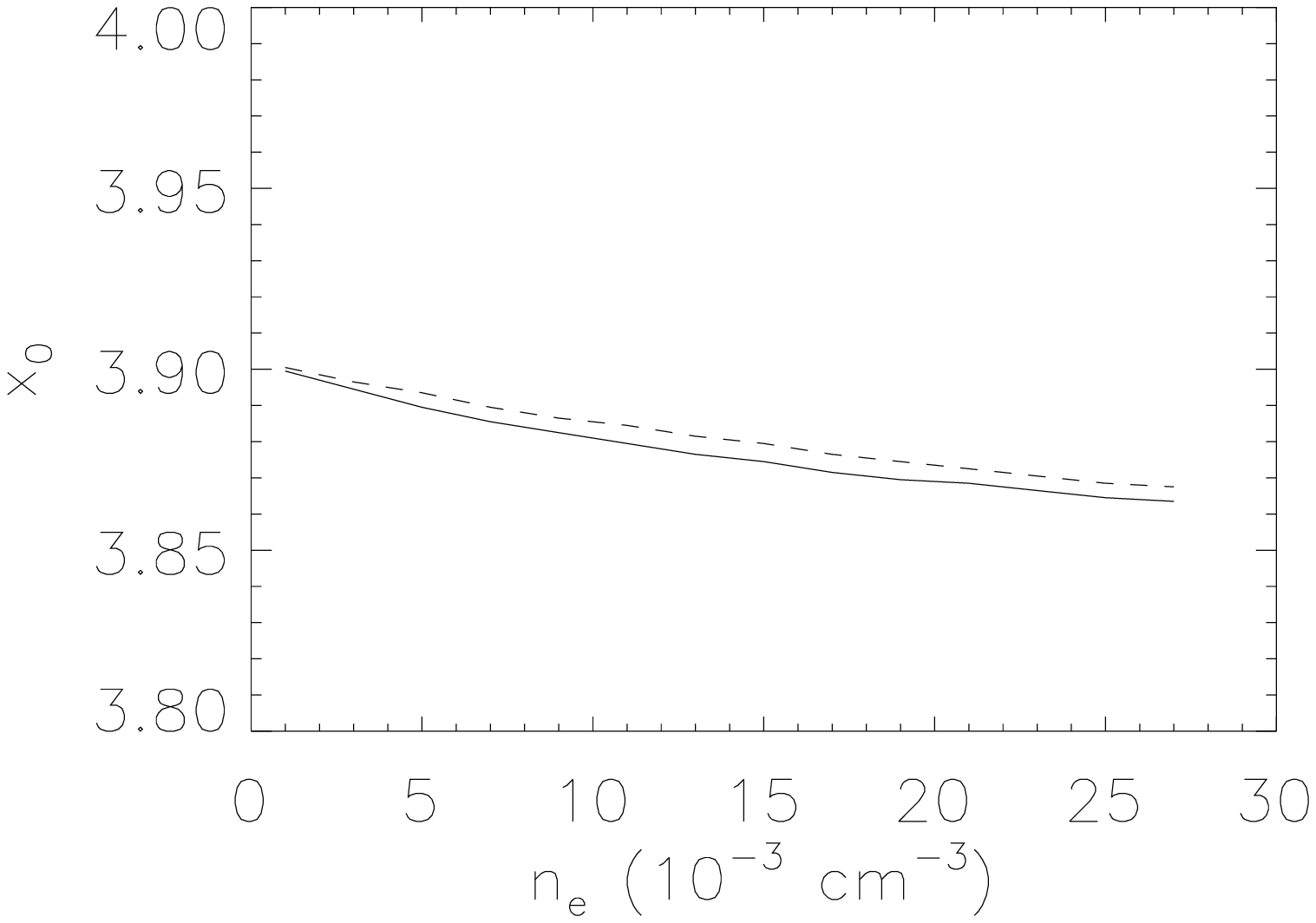,width=9.cm,height=8cm}
  \psfig{file=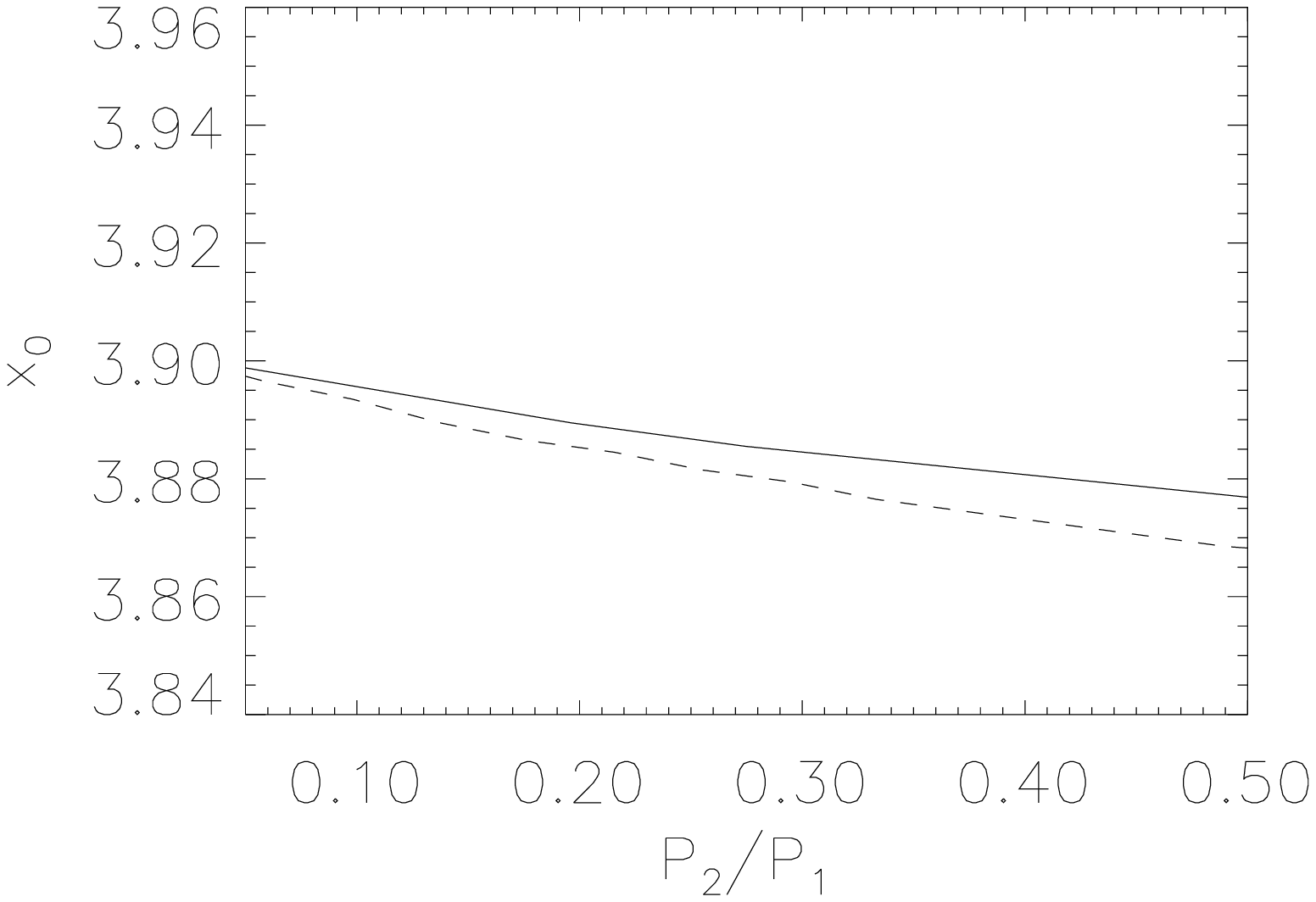,width=9.cm,height=8cm}
}
  \caption{\footnotesize{The behaviour of the zero of the total SZ effect for a combination of two thermal
  populations with  $k_B T_{e,1}=8.5$ keV and
  $k_B T_{e,2}=1$ keV (solid line) and $k_B T_{e,1}=8.5$ keV and
  $k_B T_{e,2}=0.5$ keV (dashed line) is shown as a function of the density $n_{e,2}$ (left panel)
  and of the pressure ratio $P_2/P_1$ (right panel).
  A value $n_{e,1}=3 \cdot 10^{-3} cm^{-3}$ is adopted here.}}
  \label{fig.zeri.tt1}
\end{center}
\end{figure}

\begin{figure}[tbp]
\begin{center}
\hbox{
  \psfig{file=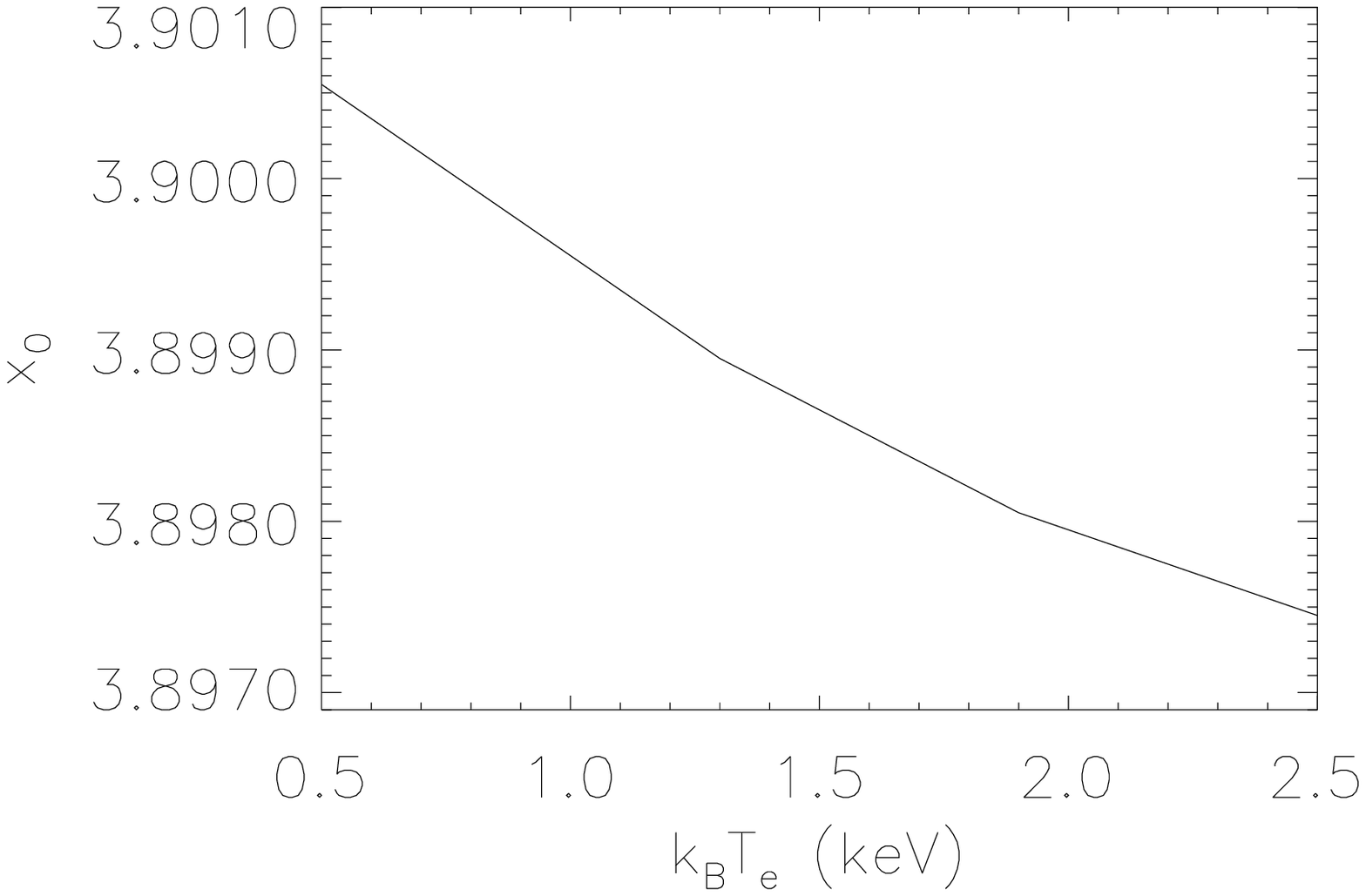,width=9.cm,height=8.cm}
  \psfig{file=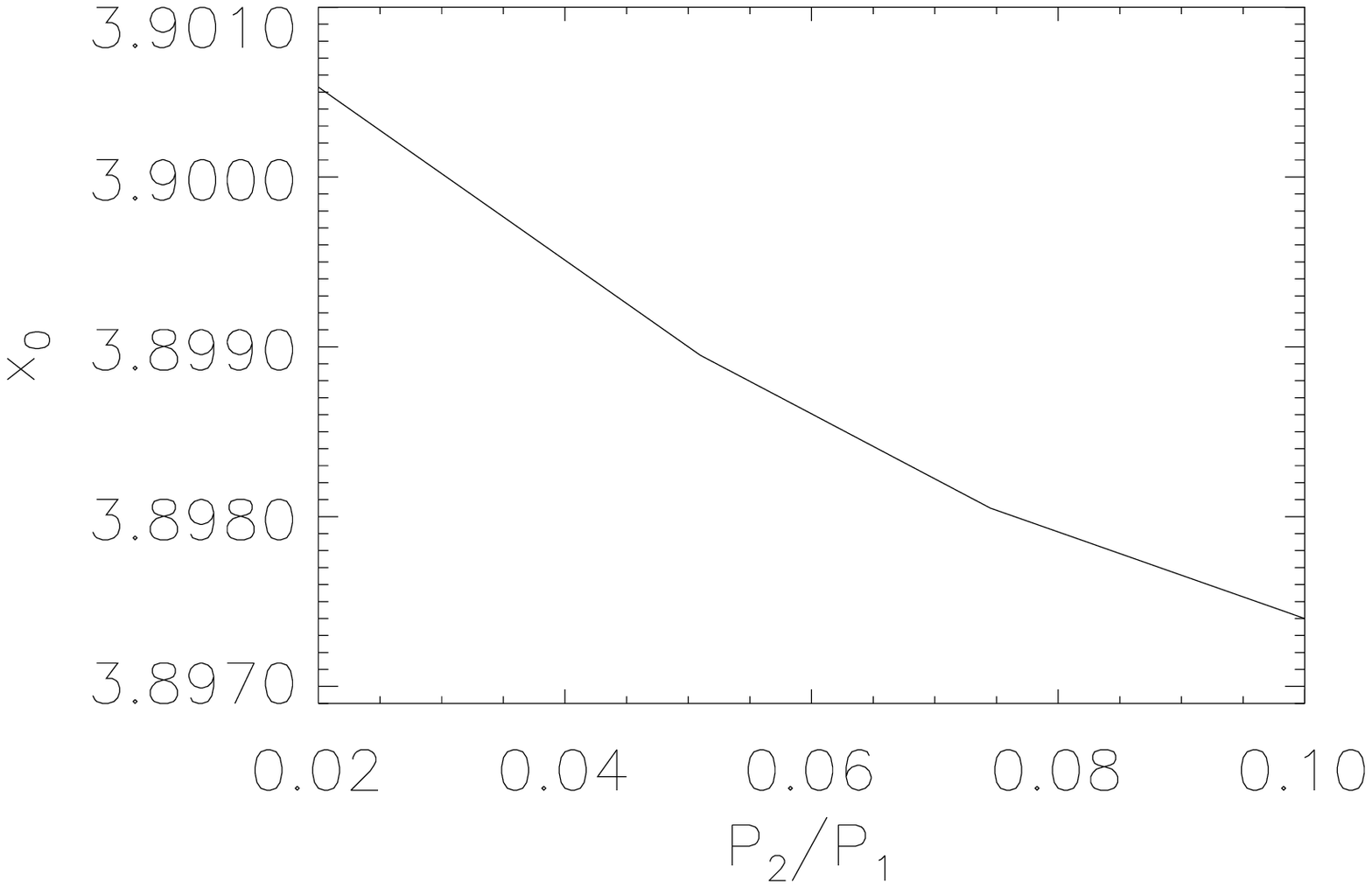,width=9.cm,height=8.cm}
}
  \caption{\footnotesize{The behaviour of the zero of the total SZ effect for a combination of two thermal
  populations with  $k_B T_{e,1}=8.5$ keV and $n_{e,2}=1\cdot10^{-3}$ cm$^{-3}$
 is shown as a function of the temperature $k_B T_{e,2}$(left panel) and of the pressure
 ratio $P_2/P_1$ (right panel). A value $n_{e,1}=3 \cdot 10^{-3} cm^{-3}$ is adopted here.}}
  \label{fig.zeritemp.tt}
\end{center}
\end{figure}

We also show in Fig.\ref{fig.zeri.tt1} the variation of the zero of the total SZ
effect as a function of the density of the cooler population and of the pressure
ratio $P_2/P_1$, respectively. As the cooler population becomes more relevant
(i.e., for increasing values of its pressure $P_2$), the position of $x_0$ moves
towards lower frequency values, as expected. In Fig.\ref{fig.zeritemp.tt} the
behaviour of $x_0$ is shown as a function of the temperature of the cooler
population and of the pressure ratio $P_2/P_1$ for a fiducial value of the
density $n_{e,2}=1\cdot10^{-3}$ cm$^{-3}$.

\section{The spatial distribution of the total SZ effect}

In this section we want to study the effect of the superposition of two distinct
electron populations on the spatial distribution of the total SZ effect in
galaxy clusters. In fact,  the thermal and the non-thermal populations have, in
general, a different spatial distribution as indicated by the spatial extension
of the X-ray emission and of the radio halo emission in many clusters (see
Giovannini et al. 2001).
 Also, there are indications that the EUV emission in some nearby clusters is
more extended than the thermal X-ray emission suggesting that there is a warm
gas and/or a population of relativistic electrons which are spatially more
extended than the X-ray emitting IC gas (Lieu et al. 1999).

Let us first consider a galaxy cluster in which  two different electron
populations are present: a thermal population, described by a relativistic
Maxwellian velocity distribution, and a non-thermal population, with an energy
spectrum described by a double power-law (see Eq.34). We consider these two
populations as independent and spatially superposed, consistently with our
analysis of Sect. 4. These assumptions are indeed reasonable since in the
cluster A2163, for instance, the radio halo emitting electrons occupy
approximately the same volume of the X-ray emitting, thermal electrons (see
Feretti et al. 2001) and, moreover, the relativistic electrons with $E_e \sim$ a
few GeV (which produce the radio-halo synchrotron emission) do not sensitively
affect the IC thermal electrons.

As for the spatial distribution of the thermal electrons we use here a
isothermal $\beta $-model (see, e.g., Cavaliere \& Fusco-Femiano, 1976),
according to which the temperature $T_e$ of the ICM is constant and the electron
density is described by a spherical distribution
\begin{equation}
\label{bmodelx} n_e(r)=n_{e0} \left( 1+ \frac{r^2}{r_c^2} \right)^{-3\beta/2} ~,
\end{equation}
where $n_{e0}$ is the central IC gas density, $r_c$ is the core radius of the
cluster and $\beta = \mu m_p v^2/k_B T_e$ is observed in the range $0.6 - 1$
(see Sarazin 1988 for a review). Under this assumption the angular dependence of
the optical depth of the thermal population is given by
\begin{eqnarray} \label{tau.theta.1pop}
\tau(\theta)&=&\sigma_T \int n_e(r) d\ell= \sigma_T n_{e0} \int
\left[ 1+ \left( \frac{r}{r_c} \right)^2 \right]^{-3\beta/2} d\ell
\nonumber \\
 & \equiv &n_{e0} \sigma_T r_c Y(\theta),
\end{eqnarray}
where $\theta$ is the angle between the centre of cluster and the direction of
observation. All the angular dependence is contained in the function
\begin{equation}
Y(\theta)=\sqrt{\pi} \frac{\Gamma( \frac{3}{2}\beta-\frac{1}{2})} {\Gamma(
\frac{3}{2}\beta)} \left[ 1+ \left( \frac{\theta}{\theta_c} \right)^2 \right]^{
\frac{1}{2} - \frac{3}{2} \beta} ~,
\end{equation}
where $\theta_c=r_c/D_A$, in terms of the angular diameter distance $D_A$ (see
Weinberg 1972).

In the following we will describe in details, for the sake of clarity, the
spatial distribution of the approximated (to first and second order in $\tau$)
expression of the SZ effect and we will give also the general expression for the
exact calculations.\\ The angular dependence of the SZ-induced spectral
distortion evaluated at first order in $\tau$ writes as
\begin{equation} \label{dist1_theta}
\Delta i(x,\theta)=\tau(\theta) [j_1(x)-j_0(x)] ~,
\end{equation}
where $\Delta i$ is in units of $ 2(k_B T_0)^3/(hc)^2$ and $\tau(\theta)$ is
given by Eq.(\ref{tau.theta.1pop}). Since the spectral distortion can be written
in the general form $\Delta i(x)=y \tilde{g}(x)$, the Comptonization parameter
writes as
\begin{eqnarray} \label{y_theta}
y(\theta)&=& \frac{\sigma_T}{m_e c^2}\int P d\ell= \nonumber \\ &=&
\frac{\sigma_T}{m_e c^2} \langle k_B T_e \rangle\int n_e(r) d\ell= \nonumber \\
&=& \frac{\langle k_B T_e \rangle}{m_e c^2} \tau(\theta)= \nonumber \\
&=&\frac{\sigma_T}{m_e c^2} P_0 r_c Y(\theta) ~,
\end{eqnarray}
where the quantity $\langle k_B T_e \rangle$ is defined in Eq.(\ref{temp.media})
and depends only on the parameters of the electron distribution (here the
central pressure is $P_0=n_{e0} \langle k_B T_e \rangle$).  The spectral shape
of the SZ effect at an angular distance $\theta$ from the cluster center can be
written, at first order in $\tau$, as
\begin{equation}
\tilde{g}(x)=\frac{\Delta i(x,\theta)}{y(\theta)}= \frac{\sigma_T n_{e0} r_c Y(\theta) }{ \frac{\sigma_T}{m_e c^2}
P_0 r_c Y(\theta)} [j_1(x)-j_0(x)]= \frac{m_e c^2}{\langle k_B T_e \rangle} [j_1(x)-j_0(x)].
\end{equation}
and it does not contain the angular dependence of the effect.
 However, at second order in $\tau$ the spectral distortion writes as
\begin{equation}
\Delta i(x,\theta)=\frac{1}{2} \tau^2(\theta) [j_2(x)-2j_1(x)+j_0(x)],
\end{equation}
and, as a consequence, the second order approximation of the function
$\tilde{g}(x)$ is
\begin{eqnarray} \label{g2_theta}
\tilde{g}(x,\theta)&=& \frac{1}{2} \frac{\sigma_T^2 n_{eo}^2 r_c^2 Y^2(\theta)
}{\frac{\sigma_T}{m_e c^2} P_0 r_c Y(\theta)} [j_2(x)-2j_1(x)+j_0(x)]= \nonumber
\\ &=& \frac{1}{2} \frac{\sigma_T m_e c^2 n_{e0}^2 r_c Y(\theta)}{P_0}
[j_2(x)-2j_1(x)+j_0(x)] ~,
\end{eqnarray}
which depends explicitly on $\theta$.
 It is important to notice that the angular dependence of the spectral
distortion is present both in the Comptonization parameter $y$ and in the
spectral shape $\tilde{g}(x)$. The dependence of the SZ effect from $\theta$ is
present at all orders $n>1$ of the approximations and hence in the full, exact
expression of the SZ effect. Thus, also for the study of the spatial
distribution of the total SZ effect, considering the first order approximation
is neither appropriate nor consistent.

For a combination of two electron populations, the total spectral distortion is
given, at first order approximation in $\tau$, by the sum of the single spectral
distortions of each population
\begin{equation}
\label{dist2pop1.theta}
\Delta i(x,\theta)=\Delta i_{th}(x,\theta) +\Delta i_{nonth} (x,\theta).
\end{equation}
This expression can be written more explicitly as
\begin{eqnarray} \label{dist2pop1.theta2}
\Delta i(x,\theta)&=&y_{th}(\theta) \tilde{g}_{th}(x)+y_{nonth} (\theta) \tilde{g}_{nonth}(x)= \nonumber \\
&=&\frac{\sigma_T}{m_e c^2}P_{0,th} r_{c,X} Y_{th}(\theta)
\tilde{g}_{th}(x)+ \nonumber \\ & &+\frac{\sigma_T}{m_e
c^2}P_{0,nonth} r_{c,rad} Y_{nonth}(\theta) \tilde{g}_{nonth}(x)=
\nonumber \\ &=&\frac{\sigma_T}{m_e
c^2}P_{0,th}[r_{c,X}Y_{th}(\theta) \tilde{g}_{th}(x)+ \nonumber \\
& &+ r_{c,rad} \bar{P} Y_{nonth}(\theta) \tilde{g}_{nonth}(x)]
\end{eqnarray}
where $\bar{P}= P_{0,nonth}/P_{0,th}$ is the pressure ratio at the cluster
center. We assume here that the spatial distribution of the thermal and of the
non-thermal electron populations are given by a $\beta$-model with core radii
$r_{c,X}$ and $r_{c,rad}$ and parameters $\beta_{X}$ and $\beta_{rad}$,
respectively.
Such an expression is also valid in the case of the combination of two thermal
populations. At higher approximation orders in $\tau$, the total spectral
distortion contains cross-correlation terms of the parameters of each population
and thus it is not possible to write it as a linear combination of the spectral
distortions of the separate electron populations. An exact derivation (or at
least the $3^{rd}$ order approximation in $\tau$) is required to describe
correctly the spatial dependence of the total SZ effect in galaxy clusters. The
general calculation of the spatial distribution of the total SZ effect can be
made using the exact expressions derived in Eqs.(21-23, 62 and 63) in which the
spatial dependence of the thermal and non-thermal electron populations are
considered.

\begin{figure}[tbp]
\vbox{
 \psfig{file=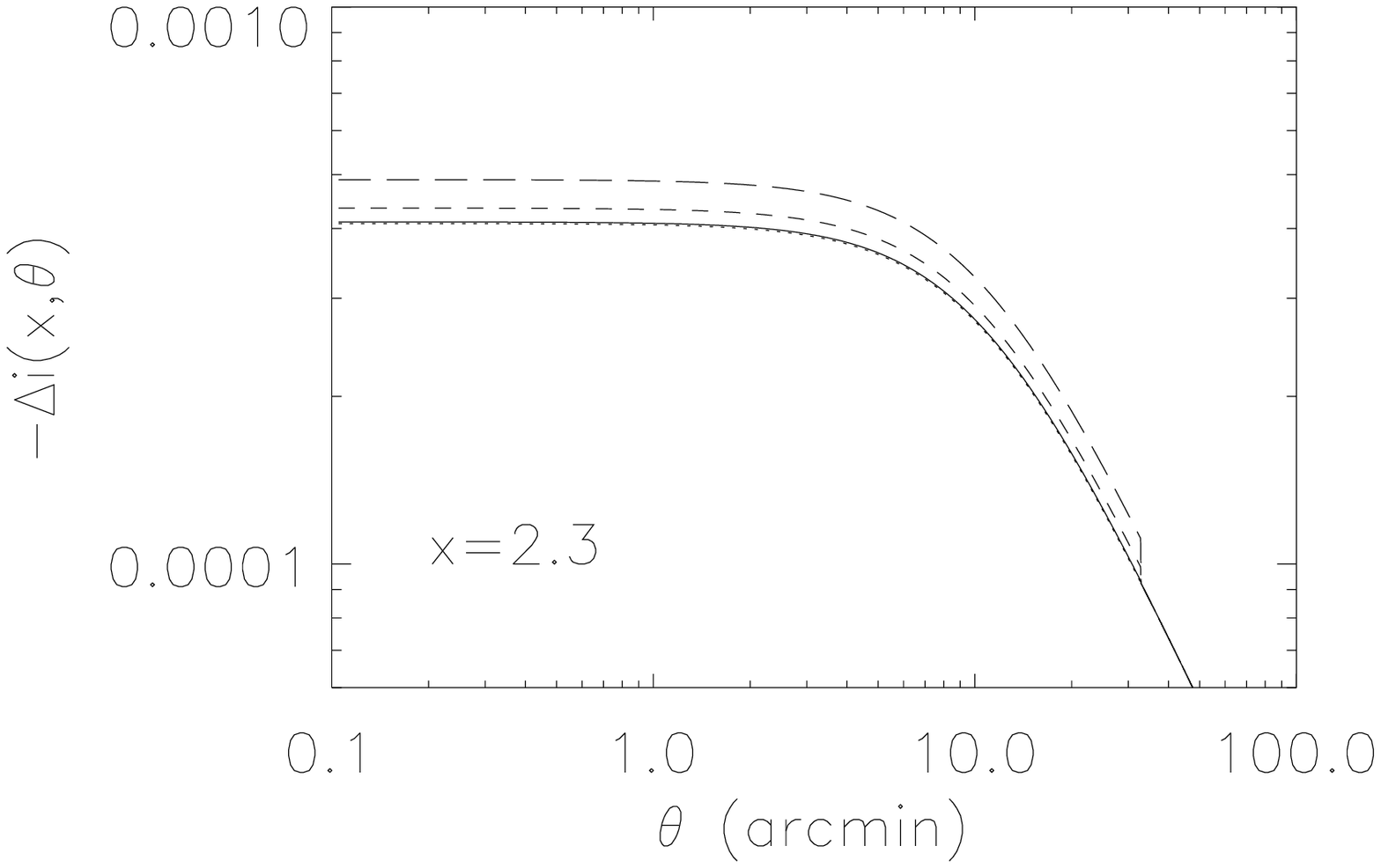,width=10.truecm,height=10.truecm}
 \psfig{file=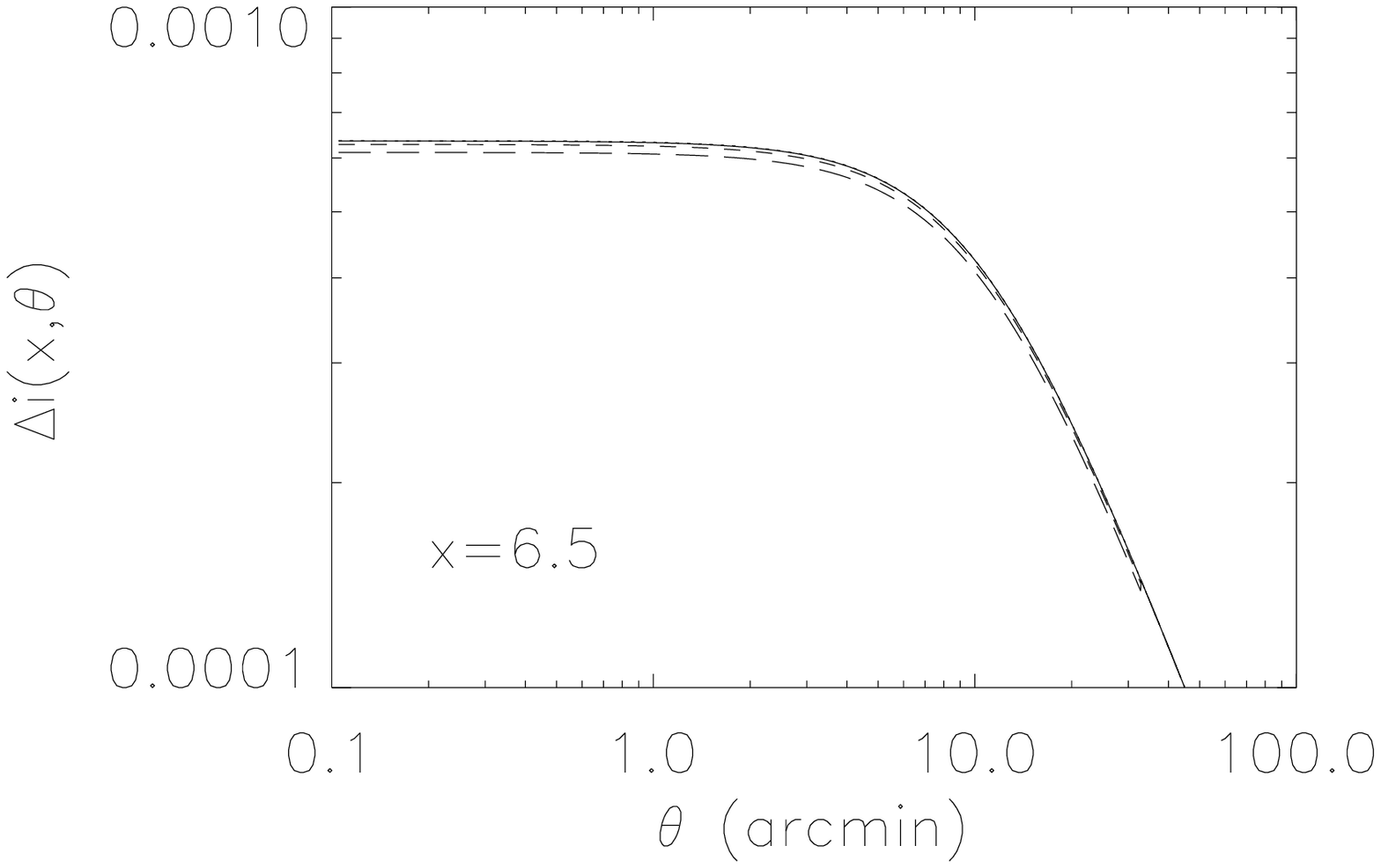,width=10.truecm,height=10.truecm}
}
  \caption{\footnotesize{The spatial dependence of the total SZ effect produced by the
  combination of the thermal and relativistic electrons in a Coma-like cluster.
  We show the thermal SZ effect (dots)
  and the total SZ effect with $\bar{P}
  =$ 0.05 (solid), 0.49 (short dashes) and 1.48 (long dashes).
  Calculations are done for an a-dimensional frequency of $x=2.3$ where
  the total SZ effect has its minimum value (upper panel) and for $x=6.5$ where
  the total SZ effect has its maximum value (lower panel).
  }}
  \label{fig.sztot_spatial}
\end{figure}

In Fig.\ref{fig.sztot_spatial} we show the spatial behaviour of the total SZ
effect in the case of a Coma-like cluster.
We use specifically the following parameters: $r_{c,X} = 0.42 \,h_{50}^{-1}$
Mpc, $r_{c,rad} =0.4 \,h_{50}^{-1}$ Mpc, $R_{X} = 4.2 \,h_{50}^{-1}$ Mpc,
$R_{halo} = 1.25 \,h_{50}^{-1}$ Mpc, $\beta _{X} =0.75$, $\beta _{rad} = 0.8$,
$n_{e0,X} = 2.89 \cdot 10^{-3} \;h_{50}^{2}$ cm$^{-3}$, $k_bT_e = 8.21$ keV. The
non-thermal electron population fitting the radio halo spectrum is taken with
spectral parameters $p_1 = 1000$ and $\alpha = 2.5$.\\
The total SZ effect shows a peculiar spatial behaviour which is particularly
evident at the frequencies $x \sim 2.3$ where it attains its minimum value and
at frequencies $x \sim 6.5$ close to its maximum. At $x \sim 2.3$ the spatial
profile of the total SZ effect is predicted to be higher than that due only to
the thermal population because the non-thermal SZ effect sums up in the region
of the radio halo which extends up to $\sim 1.3 h_{50}^{-1}$ Mpc (see
Fig.\ref{fig.sztot_spatial}). At larger distances from the cluster center, the
SZ effect is dominated by the thermal component and resemble its spatial
behaviour. An opposite behaviour is expected at frequencies higher than that of
the zero of the total SZ effect. For instance at $x \sim 6.5$
the spatial profile of the total SZ effect is lower than that of the single
thermal effect because the spectrum of the total SZ effect is lower than that of
the thermal case. So, a depression in the spatial profile of the total SZ effect
is expected at radii smaller than the radio halo extension $R_{halo}$. At
distances larger than the radio halo extension, the total SZ effect is again
dominated by the thermal component.
The amplitude of the non-thermal SZ contribution depends from the pressure ratio
$\bar{P} = P_{rel}/P_{th}$ as shown in Fig.\ref{fig.sztot_spatial}. Such spatial
features of the total SZ effect should be detectable with sensitive instruments
with good spatial resolution and narrow spectral frequency band. Such
instrumental capabilities can also be able to detect the transition between the
ambient thermal SZ effect and the non-thermal SZ effect expected in radio relics
and in the radio lobes of active galaxies found in the cluster environment.

\section{Application to specific galaxy clusters}

The non-thermal SZ effect may already have been observed as a part of the SZ
signal from some hot galaxy clusters which possess also non thermal phenomena
like radio halos or relics.\\
 In the Coma cluster, in fact, Herbig et al.
(1995) detected a strong SZ effect at the level of $y \sim 10^{-4}$ from an
observation at low frequencies $\nu_r \sim 30$ MHz. Since the Coma cluster has a
bright and extended radio halo (see, e.g., Deiss et al. 1997), it is well
possible that some fraction of the SZ signal might be contributed by the
non-thermal SZ effect. However, to disentangle the non-thermal SZ effect from
the thermal one a detailed spectral coverage with reasonably good sensitivity in
the frequency range from $x \sim 2.5 $ up to $x \simgt 8$ is needed, with
particular care to the region $x \sim 3.8-4$. At the time of the acceptance of
this paper, the available SZ observations for Coma at these relevant frequencies
are still lacking and we will discuss this case in more details in a further
paper (Colafrancesco \& Marchegiani 2002).\\
 In the following, nonetheless, we will discuss specifically the case of A2163 for
which there are observations at different frequencies which cover the
interesting part of the SZ spectrum and can be used to put constraints on the
presence and on the nature of the non-thermal SZ effect in this radio-halo
cluster.

\subsection{The case of A2163}

The cluster A2163 ($z =0.203$) is one of the hottest ($k_BT_e \sim 12.5$ keV)
clusters which possesses a giant radio halo (Herbig \& Birkinshaw 1994; Feretti
et al. 2001) with diameter size of $\sim (2.9 \pm 0.1) h^{-1}_{50}$ Mpc which is
centered on the peak of the X-ray emission. The slope of the power-law
synchrotron spectrum is $\alpha_r \approx 1.6 \pm 0.3$ and has been estimated
from radio data at $1.365$ and $1.465$ GHz (Feretti et al. 2001); this is
consistent with the results obtained by Herbig \& Birkinshaw (1994) in the
frequency range 10 MHz -- 10 GHz. There is, however, no evidence of hard X-ray
excess due to a non-thermal component in the BeppoSAX data of this cluster
(Feretti et al. 2001).

A strong SZ effect has been observed in A2163 at different frequencies. In
particular the SZ effect spectrum has been observed from BIMA at 28.5 GHz
(LaRoque et al. 2002), from DIABOLO at 140 GHz (Desert et al. 1998) and from
SuZIE at 140, 218 and 270 GHz (Holzapfel et al. 1997; these data are
dust-corrected in LaRoque et al. 2002) thus including and bracketing the null of
the thermal SZ effect. Considering a pure thermal SZ effect, the previous data
on A2163 are fitted with a Compton parameter $y_{th} = (3.65 \pm 0.40) \cdot
10^{-4}$ and with the addition of a kinematic SZ effect whose amplitude
corresponds to a positive peculiar velocity $V_p = 415^{+920}_{-765}$ km
s$^{-1}$ (Carlstrom et al. 2002) whose large uncertainties, however, makes it
consistent with a zero value. Note that, in principle, the addition of a
positive-velocity kinematic SZ effect contributes to provide a deeper negative
SZ signal in the region between the minimum and the zero of the SZ effect.

Here, we re-analyzed the data on A2163 trying to put constraints on the possible
presence of a non-thermal SZ effect by fitting the available data with a
combination of a thermal and non-thermal SZ effect.

The thermal population in A2163 has a temperature of $k_BT_e=12.4\pm 0.5$ keV
and a central density of $n_{e,th}\simeq6.82\cdot10^{-3}$ cm$^{-3}$. The
parameters describing the spatial distribution of the IC gas are $r_c=0.36 \,
h_{50}^{-1}$ Mpc and $\beta=0.66$ (Elbaz et al. 1995; Markevitch et al. 1996).
With these values the optical depth towards the cluster center is
$\tau_{0,th}=1.56\cdot10^{-2}$ and the central Comptonization parameter is
$y_{0,th}=3.80\cdot10^{-4}$.

\begin{figure}[htp]
\begin{center}
   \psfig{file=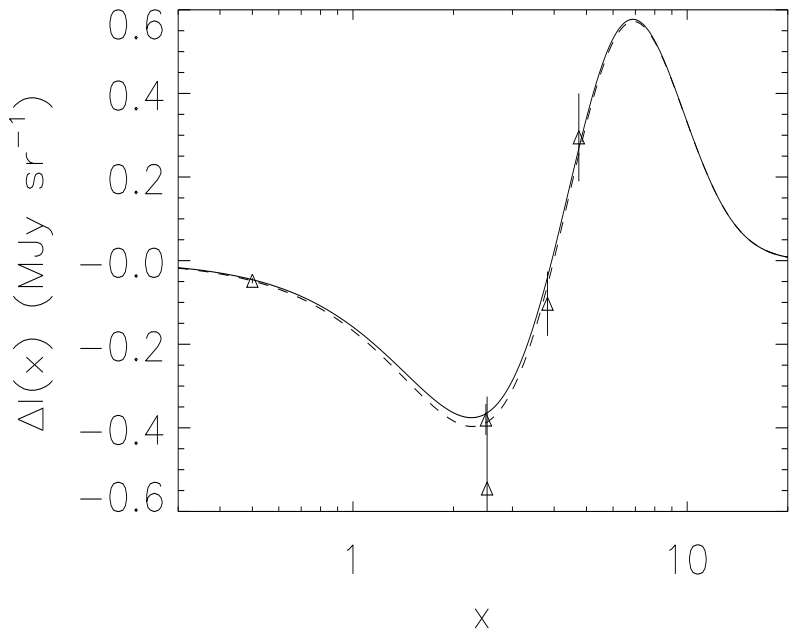,width=12cm,height=10cm}
  \caption{\footnotesize{Theoretical expectations for the spectrum of the SZ effect
  in A2163. We show the fit to the available data yielded by a thermal population
  (solid curve) and the expectations obtained from a combination of thermal and non-thermal populations
  with  $p_1=100$ for a value of the pressure ratio
  $P_{rel}/P_{th} = 0.29$ (dashed curve), which provides the best fit.
  }}
  \label{fig.a2163-2}
\end{center}
\end{figure}

We show in  Fig.\ref{fig.a2163-2} the relativistic, thermal SZ effect expected
for A2163 compared to the available data. The thermal SZ effect fits the data
with a $\chi^2=1.71$ and hence it is statistically acceptable. However, we will
show in the following that the inclusion of a non-thermal component of the SZ
effect is able to improve sensitively the fit to the data.

We consider a non-thermal population with a double power-law spectrum with
parameters $\alpha_1=0.5$, $\alpha_2=2.5$, $p_{cr}=400$, $p_2\rightarrow\infty$
and we set $p_1$ and the density \npt, with $\tilde{p}_1=100$, as a free
parameters. We also assume that the spatial distribution of the non-thermal
population is similar to that of the thermal population as indicated by the
extension of the radio halo in A2163 (see, e.g., Feretti et al. 2001). In our
calculations, we fix $p_1$ and we search for the value of \npt which minimizes
the $\chi^2$. We show in Table \ref{tab.a2163-2} and in Fig.\ref{fig.a2163-2}
the results of our calculations. We see that the data are better fitted with
values of $p_1$ which correspond to the electrons with $p_1\simgt 10^2$; we see
also that these electrons can produce a detectable non-thermal SZ effect with a
value of the pressure (\prapp$\sim0.3$) and the density  ratio $\bar{n}\sim270$
which do not imply a strong influence on the dynamical and thermal state of the
IC gas. A model in which $p_1=100$ and $n_{e,rel}(\tilde{p}_1)\approx 2.5 \cdot
10^{-5}$ cm$^{-3}$ best fits the data with a $\chi^2_{min} = 1.0534$, much lower
than the $\chi^2_{min}=1.71$ yielded by the single thermal population.
 The previous parameters point to a non-thermal population which carries a non negligible
 pressure contribution $P_{rel} \approx 0.29 P_{th}$ and which has a spectrum not
extended at momenta lower than $p_{min} \simeq 100$, corresponding to $E_{min}
\approx 50$ MeV.\\
 This result does not imply that the fit with a combination of thermal plus
non-thermal populations is statistically excluding the single thermal population
model. However, our analysis shows that: {\it i)} a non-thermal SZ effect is
produced by relativistic electrons producing radio halo emission has to be
present in A2163. The $\chi^2$ analysis indicates that its amplitude could be
appreciable (see Fig.27) and corresponds to a pressure in relativistic particles
$P_{rel} \approx 0.3 P_{th}$; {\it ii)} the possibility of having SZ
observations with better precision can offer the possibility to disentangle
between the thermal and any non-thermal component of the SZ effect; {\it iii)}
the detection of a non-thermal SZ effect can set strong constraints on the
nature of the non-thermal population and on its feedback on the thermal one.\\
 In these respects, it is appealing that the physical characteristics of the
 non-thermal population can be constrained through a detailed study of the SZ
 effect observed in the same galaxy cluster.
 Once the value of $p_1$ has been set, the quantity \npt can be constrained
 with a quite good precision and vice-versa.
 This example shows one of the potential uses of the SZ effect to obtain
 information on the properties of different electronic populations which are
 residing in the atmospheres of galaxy clusters.
\begin{table}[htb]
\begin{center}
\begin{tabular}{|*{5}{c|}}
\hline
 $p_1$ & \npt (cm$^{-3}$)& $\bar{P}$ & $\bar{n}$ & $\chi^2$\\
\hline
 0.1 & $1.657\cdot10^{-5}$ & 0.25 & 260 & 1.0536718\\
 1 & $1.664\cdot10^{-5}$ & 0.25 & 266 & 1.0536183\\
 10 & $1.793\cdot10^{-5}$ & 0.26 & 270 & 1.0534818\\
 100 & $2.527\cdot10^{-5}$ & 0.29 & 270 & 1.0534685\\
 1000 & $7.991\cdot10^{-4}$ & 0.38 & 270 & 1.0534679\\
 10000 & $2.527\cdot10^{-2}$ & 0.46 & 270 & 1.0534679\\
 \hline
 \end{tabular}
 \end{center}
 \caption{\footnotesize{For each value of $p_1$ are indicated
 the value of the density $n_{e,rel}( \tilde{p}_1=100)$ which provides the best fit and the
 corresponding values of pressure ($\bar{P}=P_{rel}/P_{th}$) and density
 ($ \bar{n}= n_{e,th}/n_{e,rel}$) ratio with the corresponding value of
 $\chi^2$ for the total SZ effect in A2163
 produced by a combination of a thermal population and a non-thermal double
 power-law population.
}}
 \label{tab.a2163-2}
 \end{table}

\begin{figure}[htp]
\begin{center}
   \psfig{file=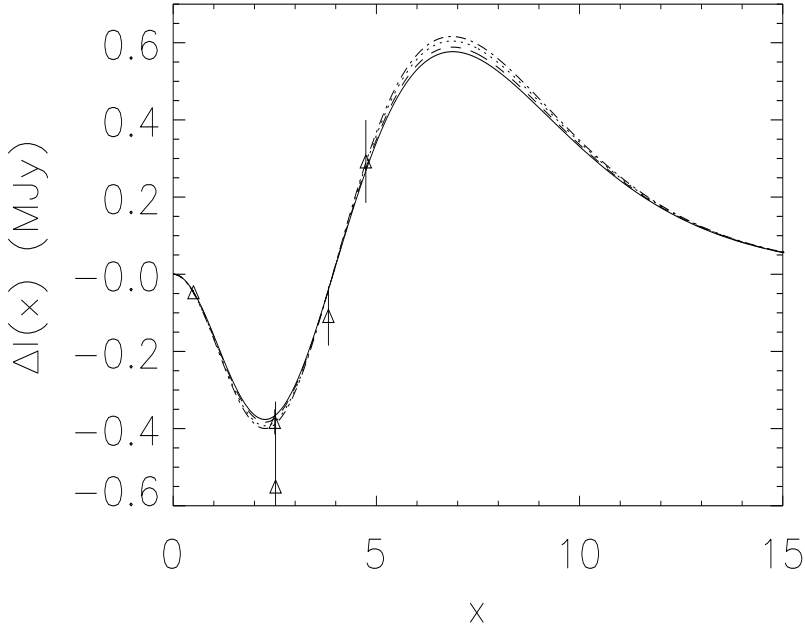,width=12cm,height=10cm}
  \caption{\footnotesize{The spectrum of the SZ effect in A2163 obtained from a
  combination of two thermal populations with $k_B T_{e,1}=12.4$ keV and
  $k_B T_{e,2}=0.5$ keV and with the same spatial distribution. We show the cases of
  $P_2/P_1$: 0 (solid line),
  $1.77\cdot10^{-2}$ (dashed line), $5.03\cdot10^{-2}$ (dotted line,
  which yields the best fit in this case),
  $5.91\cdot10^{-2}$ (dot-dashed line).}}
  \label{fig.a2163-3}
\end{center}
\end{figure}

\begin{figure}[htp]
\begin{center}
   \psfig{file=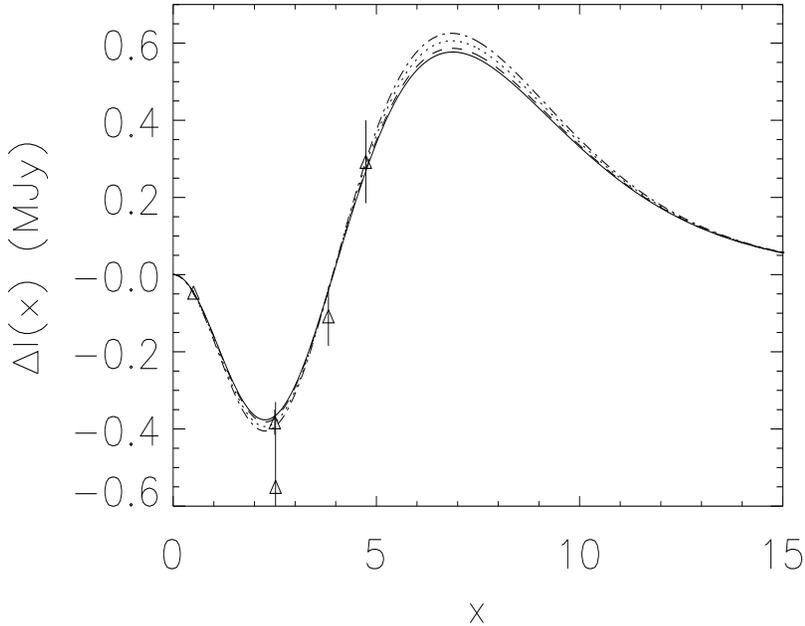,width=12cm,height=10cm}
  \caption{\footnotesize{The spectrum of the SZ effect in A2163 obtained from a
  combination of two thermal populations with $k_B T_{e,1}=12.4$ keV and
  $k_B T_{e,2}=0.5$ keV and with different spatial distributions as explained in the text. We show the cases of
  $P_2/P_1$: 0 (solid line),
  $5.91\cdot10^{-3}$ (dashed line), $2.09\cdot10^{-2}$ (dotted line, which yields the best fit),
  $2.96\cdot10^{-2}$ (dot-dashed line).}}
  \label{fig.a2163-4}
\end{center}
\end{figure}

Using our general approach described in the previous sections, we also
considered the calculation of the total SZ effect produced by a combination of
two thermal electron populations, with the warm population temperature in the
range $k_BT \sim 0.1-1$ keV and density in the range $n_{e,2} \sim
10^{-2}-10^{-4}$ cm$^{-3}$. We also considered two different spatial
distributions of the hot and warm populations: in the first case the two
populations have the same spatial extent and in the second one the cooler
population is spatially more extended than the hotter one. In this last case we
use parameters $r_{c,2}=1.5 \, r_{c,1}$ and  $\beta_2=0.5$.
\begin{table}[htb]
\begin{center}
\begin{tabular}{|*{5}{c|}}
\hline
 $k_BT_{e,2}$ (keV)& $n_{e,2}$ (cm$^{-3}$) & $P_2/P_1$ & $n_{e,1}/n_{e,2}$ & $\chi^2$\\
\hline
 0.1 & $3.922\cdot10^{-2}$ & $4.64\cdot10^{-2}$ & 0.17 & 1.5633934\\
 0.5 & $8.508\cdot10^{-3}$ & $5.03\cdot10^{-2}$ & 0.80 & 1.2395049\\
 1 & $4.222\cdot10^{-3}$ & $4.99\cdot10^{-2}$ & 1.61 & 1.2307970\\
 \hline
 \end{tabular}
 \end{center}
 \caption{\footnotesize{ For each value of $k_B T_{e,2}$ are indicated
 the value of the density $n_2$ which provides the best fit and the
 corresponding values of pressure and density
 ratio with the corresponding value of $\chi^2$ for the total SZ effect in A2163
 produced by a combination of a two thermal populations with the same spatial distribution.
}}
 \label{tab.a2163-3}
 \end{table}
The results of the fit to the data of A2163 are reported in
Table\ref{tab.a2163-3} and in Fig.\ref{fig.a2163-3}, and in
Table\ref{tab.a2163-4} and Fig.\ref{fig.a2163-4}. The available SZ observations
allow also in this case to set constraints on the parameters $n_{e,2}$ and
$T_{e,2}$ A best fit which minimizes the $\chi^2_{min}$ is obtained for values
$k_BT_{e,2}=0.5-1$ keV, while the case $k_BT_{e,2}=0.1$ keV seems to be
disfavoured by the data.

Finally, we want to remark that, in the set of models we considered in this
section, the best fit to the data (with a $\chi^2_{min} \approx 1.05$) is
obtained from a combination of thermal and non-thermal populations, a case which
seems to be favoured by the present data on A2163.
\begin{table}[htb]
\begin{center}
\begin{tabular}{|*{5}{c|}}
\hline
 $k_BT_{e,2}$ (keV)& $n_{e,2}$ (cm$^{-3}$) & $P_2/P_1$ & $n_{e,1}/n_{e,2}$ & $\chi^2$\\
\hline
 0.1 & $1.624\cdot10^{-2}$ & $1.92\cdot10^{-2}$ & 0.42 & 1.5911710\\
 0.5 & $3.531\cdot10^{-3}$ & $2.09\cdot10^{-2}$ & 1.93 & 1.2504161\\
 1 & $1.752\cdot10^{-3}$ & $2.07\cdot10^{-2}$ & 3.89 & 1.2412285\\
 \hline
 \end{tabular}
 \end{center}
 \caption{\footnotesize{ For each value of $k_B T_{e,2}$ are indicated
 the value of the density $n_2$ which provides the best fit and the
 corresponding values of pressure and density
 ratio with the corresponding value of $\chi^2$ for the total SZ effect in A2163
 produced by a combination of a two thermal populations with different spatial distributions.
}}
 \label{tab.a2163-4}
 \end{table}
Future SZ observations of A2163 with higher spectral resolution and sensitivity
will allow to set stronger constraints on the non-thermal population. In
particular, the possibility to measure accurately the frequency location of
$x_0$ could yield direct information on the value of the pressure in
relativistic particles, $P_{rel}$, confined in the cluster atmosphere and on
their energy distribution.

\section{Discussion and conclusions}

Since many galaxy clusters with available  SZ measurements show also the
presence of an extended radio halo and other non-thermal phenomena, it is
relevant to understand quantitatively the relevance of the non-thermal SZ effect
in galaxy clusters and to disentangle it from the true thermal SZ effect which
is extensively used for cosmological studies.
It is also relevant to use the limits on the non-thermal SZ effect to put
constraints on the spectrum of the non-thermal population which is responsible
for the high-energy phenomena occurring in galaxy clusters. This proves to be a
powerful and complementary analysis of the specific studies of non-thermal
phenomena carried out at other wavelenghts.

In such a context, we have shown in this paper that the spectra of the thermal
and non-thermal SZ effects are distinctly different and that the overall
spectrum of the SZ effects measures the energy densities in the thermal and in
the non-thermal electron populations, separately.
We have calculated here the spectral shape of the non-thermal SZ effect in the
Thomson limit using an exact approach which makes use of the full relativistic
formalism and goes beyond the Kompaneets and the single scattering
approximation. Moreover we evaluated, for the first time in a consistent way,
the total SZ effect arising from a combination of different electron populations
residing in the same galaxy cluster: we considered specifically the cases of
{\it i)} a thermal plus a non-thermal electron populations and {\it ii)} two
thermal populations with different temperatures and densities. Such a derivation
can be obtained, nonetheless, for an arbitrary set of different electron
populations.

We have shown {in our approach} that the spectral distortion of the CMB due to
both the thermal and the non-thermal SZ effect can be written in a general,
single form which depends separately on the Comptonization parameter and on the
spectral shape of the specific SZ effect. Such an expression is valid for both
thermal and non-thermal SZ effects produced by single electron populations as
well as for the overall SZ effect produced by a combination of different
electron populations. We have also shown that the spectral dependence of the SZ
effect, contained in the function $\tilde{g}(x)$, cannot be separated from the
dependence on the cluster parameters even at the first order approximation in
$\tau$, where $\tilde{g}(x)$ depends on the cluster (effective) temperature. At
higher-order approximation in $\tau$, $\tilde{g}(x)$ depends, moreover, also on
$\tau$. In the exact approach, the spectral shape of the SZ effect depends
strongly on the cluster parameters themselves and specifically from the pressure
and the optical depth of the considered electron population.

We have calculated in details the precision at which each approximation order in
$\tau$ reproduces the exact form of the SZ effect: we have shown that the third
order approximation in $\tau$ yields a precision $\simlt 1 \%$ at each
interesting frequency for both the non-thermal and the thermal SZ effect.\\
We have shown that the main spectral differences for the non-thermal SZ effect
with respect to the thermal SZ effect are a maximum of the spectral shape which
is moved to higher and higher frequencies for increasing values of the lower
cutoff in the momentum distribution, $p_1$. As a consequence, also the null of
the non-thermal SZ effect, $x_0$, is moved to higher frequencies for increasing
values of $p_1$. Since the pressure $P_{rel}$ of the non-thermal electron
distribution depends on $p_1$, the position of $x_0$ decreases with increasing
values of $P_{rel}$ for a fixed normalization of the non-thermal electron
density.\\
We have shown in this paper that the shape of the exact non-thermal SZ effect is
different from the 1$^{st}$ order approximation estimate used so far by
different authors (Birkinshaw 1999, Ensslin \& Kaiser 2000, Blasi et al. 2000,
Shimon \& Rephaeli 2002). We have shown that the 3$^{rd}$ order approximation is
required to reproduce the exact results within a $\sim 1 \%$ precision at all
relevant values of $x$.\\
We have also shown that the SZ effect from a  single non-thermal population in
clusters with non-thermal phenomena is not a realistic description of the
effect. So, we have derived for the first time a consistent expression for the
total SZ effect produced by a combination of a thermal and a non-thermal
populations, like is the realistic case in many galaxy clusters. The shape and
the amplitude of the total SZ effect is quite different from that suggested in
previous papers (Ensslin \& Kaiser 2000, Blasi et al. 2000). The main difference
in the derivation of the overall SZ effect is that it is not just given by the
sum of the separate thermal and non-thermal contributions (as is the case for
the first order approximation in $\tau$) but it is given by a properly weighted
(according to the optical depths of each separate population) combination of the
{thermal and non-thermal} contributions, as discussed in Sect. 4. Also, we
considered in our approach the effect of multiple scattering and higher ($n>1$)
order approximation terms in the analytical derivation of the overall SZ effect.
Moreover, we also discussed not only the case of {\it i)} a single power-law
electron spectrum (as was done also by the previous authors) but also the more
consistent case of a {\it ii)} double power-law electron spectrum which is able
to fit both the radio halo and the hard X-ray spectra in clusters. We showed
that, while in the case {\it i)} low-energy electrons are the dominant source of
non-thermal Compton scattering and produce an increase of SZ signal in the
region $x \simgt 7$ (where the thermal SZ effect has its maximum), in the case
{\it ii)} high-energy electrons are the dominant source of non-thermal SZ effect
and produce very large shifts of CMB photon frequencies producing a deprivation
of photons also in the region of the maximum of the thermal SZ effect ($x \sim
7$).

A relevant result of our analysis shows that the location of the zero of the
total SZ effect depends on the pressure ratio $\bar{P}= P_{rel}/P_{th}$ between
the relativistic and thermal pressures of the two different electron
populations. It increases non-linearly with $\bar{P}$ up to values $x_0 \sim
4.2$ for $\bar{P}$ up to values $\approx 1.5$. This is a unique and relevant
feature of the non-thermal SZ effect in clusters since it yields, in principle,
a direct measure of the total pressure in relativistic non-thermal particles in
the cluster atmosphere, an information which is not easily accessible from the
study of other non-thermal phenomena like radio halos and/or EUV or hard X-ray
emission excesses. Thus, the detailed observations of the frequency shift of
$x_0$ in galaxy clusters provides unambiguously a constraint to the relativistic
particle content in the cluster atmosphere. In addition, such a measurement is
crucial to determine the true amount of kinematic SZ effect in clusters since it
is usually estimated from the residual SZ signal at the location of the zero of
the thermal (relativistic) SZ effect. Due to the steepness of the SZ spectral
shape in the region of the null of the overall SZ effect,  the possible
additional non-thermal SZ signal must be determined precisely in order to derive
reliable limits for the kinematic SZ effect. This fact can severely limit the
possibility to measure effectively the kinematic SZ effect in radio-halo
clusters. An accurate subtraction of the non-thermal SZ effect needs a specific
observational strategy with measurements in at least three frequency ranges $x
\sim 2-3$, $3.8 - 4.5$ and $\sim 6-8$.

We verified that our approach is consistent with the exact derivation of the
monochromatic redistribution function containing both the relativistic and
quantum corrections already given by Fargion et al. (1997) and Sazonov \&
Sunyaev (2000) in the Thomson limit and in the single scattering limit. In the
Thomson limit, nonetheless, we have generalized the full derivation of the SZ
effect to the case of multiple scattering and to any approximation order in
$\tau$.

After the submission of our paper, Shimon \& Rephaeli (2002) presented a
computation of the SZ effect produced by non-thermal electron populations in a
few clusters like Coma and A2199. However, we want to stress that their
derivation of the non-thermal SZ effect is still approximate since they consider
only the first order approximation in $\tau$ (in which case the total SZ effect
is the sum of the thermal and non-thermal effects) and the single scattering
limit. We have shown in our paper that such an approximated description is
neither complete nor precise since at least the third order approximation
calculation in $\tau$ is required to reproduce the exact derivation of the total
SZ effect within $\sim \%$ accuracy level.

We have also generalized our derivation of the total SZ effect in galaxy
clusters to the case of a combination of different thermal electron populations
(see Sect. 5). Any additional cool IC gas component produces a tightening of the
photon redistribution function, an increase in the total optical depth and hence
a substantial change in the spectral distortion at the minimum and at the
maximum of the SZ effect. The location of the zero of the SZ effect $x_0$ also
decreases in frequency due to the presence of the cooler component which
decreases the pressure ratio $P_2/P_1$. Thus, the possible detection of an
additional cold component in the cluster atmosphere through observations of the
total (thermal plus thermal) SZ effect allows to test the possible thermal
origin of the EUV excess observed in several nearby clusters (Lieu et al. 1999,
2000). Our analysis of the cluster A2163 does not indicate a relevant role of
the warmer component to the total pressure of the cluster atmosphere,
consistently with the present indication of the absence of a strong EUV excess
in this cluster. We will address the specific analysis of other clusters
elsewhere.

Beyond the relevance of the study of the non-thermal SZ effect as a bias for the
cosmologically relevant thermal SZ effect, the non-thermal SZ effect has also a
crucial astrophysical relevance as a barometer for the presence of any
additional population of electrons with both a non-thermal or a thermal energy
spectrum. In fact, the non-thermal SZ effect actually measures the total
pressure of the non-thermal electron population and hence yields constraints to
its energy spectrum. Analogously, the additional thermal SZ effect produced by a
cooler thermal component provides information on its temperature and density.
Specifically, we have shown that combining information from the cluster
radio-halo observation and non-thermal SZ measurements, one is able to determine
the shape and the extension of the relativistic electron spectrum. Using the
available observation of one of the best example of radio-halo clusters with
multifrequency observations of the SZ effect, A2163, we applied our method and
derived a limit on the low-energy cutoff of the relativistic electron spectrum
of $E_{min} \sim 50$ MeV, as well as constraints on the momentum spectrum which
is  required to be flat enough ($f_e \sim p^{-0.5}$) below $p \sim 400$ to avoid
destructive feedback effects on the thermal IC gas.

Relativistic electrons with such flat energy spectrum do not produce a relevant
extra heating and/or extra X-ray emission with respect to the thermal IC gas in
A2163. Also the IC energy losses of such relativistic electrons do not yield a
substantial hard X-ray emission, in agreement with the available limit on A2163
obtained from BeppoSAX observations.

We have finally shown that the presence of a non-thermal SZ effect also
influence the spatial profile of the total SZ effect of a typical radio-halo
cluster. In fact, the region occupied by the radio halo shows an increment
(decrement) of the signal at frequencies near the minimum (maximum) of the SZ
effect. This happens as a consequence of the difference in the spectral shapes
of the non-thermal SZ effect with respect to the thermal one.

The specific spectral and spatial features of the non-thermal SZ effect makes it
possible to detect it through a multifrequency observation with high sensitivity
and narrow-band detectors. The optimal observational strategy is to observe
galaxy clusters in the frequency range $x \sim 2 - 8$ where the peculiar
spectral features allow clearly to disentangle the non-thermal SZ effect from
the thermal one. The PLANCK surveyor experiment has the capabilities to detect
and map the non-thermal SZ effect in a large number of nearby radio-halo
clusters. However, dedicated experiment with high sensitivity and narrow band
spectral coverage are also adequate to detect the non-thermal SZ effect in
radio-halo galaxy clusters.

\vskip 1.truecm \noindent {\bf Acknowledgments.} We thank the Referee, N. Itoh,
for several useful comments and suggestions which allow us to improve the
presentation of our results.

\begin{appendix}
\section{The total SZ effect for a combination of two electron populations:
analytic approximations}

In this Appendix, we derive detailed expressions for $I_{tot}(x)$ which are
approximated at first and second order in $\tau$ according to the general
expression:
\begin{equation} \label{I2}
I(x)=I_0(x)+\tau[J_1(x)-J_0(x)]+\frac{1}{2}\tau^2[J_2(x)-2J_1(x)+J_0(x)].
\end{equation}
We remind the reader that the function $J_1(x)$ is given by Eq.(18):
\begin{equation} \label{j1}
J_1(x)=\int_{-\infty}^{+\infty} I_0(xe^{-s}) P_1(s) ds,
\end{equation}
where $P_1(s)$ is given by Eq.(\ref{p1s}):
\begin{equation}
P_1(s)=\int_0^\infty dp f_e(p) P_s(s;p).
\end{equation}
Inserting Eq.(\ref{ftot}) in the previous Eq. for $P_1(s)$, we obtain:
\begin{eqnarray} \label{p1stotapp}
P_1(s)&=&c_A \int_0^\infty dp f_A(p) P_s(s;p)+ c_B \int_0^\infty dp f_B(p) P_s(s;p)
\nonumber \\ & \equiv & c_A
P_{1A}(s) + c_B P_{1B}(s).
\end{eqnarray}
Inserting Eq.(\ref{p1stotapp}) in Eq.(\ref{j1}), we also obtain the expression:
\begin{eqnarray} \label{j1tot}
J_1(x)&=&c_A\int_{-\infty}^{+\infty} I_0(xe^{-s}) P_{1A}(s) ds+ c_B \int_{-\infty}^{+\infty} I_0(xe^{-s})
P_{1B}(s) ds \nonumber
\\
  & \equiv & c_A J_{1A}(x) + c_B J_{1B}(x).
\end{eqnarray}
Inserting eqs.(\ref{j1tot}), (\ref{tautot}) and (\ref{coeffab}) in
Eq.(\ref{I2}), we derive, at first order in $\tau$, the expression:
\begin{eqnarray} \label{ord1tot}
\Delta I_{tot}^{(1)}(x)&=& (\tau_A + \tau_B)[c_A J_{1A}(x)+c_B J_{1B}(x)-J_0(x)] \nonumber \\
 & =&\tau_A J_{1A}(x)+ \tau_B J_{1B}(x)-\tau_A J_0(x)-\tau_B
 J_0(x) \nonumber \\
 &=&\tau_A[J_{1A}(x)-J_0(x)]+\tau_B[J_{1B}(x)-J_0(x)].
 \end{eqnarray}
As expected, the SZ effect at first order in $\tau$ is given
by the sum of the separate SZ effects produced by the
two electron distributions, respectively.

\noindent To evaluate the effect at second order in $\tau$,
it is necessary to calculate the expression:
 \begin {equation} \label{j2}
 J_2(x)=\int_{-\infty}^{+\infty} I_0(xe^{-s}) P_2(s) ds.
 \end{equation}
The function $P_2(s)$ can be derived by Eq.(\ref{pn.generica}) using Eq.
(\ref{p1stotapp}):
\begin{eqnarray} \label{p2stot}
P_2(s)&=&P_1(s)\otimes P_1(s) \nonumber \\ &=&[c_A P_{1A}(s)+c_B P_{1B}(s)]
\otimes [c_A P_{1A}(s)+c_B P_{1B}(s)] \nonumber \\ &=&c_A^2 P_{1A}(s)\otimes
P_{1A}(s)+2c_Ac_B P_{1A}(s) \otimes P_{1B}(s) + c_B^2 P_{1B}(s) \otimes
P_{1B}(s) \nonumber \\ & \equiv & c_A^2 P_{2AA}(s)+2c_Ac_B P_{2AB}(s)+c_B^2
P_{2BB}(s).
\end{eqnarray}
Inserting this expression in Eq.(\ref{j2}), we obtain:
\begin{eqnarray} \label{j2tot}
J_2(x)&=&c_A^2 \int_{-\infty}^{+\infty} I_0(xe^{-s}) P_{2AA}(s) ds+2c_A c_B
\int_{-\infty}^{+\infty} I_0(xe^{-s}) P_{2AB}(s) ds+ \nonumber \\ & &+ c_B^2
\int_{-\infty}^{+\infty} I_0(xe^{-s}) P_{2BB}(s) ds \nonumber \\ &\equiv& c_A^2
J_{2AA} (x) +2c_Ac_B J_{2AB}(x)+c_B^2 J_{2BB}(x).
\end{eqnarray}
From Eq.(\ref{I2}) we derive the expression of the second order correction to
the total distorted spectrum
\begin{eqnarray} \label{ord2tot}
\Delta I_{tot}^{(2)}(x)&=&\frac{1}{2}(\tau_A+\tau_B)^2
[c_A^2J_{2AA}+2c_Ac_BJ_{2AB}+c_B^2J_{2BB} -2c_AJ_{1A}+\nonumber
\\
 & &-2c_B J_{1B}+J_0]= \nonumber \\
&=&\frac{1}{2}\{\tau_A^2 J_{2AA} +2\tau_A\tau_B J_{2AB} -2\tau_A(\tau_A+\tau_B)
J_{1A} -2\tau_B(\tau_A+\tau_B) J_{1B}+\nonumber \\
 & &+(\tau_A^2+2\tau_A\tau_B+\tau_B^2)J_0\}=\nonumber\\
 &=&\frac{1}{2}\{\tau_A^2[J_{2AA}-2J_{1A}+J_0]+
 \tau_B^2[J_{2BB}-2J_{1B}+J_0]+\nonumber\\
  & &+2\tau_A\tau_B[J_{2AB}-J_{1A}-J_{1B}+J_0]\}.
\end{eqnarray}
Notice that at second order in $\tau$ there is an additional term describing the
probability that a CMB photon can suffer first a scattering from the electrons
of the distribution $f_A$ and then another scattering from the electrons of the
distribution $f_B$.

In a similar approach, we can derive the further corrections to the total
distorted spectrum evaluated at any order in $\tau$; these will contain a larger
and larger number of cross-scattering terms. In a completely analogous way, we
can also consider the case of three and more electron distributions.

\end{appendix}

\end{document}